\pdfoutput=1
\documentclass[aip, amsmath,amssymb,reprint]{revtex4-1}
\usepackage[latin9]{inputenc}
\usepackage[active]{srcltx}
\usepackage{color}
\usepackage{verbatim}
\usepackage{amsmath}
\usepackage{amssymb}
\usepackage{graphicx}
\usepackage{esint}
\usepackage[unicode=true,pdfusetitle,
 bookmarks=true,bookmarksnumbered=false,bookmarksopen=false,
 breaklinks=true,pdfborder={0 0 0},pdfborderstyle={},backref=false,colorlinks=true]
 {hyperref}
\hypersetup{
 colorlinks,linkcolor=red,citecolor=blue}
\usepackage{breakurl}

\makeatletter

\providecommand{\tabularnewline}{\\}

\usepackage{amsfonts}

\makeatother

\begin{document}

\title{High-acoustic-index-contrast phononic circuits: numerical modeling}

\author{Wance Wang}
 \altaffiliation[Now at ]{Department of Physics, University of Maryland, College Park.}
\author{Mohan Shen}
\author{Chang-Ling Zou}
\affiliation{Department of Electrical Engineering, Yale University, New Haven, Connecticut 06511, USA\looseness=-1}
\affiliation{Depratment of Optics, University of Science and Technology of China, CAS, Hefei, Anhui 230026, China\looseness=-1}
\author{Wei Fu}
\affiliation{Department of Electrical Engineering, Yale University, New Haven, Connecticut 06511, USA\looseness=-1}
\author{Zhen Shen}
\affiliation{Department of Electrical Engineering, Yale University, New Haven, Connecticut 06511, USA\looseness=-1}
\affiliation{Depratment of Optics, University of Science and Technology of China, CAS, Hefei, Anhui 230026, China\looseness=-1}
\author{Hong X. Tang}%
 \email{hong.tang@yale.edu}
\affiliation{Department of Electrical Engineering, Yale University, New Haven, Connecticut 06511, USA\looseness=-1}

\date{\today}

\begin{abstract}
\rightskip=\leftskip
We numerically model key building blocks of a phononic integrated circuit that enable phonon routing in high-acoustic-index waveguides. Our particular focus is on Gallium Nitride-on-sapphire phononic platform which has recently demonstrated high acoustic confinement in its top layer without the use of suspended structures. We start with systematic simulation of various transverse phonon modes supported in strip waveguides and ring resonators with sub-wavelength cross-section. Mode confinement and quality factors of phonon modes are numerically investigated with respect to geometric parameters. Quality factor up to $10^{8}$ is predicted in optimized ring resonators. We next study the design of the phononic directional couplers, and present key design parameters for achieving strong evanescent couplings between modes propagating in parallel waveguides. Last, interdigitated transducer electrodes are included in the simulation for direct excitation of a ring resonator and critical coupling between microwave input and phononic dissipation. Our work provides comprehensive numerical characterization of phonon modes and functional phononic components in high-acoustic-index phononic circuits, which supplements previous theories and contributes to the emerging field of phononic integrated circuits. 
\end{abstract}

\maketitle

\section{Introduction}

Surface acoustic wave (SAW) devices are widely used in electronic circuits, finding applications such as filters and oscillators in communication devices \cite{campbell1998surface} and a range of sensing applications \cite{Lao1980,Friend2011,Campbell1989}. Recently, SAWs are also exploited for coherent control of various quantum systems, including the electron quantum dots \cite{Mcneil2011,Hermelin2011,Chen2015}, electron spins in diamond \cite{Golter2016}, superconducting qubits \cite{Gustafsson2014,Satzinger2018n,Bienfait2019s}, and integrated photonic devices \cite{Fuhrmann2011,Tadesse2014,Uan2015,Tadesse2015}. Thus, SAW provides a promising platform for hybrid quantum systems \cite{Kurizki2015,Schuetz2015,Shumeiko2016,Clerk2020np}, where the traveling SAW phonon can serve as a quantum bus to facilitate quantum state transfer. 

The coupling strength between SAW and matter/photons is enhanced with reduced mode volume of the SAW. The enhanced interactions such as the strong coupling regime of cavity quantum acoustodynamics \cite{Manenti2017nc,Moores2018prl} not only increases coherent coupling rates between quantum bits, but also improve the sensitivity of SAW-based measurements \cite{Noguchi2017prl}. Also, the full potential of phononic systems can be only unraveled when maneuverability of phonon becomes comparable to its electrical and optical counterparts. Therefore, a strong confinement of the SAW to the diffraction limit and the long lifetime phononic resonators are in high demand. To that end, phonon waveguides \cite{Yen1972,Oliner1976,Oliner1978} and ring resonators \cite{Knox1970,Sandy1976} based on SAW were proposed and experimentally studied in the 1970s. However, in the following several decades, the experiments and detailed theoretical studies of the confinement of itinerant phonons in microstructures were less active, especially for the phonon resonators of radiating waves in nature. Only in recent years, due to the advances of nano-fabrication technologies, ultra-low loss phononic waveguides and resonators are achieved with with bulk acoustic wave \cite{Goryachev2014aipcp,Carvalho2019apl} and with SAW \cite{Biryukov2007,Biryukov2009,Maznev2009,Manenti2016,Liu2009,Boucher2014,wg_Fan2016,wg_Mohammadi2011,wg_PhysRevLett.121.040501,wg_PnC_fang2016optical,wg_suspended_wg_hatanaka2014phonon,Xu2018apl}, pushing the study of phononic devices, including SAW-related waveguides and resonators back to the frontier of researches as a strong contender for advanced phononic circuits. 

Interdigital transducers (IDTs) convert the electrical RF signal into SAW or \textit{vice versa} by employing the piezoelectric materials. In most applications, IDTs are fabricated on a uniform film, and the lateral size of IDT is much larger than the wavelength of SAW to excite and collect the quasi-collimated SAW. As a primary source or receiver of SAWs, the scaling of IDTs and their integration with other SAW devices is also in critical demand in the development of phononic circuits. 

A variety of piezoelectric material platforms have been explored for
SAW devices, including ST-X quartz \cite{Manenti2017nc, Noguchi2017prl},
zinc oxide \cite{Golter16prl, Huang10apl, Magnusson15apl}, lithium
niobate \cite{Satzinger2018n, Ung17lc, Shao19pra}, lithium tantalate
\cite{Fu17apl}, aluminum nitride \cite{Chu17s, Aubert10apl, Tadesse14nc, Fujii13ituffc}, gallium arsenide \cite{Gustafsson2014, Moores2018prl, Okamoto13np, Metcalfe10prl}, and gallium nitride \cite{Xu2018apl, Valle19apl, Wang1522iicmemsm, Fu2019nc}. Together with substrate material such as sapphire, diamond, and silicon, various combinations of high-acoustic-index-contrast
SAW devices can provide better confinement and thus low loss, as well as the dispersion can be engineered. Table \ref{tab:material} summarizes common piezoelectric materials and substrates. Among these piezo-materials,
$\mathrm{LiNbO_{3}}$ and $\mathrm{LiTaO_{3}}$ possess much larger
electromechanical coupling coefficients ($K^{2}$) than other materials,
but they have poor temperature stabilities, whereas quartz is in a reverse
case with weak coupling but good temperature stability at room temperature
\cite{SAW-book}. $\mathrm{ZnO}$ has wide bandgap \cite{Pang13saap}
and good film stoichiometry \cite{Fu17pms}  in addition to relatively
large $K^{2}$ \cite{Fu19apl}. $\mathrm{ZnO/diamond}$ is also a quantum
interface between SAW and nitrogen-vacancy center in diamond \cite{Golter16prl}.
$\mathrm{AlN}$ has high refractive index and high phase velocity
\cite{Fu2019nc, Deger98apl} which is in favor of high-frequency devices, but in turn, the contrast of acoustic index will be low. It also finds applications in high-temperature environments \cite{Aubert10apl}. Semiconductor
material $\mathrm{GaN}$ and $\mathrm{GaAs}$ are compatible with
standard wafer-scale fabrication, $\mathrm{GaAs}$ is also naturally
suited for $\mathrm{InGaAs}$ quantum dots \cite{Metcalfe10prl} which
benefit from the telecom range. The availability of $\mathrm{GaN}$ high-performance amplifiers enables monolithic integration with devices.
For the layered system to our knowledge, $\mathrm{ZnO/diamond}$,
$\mathrm{ZnO/sapphire}$, $\mathrm{ZnO/SiO_{2}}$ \cite{Nakahata03sst,
Golter16prl, Huang10apl}, $\mathrm{AlN}/\mathrm{sapphire}$, $\mathrm{AlN}/\mathrm{SiO_{2}}$
\cite{Aubert10apl, Tadesse14nc}, and $\mathrm{GaN/sapphire}$ \cite{Fu2019nc}
have been explored. Diamond the highest known SAW phase velocity but
its fabrication requires chemical vapour deposition \cite{Nakahata03sst},
and it suffers from larger propagation loss. The contrast of $\mathrm{SiO_{2}}$-based
platform isn't high. Concluding all factors, $\mathrm{GaN/sapphire}$
platform stands out as a high-performance and ready-access choice.

\begin{table}
\begin{tabular}{c|ccc}
\hline 
Material & $V_{l}$ ($\mathrm{m\cdot s^{-1}}$) & $V_{t}$ ($\mathrm{m\cdot s^{-1}}$) & Piezoelectricity\tabularnewline
\hline 
Quartz \cite{Pohl02rmp} & 5700 & 3158 & Weak\tabularnewline
$\mathrm{ZnO}$ \cite{Azuhata03jap} & 5790 & 2700 & Strong\tabularnewline
$\mathrm{LiNbO_{3}}$ \cite{Warner67jasa} & 7316 & 4795 & Strong\tabularnewline
$\mathrm{AlN}$ \cite{GaN-Poisson+Youngs} & 10,169 & 6369 & Good\tabularnewline
$\mathrm{GaN}$ \cite{GaN-Poisson+Youngs} & 7350 & 4578 & Good\tabularnewline
Diamond \cite{Flannery03sst} & 18,000 & 12,000 & No\tabularnewline
Sapphire \cite{Auld1973} & 10,658 & 5796 & No\tabularnewline
$\mathrm{SiO_{2}}$ \cite{Hopcroft10jms} & 8433 & 5843 & No\tabularnewline
\hline 
\end{tabular}

\caption{\label{tab:material}SAW waveguide and substrate materials. $V_{l},\,V_{t}$ are longitudinal and transverse wave speed, respectively.}
\end{table}

In this paper, we numerically investigate the properties of integrated phononic waveguides and resonators based on GaN-on-sapphire platform. Different from $\Delta V/V$ confinement effect induced by the metal electrodes \cite{Biryukov2009,Biryukov2007,Maznev2009,Manenti2016,Hughes1972}, we consider the pure geometric confinement of phonon on the surface of the chip, thus the metal-induced loss is eliminated.  We prove that the phononic ring resonator can have exceptionally high quality factors ($Q$) with the radius of tens of microns, in which the whispering gallery modes (WGMs) can be excited either by the evanescent acoustic field through the phononic waveguide coupling or by direct IDT excitation. These waveguide and resonator structures will be basic elements in future phononic integrated circuits of both fundamental and practical interests \cite{Fu2019nc,shen2017rosi}.

\section{Waveguide}

\begin{figure}
\begin{centering}
\includegraphics[width=1\columnwidth]{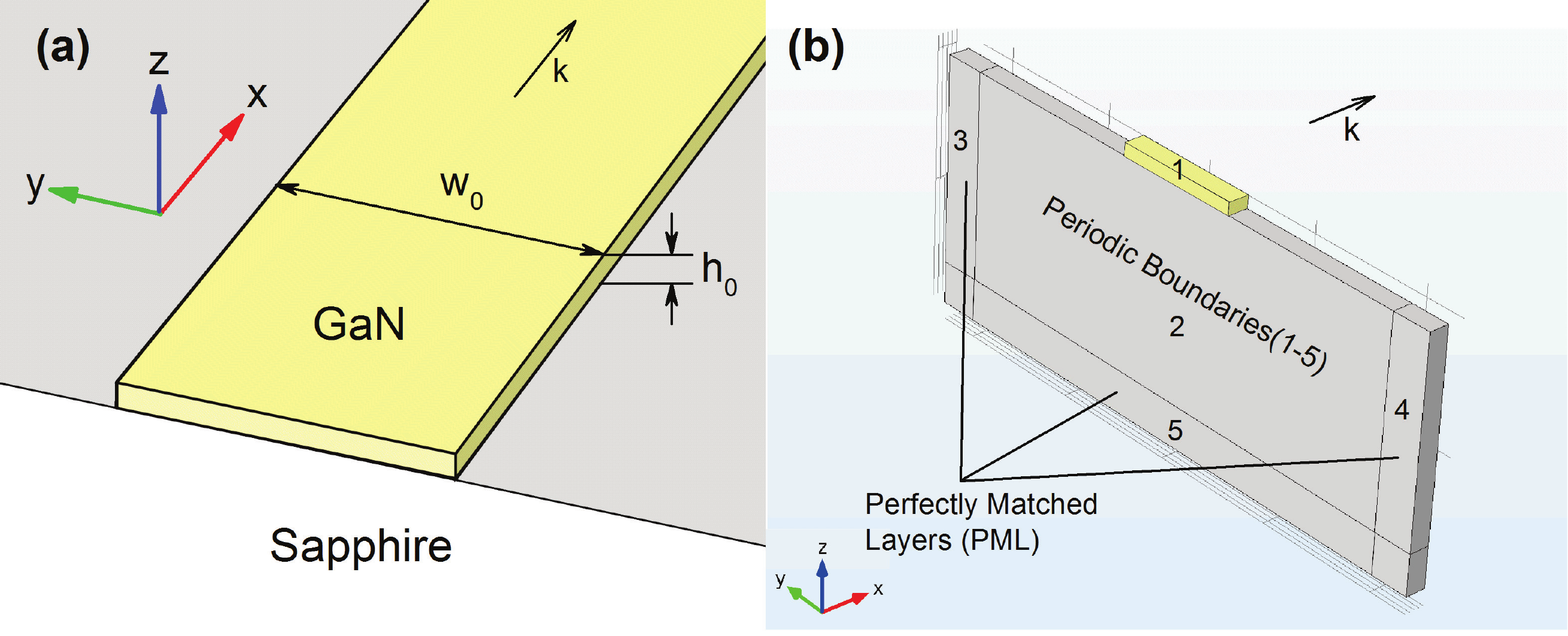}
\par\end{centering}
\protect\caption{(a) Schematic illustration of an unsuspended strip phononic waveguide. $w_{0}$ and $h_{0}$ represent the width and height of the GaN strip, respectively. (b) The geometry setup in COMSOL simulation. The phononic waveguide modes propagating in the uniform structure along $\protect\overrightarrow{x}$ with wavevector $\protect\overrightarrow{k}$ are calculated by applying a periodic boundary condition, and PMLs are employed to absorb radiative acoustic waves in substrate.}

\label{fig:shape} 
\end{figure}

Efficient confinement and guiding of phonons is essential for scalable operation of phononic devices. In the past decades, SAW has been mostly employed in applications that utilize its vertical confinement at the surface of substrates. However, the lateral confinement to the SAW is seldom studied. Alternatively, three-dimensional confinement of phonon has been mostly achieved in suspended phononic microstructures, which imposes practical constraints in fabrication yield and structural robustness. Here, we numerically study a phononic architecture based on unsuspended phononic waveugides and resonators. Shown in Fig.$\,$\ref{fig:shape}(a) is a typical strip waveguide. The basic requirement for phonon confinement is that the speeds of both transverse and longitudinal waves in strip material are slower than those in the substrate \cite{Oliner1976}. Thus, we choose the material platform of GaN-on-Sapphire satisfying these requirements, with parameters shown in Table$\,$\ref{table-1}. An equivalent condition is also drawn in Ref.$\,$\cite{Tiersten1969} where the waveguide layer material needs to be heavier and less stiff. More detailed analyses about the requirements for the existence of Love wave can be found in Ref.$\,$\cite{Microwave-I,Microwave-II,Tiersten1969}.

Conventional Rayleigh waves have been widely studied and applied in
SAW devices, while half-space single medium also supports shear-horizontal
(SH) waves. In layered systems, in addition to layered Rayleigh waves,
there are also Love waves \cite{SAW-book}, which is first discovered
by A.E.H. Love \cite{Love1911}. Love waves have a shear component with
displacements in the surface, which can be regarded as modified forms
of the SH wave. SH waves and Love waves are now attracting interests
in spin--orbit interactions of phonons \cite{Fu2019nc} and sensing applications \cite{Pang13saap, Fu17apl}.

The basic properties of confined guiding modes are studied numerically by three-dimensional finite-element method (COMSOL Multiphysics v5.2). The waveguide geometrical parameters, namely its width $w_{0}$ and height $h_{0}$ marked in Fig.$\,$\ref{fig:shape}(a), determine the properties of the waveguide modes such as frequency, confinement, loss, and energy distributions. In this paper, all the simulations are carried out using periodic boundary conditions according to the translational symmetry of the structure, and perfectly matched layers (PMLs) are employed for studying the radiative loss, with the detailed numerical model illustrated in Fig.$\,$\ref{fig:shape}(b). To reveal the basic behaviors of the phononic waveguides, the anisotropy of GaN and sapphire are neglected. 

\begin{table}
\begin{tabular}{ccc}
\hline 
Material & GaN \cite{GaN-density,GaN-Poisson+Youngs} & Sapphire \cite{Sapphire}\tabularnewline
\hline 
Density ($\mathrm{g\cdot cm^{-3}}$) & 6.15  & 3.98\tabularnewline
Young's modulus ($\mathrm{GPa}$) & 305  & 345\tabularnewline
Poisson ratio  & 0.183  & 0.29\tabularnewline
Longitudinal wave speed ($\mathrm{m/s}$) & 7350.0 & 10658.0\tabularnewline
Transverse wave speed ($\mathrm{m/s}$) & 4578.3 & 5796.4\tabularnewline
\hline 
\end{tabular}
\caption{Elastic properties of GaN and Sapphire used in the simulation.}
\label{table-1}
\end{table}

\begin{figure}
\includegraphics[width=0.8\columnwidth]{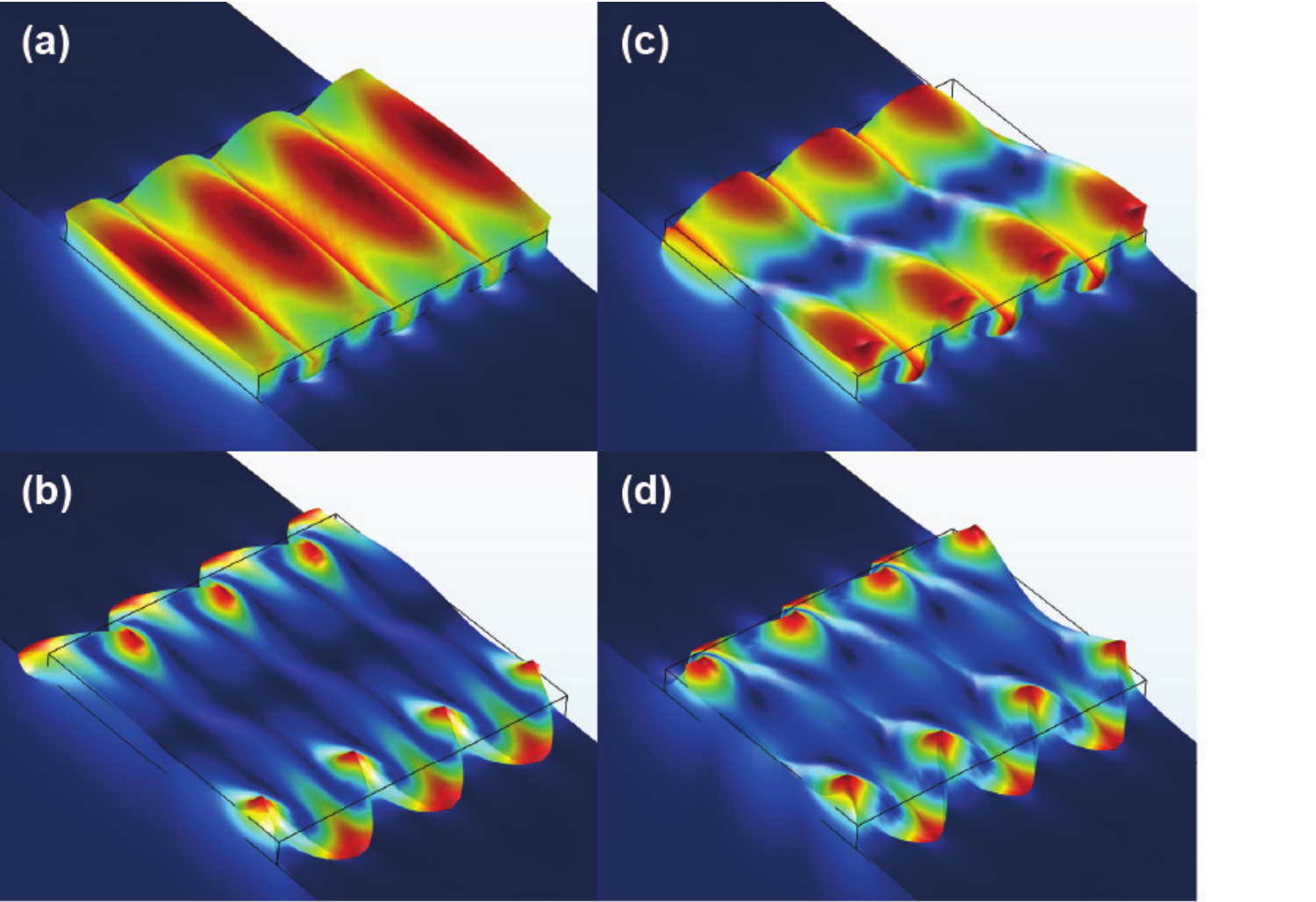}
\includegraphics[width=0.8\columnwidth]{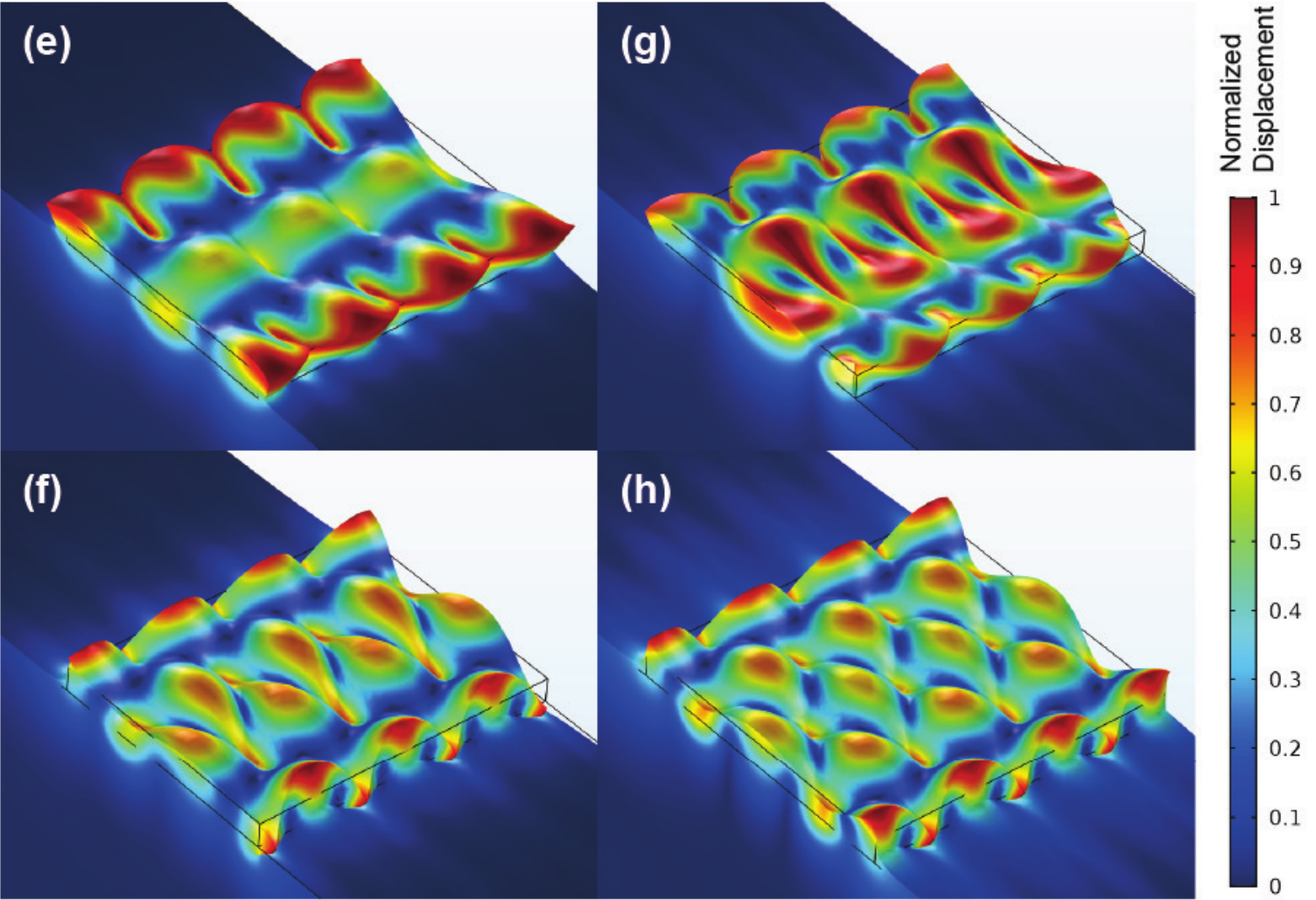}
\caption{Simulated displacement field profile for low-order phonon modes supported in an unsuspended strip waveguide. The color mappings show the strength of displacements $|\vec{x}|$ normalized by their own maximum $\max{(|\vec{x}|)}$, which applies to all other figures of displacement field in this paper. Displacement amplitude is internally excited by the simulation software, so only relative strength $|\vec{x}|/\max{(|\vec{x}|)}$ has significance. $\max{(|\vec{x}|)}$ is extracted among the entire field of each mode, and every following figure is normalized to their own maximum. But since the color map setups are the same, we let them share one common color bar for simplicity. (a) Symmetric quasi-Rayleigh mode \emph{a}. (b) Anti-symmetric quasi-Love mode \emph{b}. (c) Anti-symmetric quasi-Rayleigh mode \emph{c}. (d) Symmetric quasi-Love mode \emph{d}. (e)-(h) Higher-order phonon modes \emph{e-h}. The italic characters label different modes and will be used throughout this paper. Wavelength $\lambda$ along guided direction $x$ are all set at $2\,\mathrm{\mu m}$. The waveguide height is fixed at and $h_{0}=0.5\,\mathrm{\mu m}$ while its width is set at $w_{0}=5\,\mathrm{\mu m}$ except that the  $w_{0}$ of mode \textit{h} is chosen to be $6\,\mathrm{\mu m}$ because that specific mode does not appear for $w_{0}=5\,\mathrm{\mu m}$.
}
\label{fig:line modes}
\end{figure}

\begin{figure*}
\includegraphics[height=10cm]{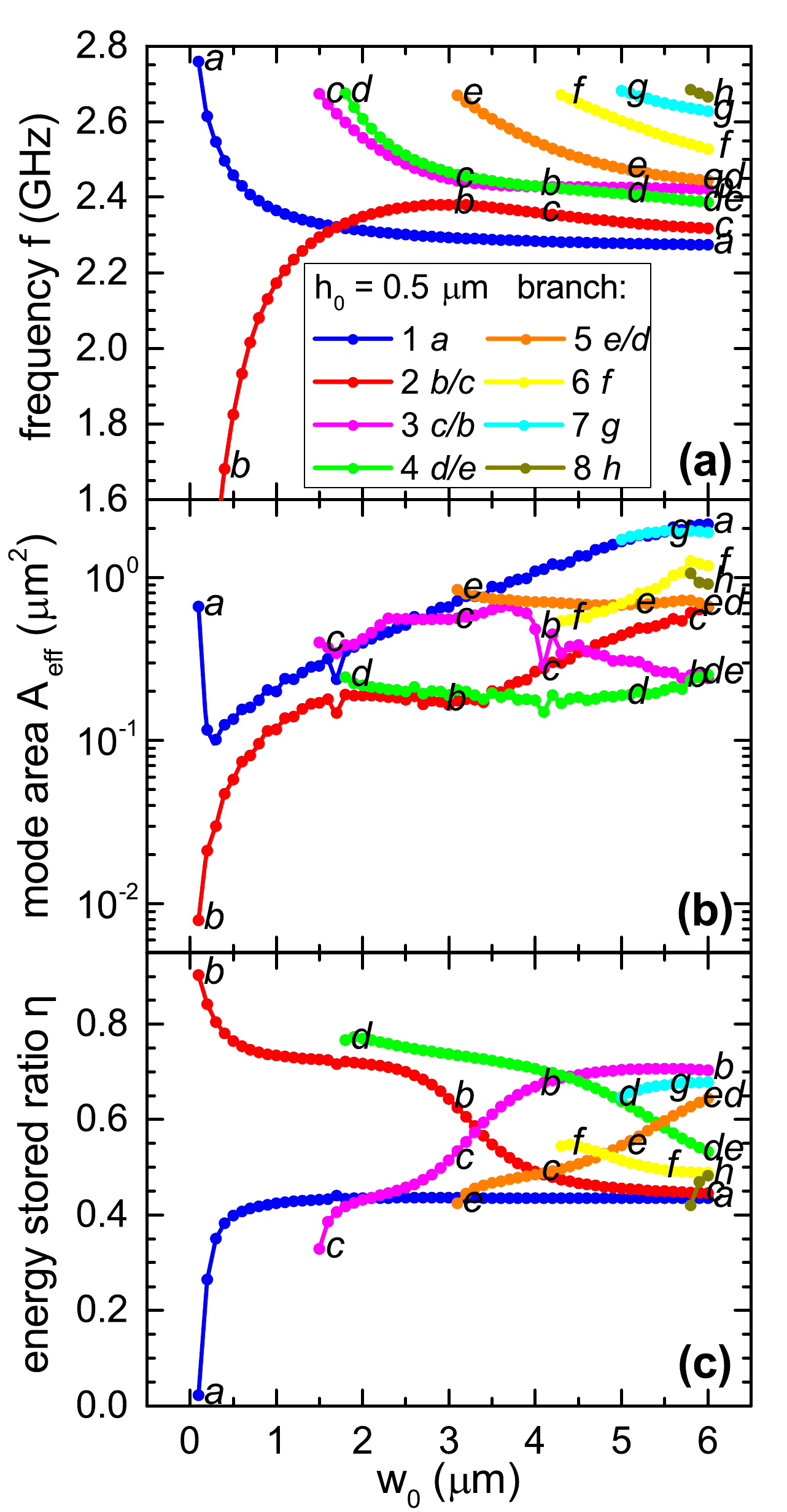}\hspace{1pt}\includegraphics[height=10cm]{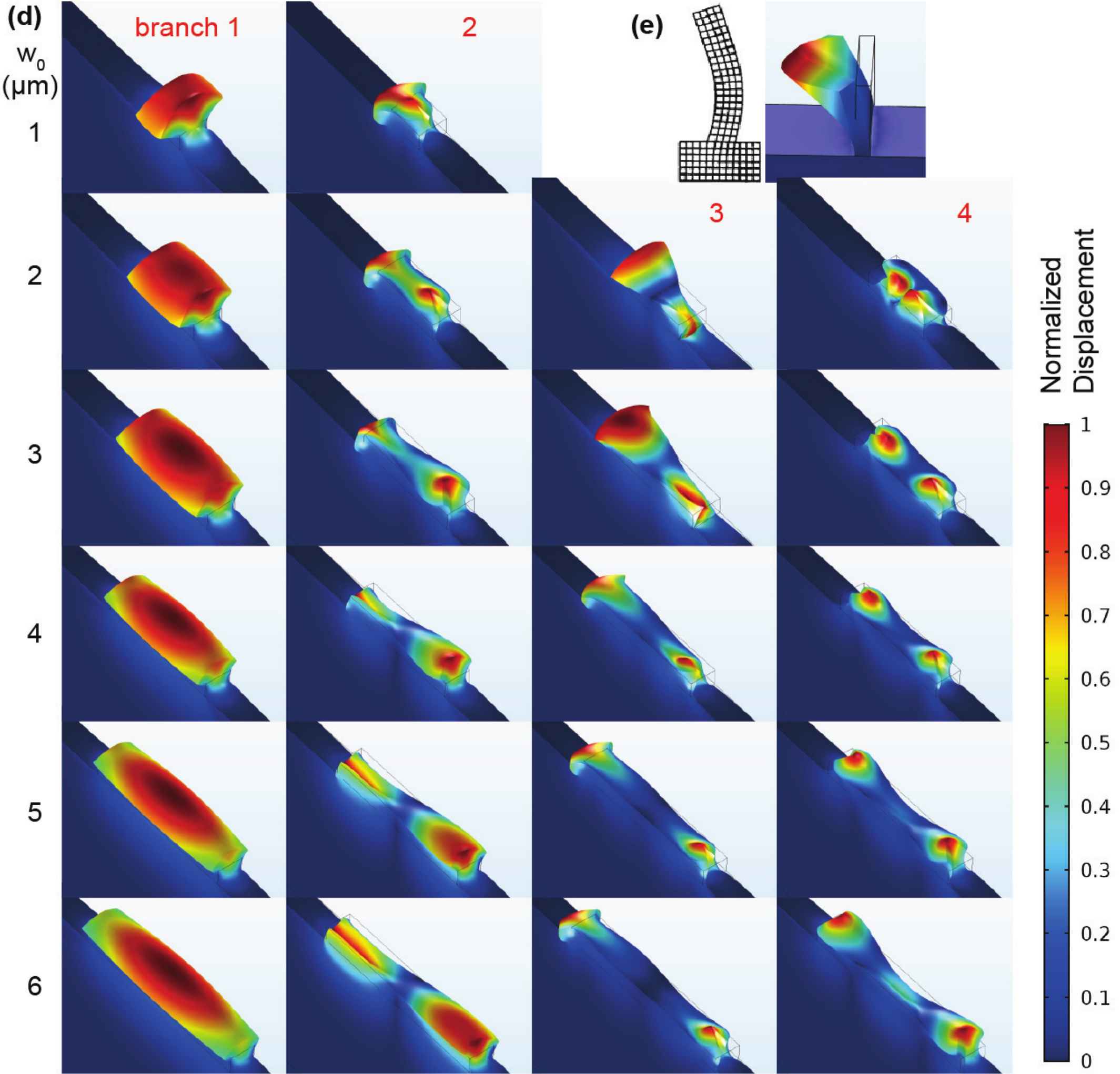}
\caption{Properties of phononic modes in a strip waveguide as the waveguide width $w_{0}$ is varied. (a)-(c) Frequency $f$, mode area $A_{eff}$, and energy confinement ratio $\eta$, versus $w_{0}$, respectively. (d) Evolution of mode hybridizations when $w_{0}$ is varied with an increment of $1\,\mathrm{\mu m}$. The color mappings show the strength of normalized displacement $|\vec{x}|/\max{(|\vec{x}|)}$ in each mode profile. Note the $\max{(|\vec{x}|)}$ is selected among all displacement field images for each mode, or equivalently, each column which plots one common mode has the same normalization. Among different columns, they use a common color mapping but their normalizations are distinct. (e) The deformation of anti-symmetrical flexural Lamb wave (left) and the simulation result (right) for a waveguide with an aspect ratio $w_{0}/h_{0}=0.2$.}
\label{fig:waveguide w0}
\end{figure*}

Shown in Figs.$\,$\ref{fig:line modes}(a)-(h) are fundamental to higher-order transverse phonon modes supported in a strip waveguide, with its width set at $w_{0}=5\,\mathrm{\mu m}$ (for mode $h$, $w_{0}=6\,\mathrm{\mu m}$ because it doesn't appear at $w_{0}=5\,\mathrm{\mu m}$) and $h_{0}=0.5\,\mathrm{\mu m}$. Wavelength along propagating direction is set at $\lambda=2\,\mathrm{\mu m}$. The color mappings show the strength of normalized displacement $|\vec{x}|/\max{(|\vec{x}|)}$ in each mode profile, which applies to all other figures of displacement field in this paper. Note that we label the modes by the corresponding characters $a,..,h$ in the following for analyzing the evolution of mode profiles against geometry parameters. Mode \emph{a} (Fig.$\,$\ref{fig:line modes}(a)) is dominated by out-of-plane (z-direction) displacement, which we refer as quasi-Rayleigh mode owing to its reminiscence of Rayleigh wave \cite{Rayleigh1885}. Mode \emph{b} has dominant in-plane (y-direction) motion and is a quasi-Love mode because of its similarity to Love wave \cite{Love1911}. The modes \emph{a} and \emph{b} can be regarded as the fundamental phononic modes in the strip with different polarizations (z- and y-polarized motion). Due to the lateral confinement and the edges, we found that the phononic modes are distinct from SAW modes. On one hand, the modes do not show pure flexural wave or in-plane shear waves. Especially, at the edge, it is difficult to identify the mode orientation because of the strong hybridization of deformation in all three orthogonal directions. On the other hand, higher-order transverse modes appear due to the lateral confinement (Figs.$\,$\ref{fig:line modes}(c)-(h)). For example, compared to fundamental mode \emph{a} and \emph{b}, in which vibrations on the strip edges are in-phase, the motion of mode \emph{c} and \emph{d} on the opposite edges are out-of-phase. In the rest of this paper, we call the symmetric (in-phase) mode \emph{a} as S-Rayleigh mode and call the anti-symmetric (out-of-phase) mode \emph{c} as A-Rayleigh mode, as well as A-Love mode \emph{b} and S-Love mode \emph{d}. We also observe that higher-order modes (Figs.$\,$\ref{fig:line modes}(e)-(h)) exhibit more complex field distributions.

For most applications, the concerned properties of guided phonon modes are \cite{Oliner1976}: dispersion, confinement, radiation loss, energy distributions. Therefore, we numerically evaluate the following properties with varying waveguide geometrical parameters.
\begin{enumerate}
\item Mode frequency $f$ for a given wavelength $\lambda$ or wavenumber $k_{x}=2\pi/\lambda$  along guided direction. 
\item Phase velocity $v_{p}=2\pi f/k_{x}$ and group velocity $v_{g}=2\pi df/dk_{x}$. 
\item Mode area 
\begin{equation}
A_{\mathrm{eff}}=\frac{1}{L}\frac{\iiintop W(x,y,z)dxdydz}{\mathrm{\text{\ensuremath{\max}}}(W(x,y,z))}\label{eq:Aeff}
\end{equation}
where $L$ is the waveguide length along the propagation direction, $W(x,y,z)$ is the elastic strain energy density. The integral and maximum are calculated in the full simulated region. The mode area is a concept borrowed from photonics (see Ref.$\,$\cite{Mode-area-optics,Mode-area-2008}) as a measure of lateral confinement, where a smaller $A_{\mathrm{eff}}$ indicates stronger confinement of phonon.

\item Quality factor 
\begin{equation}
Q=2\pi f\times\frac{\mathrm{Energy\:stored}}{\mathrm{Power\:loss}}.
\end{equation}
In this paper, we only calculate the radiation loss into substrate. A discussion of other loss mechanisms can be found in the conclusion Sec.$\,$VI.

\item Energy confinement ratio
\begin{equation}
\eta=\frac{\iiintop_{\mathrm{strip}}W(x,y,z)dxdydz}{\iiintop_{\mathrm{all}}W(x,y,z)dxdydz}
\end{equation}
indicates the ratio between elastic strain energy stored in the phononic waveguide and the total elastic strain energy.

\end{enumerate}
\begin{figure}
\includegraphics[width=8.8cm]{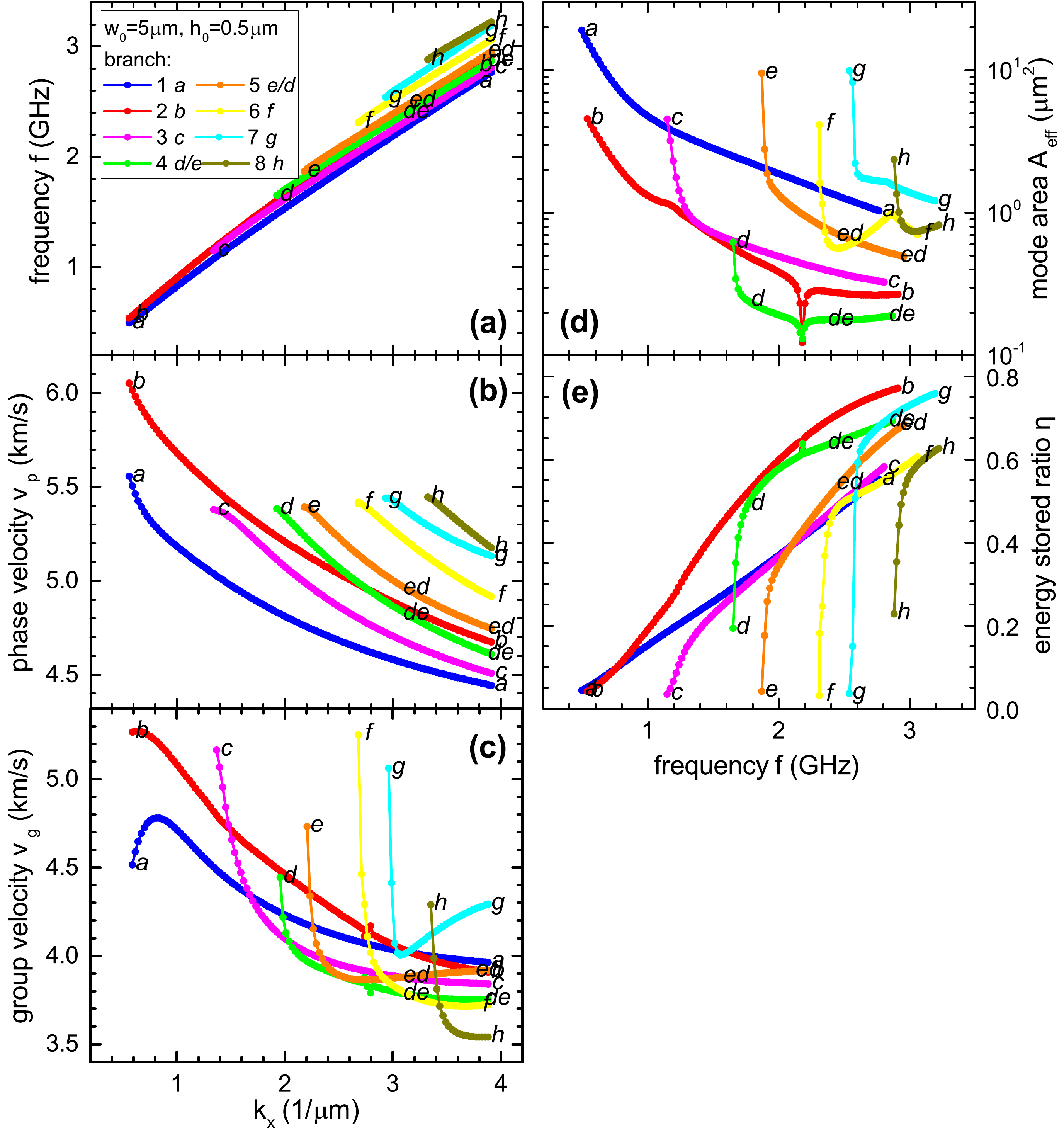}
\caption{Dispersion characteristics for phononic modes on strip waveguide, with geometry parameters $w_{0}=5\,\mu\mathrm{m}$ and $h_{0}=0.5\,\mathrm{\mu m}$
(aspect ratio $w_{0}/h_{0}=10$). (a) Frequency $f$ versus guide wavenumber $k_{x}$. (b) and (c) Phase velocity and group velocity versus $k_{x}$, respectively. (d) and (e) Mode area $A_{eff}$ and energy confinement ratio $\eta$ versus $f$, respectively.}
\label{fig:line dispersion}
\end{figure}

Figures$\,$\ref{fig:waveguide w0}(a)-(c) show the modal frequency ($f$), area ($A_{eff}$), and energy storage ratio ($\eta$) of 8 phonon modes versus the width $w_{0}$ of strip waveguide, with a fixed height $h_{0}=0.5\,\mathrm{\mu m}$ and wavelength along propagation direction $\lambda=2\,\mathrm{\mu m}$. To clarify complicated mode hybridization, we now explain the notations and colors of plots that are adopted in this paper. By varying the waveguide geometry by small increments, the mode properties vary smoothly, thus we can track the frequencies or other parameters as continuous curves, which is called ``branch'' and labeled by different colors. As shown in the frequency plot (Fig.$\,$\ref{fig:waveguide w0}(a)), each branch is assigned with one specific color consistent with all other properties plots. As explained above, due to the strong mode hybridization in the strip waveguides, it is very difficult to identify mode orders or polarizations by branches. Therefore, we only label the modes in the regime that can be clearly distinguished according to the modes shown in Fig.$\,$\ref{fig:line modes}. Sections labeled with two characters indicate the modes are experiencing hybridization. 

The model frequency features versus width $w_0$, as shown in Fig.$\,$\ref{fig:waveguide w0}(a), indicates a transition between single-mode and multi-mode waveguide. At large $w_{0}$ values, all phonon mode branches asymptotically approach slab phonon modes in GaN-on-sapphire thin films. When $w_{0}$ becomes comparable to $\lambda$, the curves branch off and the behaviors of each branch can be explained as follows. Since the lateral confinement imposes an approximate quantization condition $k_{y}\sim\frac{n\pi}{w_{0}}$, an increasing $w_{0}$ results in a reduction of total phonon wavenumber as $k=\sqrt{k_{x}^{2}+k_{y}^{2}}$ for a fixed $k_{x}$, and hence a monotonic reduction of frequency approaching a constant. This intuitive interpretation is valid for most modes, however, fails for the A-Love modes \emph{b}. According to the mode area and energy confinement ratio given in Figs.$\,$\ref{fig:waveguide w0}(b) and (c), the confinement of A-Love mode \emph{b} is excellent at small $w_{0}\ll \lambda$ (rapid reducing of mode area), and thereby almost all wave energy is confined in the strip. It could be interpreted as the effective boundary becomes air-GaN for the A-Love mode \emph{b} as opposed to sapphire-GaN for the other modes, and consequently, it can be treated as an anti-symmetrical flexural Lamb wave of a plate normal to $y$ direction (Fig.$\,$\ref{fig:waveguide w0}(e)) \cite{Lagasse1973}. For a plate mode, the vibration frequency $f\propto\sqrt{\frac{D}{\rho w_{0}}}$,
in which bending stiffness $D=\frac{w_{0}^{3}E}{12(1-\nu^{3})}$ \cite{Reddy2006}
($\rho,\,E,\,\nu$ are density, Young's modulus and Poisson ratio, respectively.). Thus the frequency of A-Love mode \emph{b} scales with $w_{0}$ and decreases for narrow waveguides.

\begin{figure*}
\vspace{-10pt}\includegraphics[height=10cm]{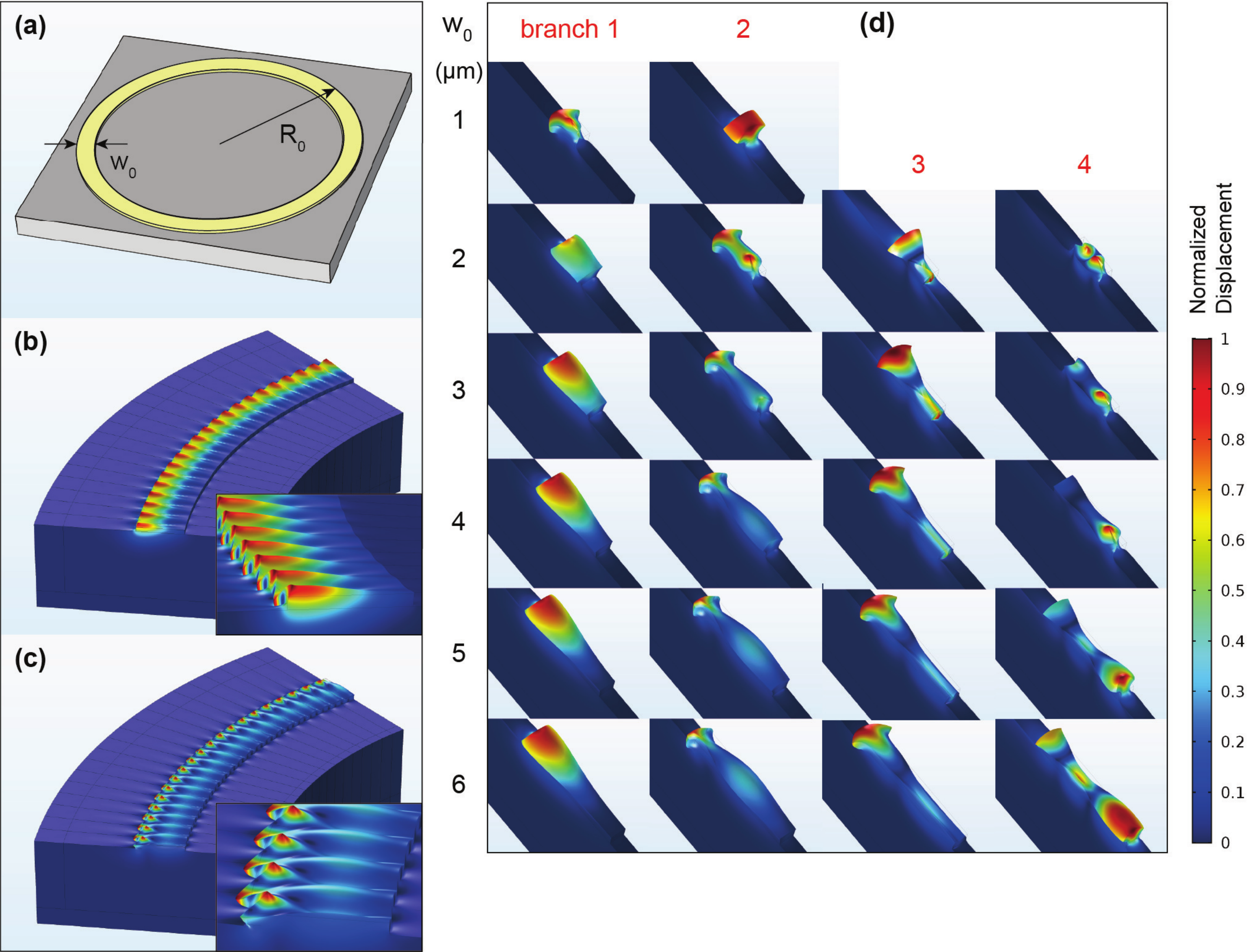}
\caption{(a) Schematic illustration of a strip ring resonator. (b) S-Rayleigh (mode \emph{a}) and (c) A-Love (mode \emph{b}) in ring resonator. The color mappings show the strength of normalized displacement $|\vec{x}|/\max{(|\vec{x}|)}$ in each mode profile. (d) The evolution of mode hybridization profiles with varying $w_{0}$. The normalization convention is the same as we used in Fig.$\,$\ref{fig:waveguide w0}(d): the $\max{(|\vec{x}|)}$ is selected among all displacement field images for each mode, or equivalently, each column which plots one common mode has the same normalization. (b-d) share a common color bar.}
\label{fig:Ring modes}
\end{figure*}

The interactions between the phonon modes lead to crossings and avoided crossings in Figs.$\,$\ref{fig:waveguide w0}(a)-(c). For the avoided crossings, the corresponding mode area and energy confinement ratio show crossing behavior, indicating the coupling between modes and the mode hybridization at these device parameters. For the crossing in frequency, the coupling between modes is negligible, so the changes in mode area and energy confinement ratio is not noticeable. Outside of the hybridization regimes, the confinement of quasi-Rayleigh modes (\emph{a} and \emph{c}) weakens (larger mode area) with the increase of width, while that of quasi-Love modes (\emph{b} and \emph{d}) remains relatively confined. At very small $w_{0}$, however, the waveguide becomes too narrow to support a fully confined mode \emph{a}, and its mode frequency rapidly approaches its cutoff frequency which is the frequency of SAW in substrate for $\lambda=2\,\mathrm{\mu m}$. The mode area of A-love mode \emph{b}, in contrast to mode \emph{a}, drops because this mode resembles anti-symmetrical flexural Lamb wave at small width, and its confinement is enhanced as width narrows \cite{Lagasse1973} (like a vibrating membrane whose energy is tightly concentrated in the waveguide region thus the influence from substrate is negligible). The quasi-Love modes (\emph{b,d}) exhibit stronger confinement than
quasi-Rayleigh modes' (\emph{a,c}) due to the much larger $\eta$ of quasi-Love modes (except the hybridization regime), as shown in Fig.$\,$\ref{fig:waveguide w0}(c).

For a strip waveguide of fixed cross-section $h_{0}=0.5\,\mathrm{\mu m}$ and $w_{0}=5\,\mathrm{\mu m}$, the dispersion characteristics shown in Fig.$\,$\ref{fig:line dispersion} are numerically calculated as a function of the mode wavenumber $k_{x}$. In Fig.$\,$\ref{fig:line dispersion}(b), the quasi-Love modes \emph{b} and \emph{d} have faster phase velocities than their corresponding quasi-Rayleigh modes and their behaviors approach the infinite surface according to analytical predictions in Ref.$\,$\cite{Microwave-I,Tiersten1973,Paul1973}. The cutoffs result in sharp rising of group velocities in Fig.$\,$\ref{fig:line dispersion}(c) at small wavevectors, and meanwhile, in Fig.$\,$\ref{fig:line dispersion}(d),
mode areas rise sharply at low frequencies, indicating a remarkable divergence of the confinement when the size of strip becomes much smaller than the wavelength. This is confirmed in Fig.$\,$\ref{fig:line dispersion}(e) that energy confinement ratio decreases, i.e. phonon energy spreads into the substrate, at small $k_{x}$.
In Fig.$\,$\ref{fig:line dispersion}(d), we also found an interesting minimum point of mode area that occur simultaneously for branches of mode \emph{b} and \emph{d} around $2.2$ GHz. A close comparison with results in Figs.$\,$\ref{fig:line dispersion}(a) and (b) reveals that these minima occur at the avoided crossing point of modes \emph{b} and \emph{d} when they are maximally hybridized. Therefore, the mode hybridization presents an unique approach to engineer the phononic modes and might find interesting phononic sensing applications.

\section{Ring resonator}

In this section, we study the properties of ring resonators formed by bending and closing strip waveguide which supports phononic WGMs. The pioneering works of the acoustic WGMs date back to the discovery by Lord Rayleigh \cite{WGM-1-Rayleigh,WGM-2-Rayleigh},
who reported the sound wave travels around a concave boundary in in St Paul\textquoteright s Cathedral due to the continuous total internal reflection. The concept is generalized to electromagnetic waves \cite{WGM-in-EM-waves,OWGM-1,OWGM-2}, which holds unique characteristics of high Q-factors, small mode volumes, and the ease of fabrication \cite{WGM-applications}. These
remarkable merits are also possessed by phononic WGMs according to our investigations. 

Due to the cylindrical symmetry of the ring resonator, the eigenmode profiles of the elastic differential characteristic equation are in the form as $u(r,z,\phi)=\psi(r,z)e^{im\phi}$, where $\left(r,z,\phi\right)$ are the cylindrical coordinators, $\psi(r,z)$ is the field distribution at the cross-section, $m$ is the angular momentum number. Utilizing the symmetry, we can numerically solve the mode profiles of a sector unit in $\phi$ direction in COMSOL, with periodic boundary condition: $u(r,z,\phi)=u(r,z,0)e^{-im\phi}$, where $m=2\pi R/\lambda$ for an azimuthal  wavelength $\lambda=2\,\mathrm{\mu m}$ along tangent direction.

\begin{figure}
\includegraphics[width=8.8cm]{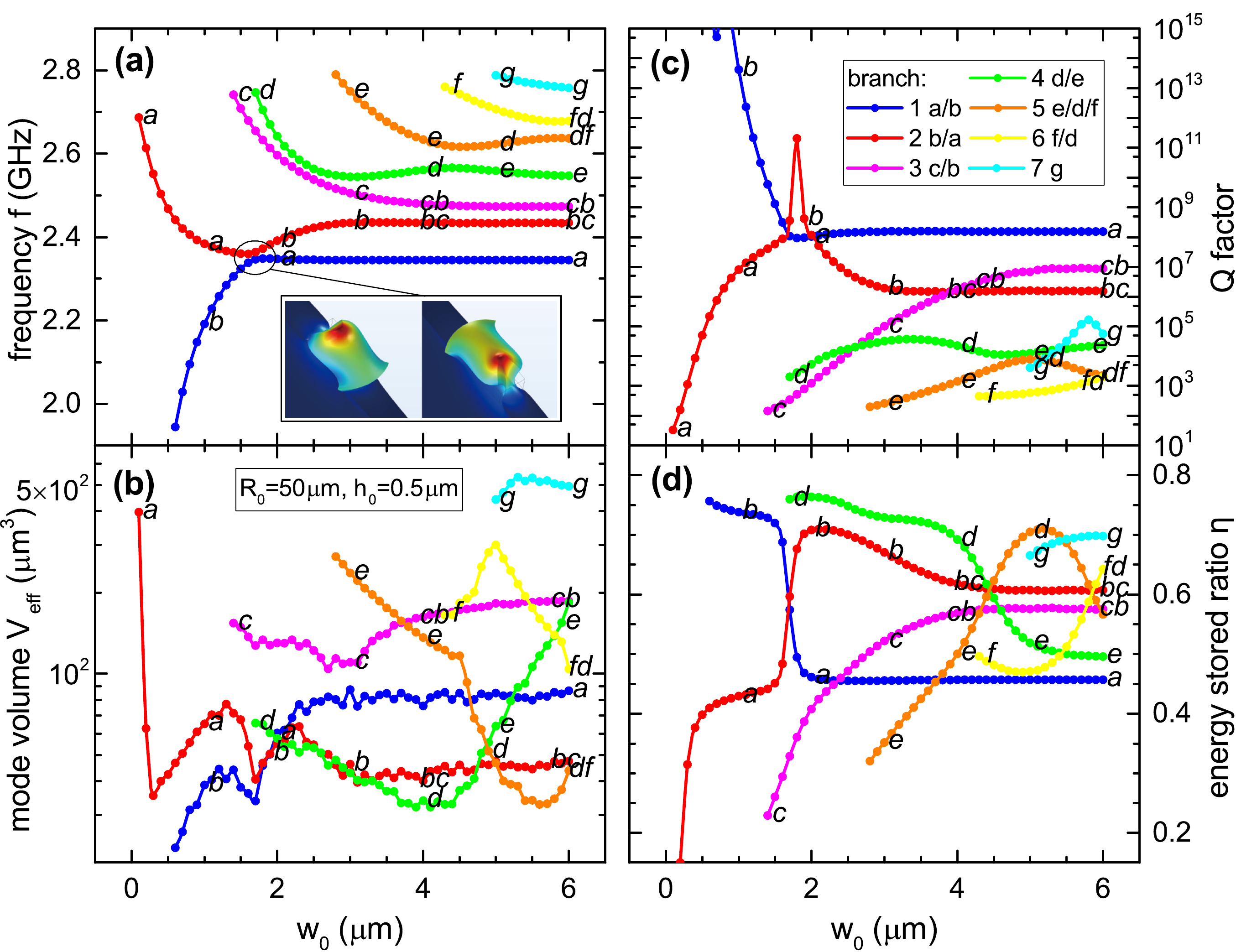}

\caption{Phononic mode properties in a ring resonator of varying width $w_{0}$. The ring radius and height are fixed respectively at $R_{0}=50\,\mathrm{\mu m}$ and $h_{0}=0.5\,\mathrm{\mu m}$. (a)-(d) show frequency, mode volume $V_{eff}$, $Q$ factor, and energy confinement ratio $\eta$, respectively. The insets of (a) are the mode profiles of hybridized modes of \emph{$(a,\,b)$} at $w_{0}=1.7\,\mathrm{\mu m}$.}
\label{fig:Ring resonator w0}
\end{figure}

\begin{figure}
\includegraphics[width=8.8cm]{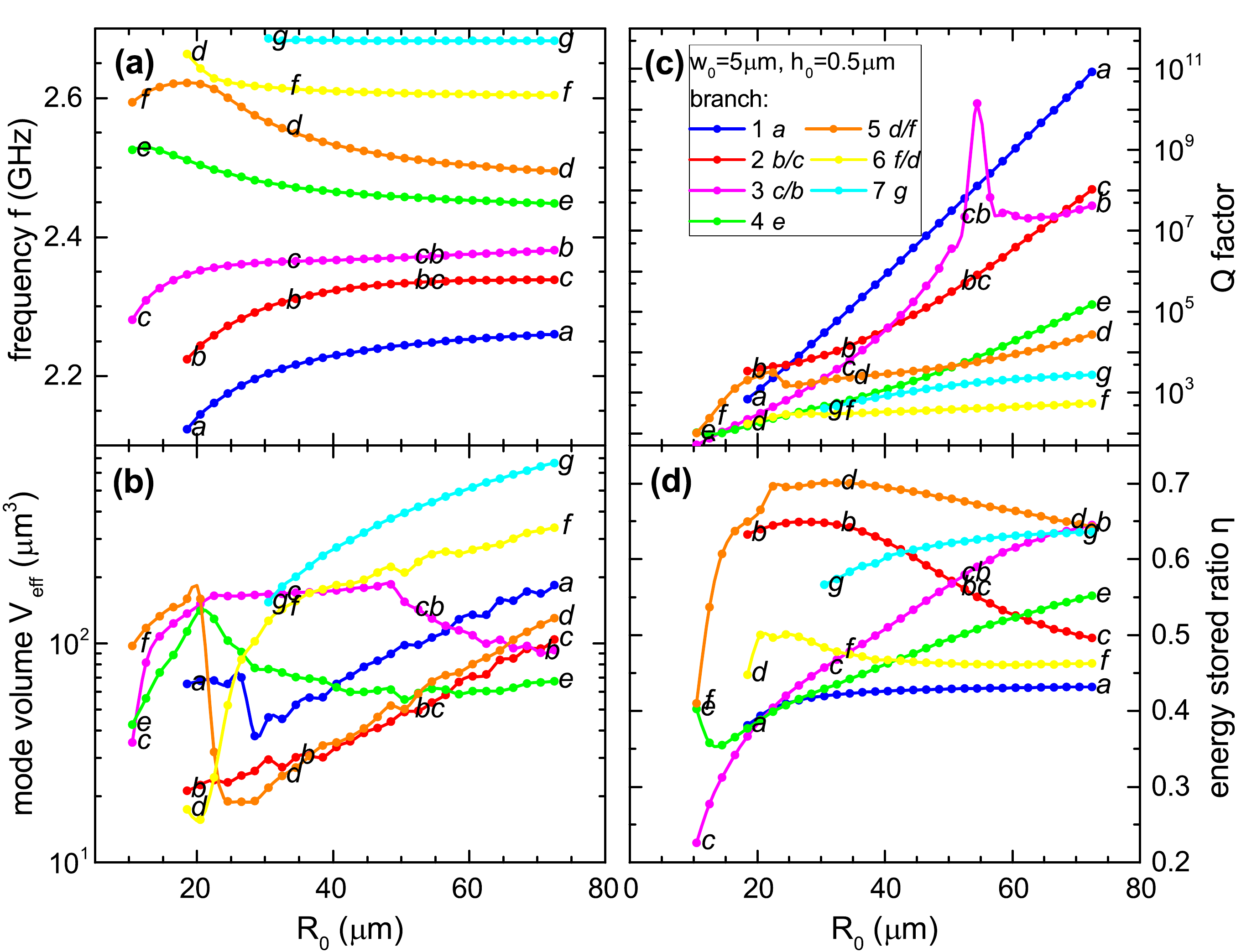}
\caption{Phononic mode properties in a ring resonator as functions of ring radius $R_{0}$ with width $w_{0}=5\,\mathrm{\mu m}$, height $h_{0}=0.5\,\mathrm{\mu m}$. (a-d) show frequency, mode volume $V_{eff}$, $Q$ factor and energy confinement ratio $\eta$, respectively.}
\label{fig:Ring resonator R0}
\end{figure}

\begin{figure}
\begin{centering}
\includegraphics[width=8.8cm]{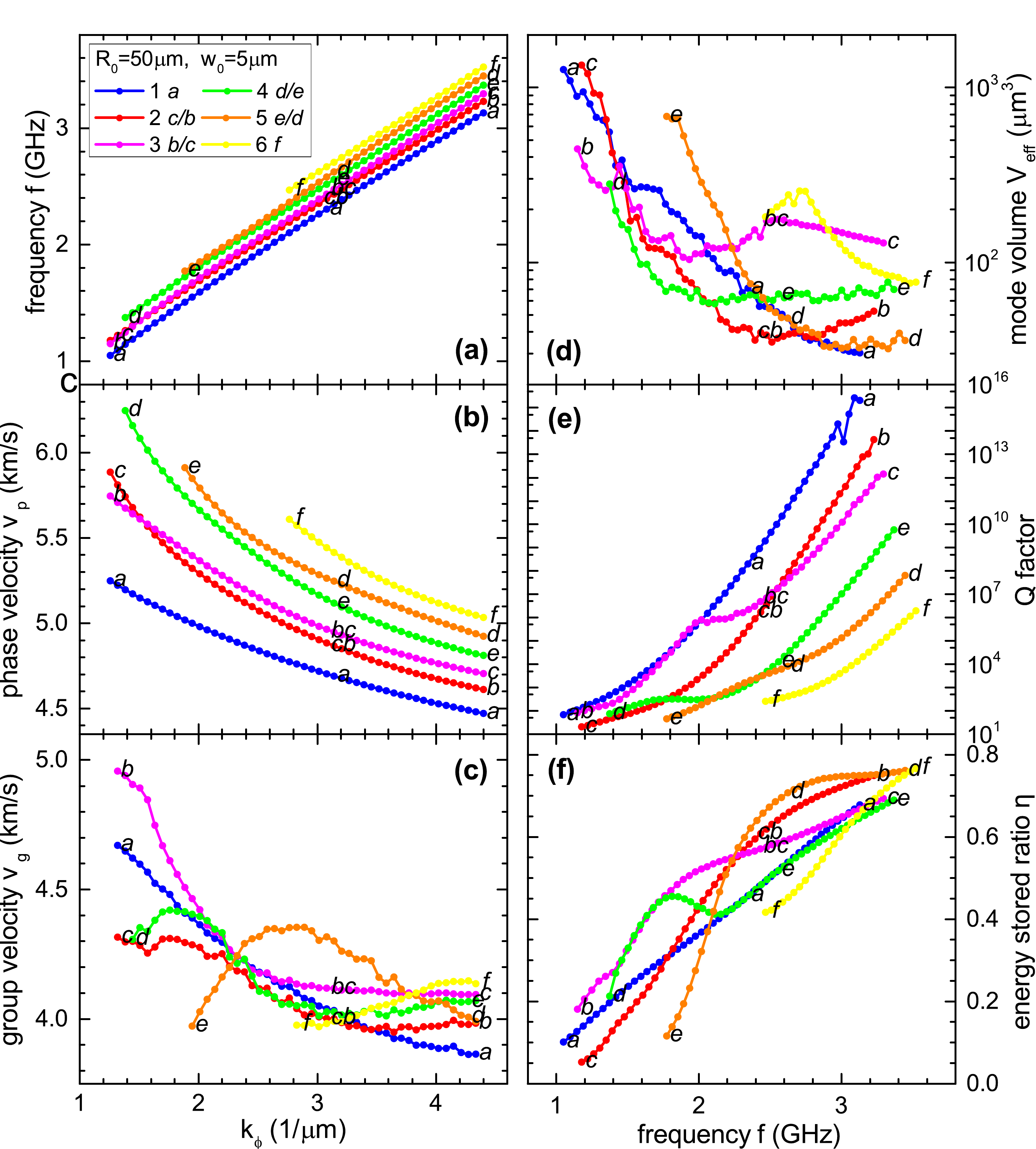}
\par\end{centering}
\protect\caption{Dispersion characteristics for modes in a ring resonator of radius $R_{0}=50\,\mathrm{\mu m}$, width $w_{0}=5\,\mathrm{\mu m}$, and height $h_{0}=0.5\,\mathrm{\mu m}$.  (a) Modal frequencies $f$ versus $k_{\phi}$---the effective propagating wavenumber defined as $m/R_{0}$, with $m$ is angular momentum number.
(b) and (c) Phase velocity and group velocity versus $k_{\phi}$. (d-f) Mode volume $V_{eff}$, $Q$ factor, and energy confinement ratio $\eta$ versus $f$, respectively.}

\label{fig:Ring dispersion}
\end{figure}

Figures$\,$\ref{fig:Ring modes} (b) and (c) display the typical quasi-Rayleigh and quasi-Love WGMs in a ring resonator. Compared to those in waveguide (Figs.$\,$\ref{fig:line modes}(a)-(b)), the displacement fields of WGMs in the ring are more concentrated at the outer rim. Figure$\,$\ref{fig:Ring modes}(d) shows the mode profiles of the first four branches in Fig.$\,$\ref{fig:Ring resonator w0}(a). In the rest of this section, except for dispersion behaviors, we study the properties of WGMs around fixed geometry parameters $\left\{ R_{0},\,w_{0},\,h_{0}\right\} =\left\{ 50\,\mathrm{\mu m},\,5\,\mathrm{\mu m},\,0.5\,\mathrm{\mu m}\right\}$.
Similar to the cases of strip waveguides, the colors of different branches are fixed in Figs.$\,$\ref{fig:Ring resonator w0}, \ref{fig:Ring resonator R0}, and \ref{fig:Ring dispersion}. Instead of the mode area, the confinement in resonator is characterized by the mode volume \cite{Mode-Volume}, 
\begin{equation}
V_{\mathrm{eff}}=\frac{\iiintop W(r,\theta,z)rdrd\theta dz}{\max[W(r,\theta,z)]}.\label{eq:Veff}
\end{equation}

As shown in Fig.$\,$\ref{fig:Ring resonator w0}(a), the dependence of mode frequencies with respect to width $w_{0}$ follows the same trend of those of strip waveguides [Fig.$\,$\ref{fig:waveguide w0}(a)]. However, comparing to the crossing in waveguide, there is avoided crossing between mode \emph{a} and \emph{b}, arising from the broken symmetry (hence mode hybridization) in the bending waveguide. The mode volume and energy confinement ratio $\eta$ are presented in Figs.$\,$\ref{fig:Ring resonator w0}(b) and (d), showing similar crossing and avoid-crossing behavior as in the waveguide case. 

For ring resonators, in contrast to the strip waveguides, there is always radiation loss due to the waveguide bending, and thus the $Q$ factors of WGMs are finite. As shown in Fig.$\,$\ref{fig:Ring resonator w0}(c),
S-Rayleigh \emph{a}, A-Rayleigh \emph{c} and A-Love \emph{b} are three modes having the highest $Q$ factors. For mode \emph{a} and \emph{b}, $Q$s are sensitive to the width at small $w_{0}$ value but almost constant when $w_{0}$ exceeds $3\,\mathrm{\mu m}$, implying a saturation of loss approaching the case of a disk resonator. We observe an abrupt increment of $Q$ when \emph{b} and \emph{a} are hybridized at around $w_{0}=1.7\,\mathrm{\mu m}$, suggesting that the mode \emph{b} and \emph{a} couple to a common leaky modes in substrate and their destructive interference suppresses the phonon loss. Thus the special width leads to a parameter-tuning type of bound state in continuum \cite{Hsu16nrm,Chen16njp}.

Figure $\,$\ref{fig:Ring resonator R0}(a) shows the properties of phononic WGMs with varying radius. For a fixed $\lambda$, the frequencies of the ring converge to the case of straight waveguide at large $R_{0}$ values with their calculated modal volumes shown in Figure $\,$\ref{fig:Ring resonator R0}(b). An optimal radius exists for each mode because the lateral confinement (i.e. effective mode area $A_{\mathrm{eff}}$) becomes better for larger $R_{0}$, while the mode volume is approximately proportional to $R_{0}$ ($V_{\mathrm{eff}}\approx2\pi R_{0}A_{\mathrm{eff}}$). 
The $Q$ factors in Fig.$\,$\ref{fig:Ring resonator R0}(c) increase exponentially with increasing $R_{0}$, indicating radiation loss to the substrate caused by bending decreases exponentially with the curvature, similar to the radiation loss of optical WGM in dielectric spheres \cite{Mode-Volume,WGM-in-EM-waves}. The increment of $Q$ when \emph{b} and \emph{c} are hybridized at around $R_{0}=54\,\mathrm{\mu m}$ can be explained with the same reason as that of \emph{b} and \emph{a} in Fig.$\,$\ref{fig:Ring resonator w0}(c).

The dispersion characteristics of ring resonator are summarized in Fig.$\,$\ref{fig:Ring dispersion}. The $k_{\phi}=m/R_{0}$ in the figure is effective propagating wavenumber along tangential direction. Most branches resemble their counterparts in the strip waveguide (Fig.$\,$\ref{fig:line dispersion}), including  $f-k_{\phi}$ relation, and the decrease of phase velocity and mode volume, and increase of ratio $\eta$ with increasing $k_{\phi}$ or $f$. We also compute $Q$ factors which grow exponentially at short wavelengths.

\section{Directional Coupler}

In an integrated phononic circuit, multiport devices transporting and routing phonons between components such as the waveguide-waveguide and waveguide-resonator coupler are indispensable circuit elements. A directional coupler consists of two closely placed parallel waveguides. Here we study the phononic coupling through the coupled-mode theory by analogy to its optical counterpart widely used in the photonic community \cite{Couple-1973,Couple-1985,Couple-1994}. The coupling between the modes in different waveguides arises from the tunneling of the elastic wave between them. The confined waveguide mode has a non-zero evanescent field that overlaps with the other waveguide, leading energy transfer in between.

The spatial mode amplitude evolution along the propagation
direction in coupled waveguides for a given input signal frequency  are described by the following equations
\begin{align}
\frac{d}{dz}\alpha_{1}(z) & =-ik_{1}\alpha_{1}(z)-ig_{12}\alpha_{2}(z),\label{eq:couple1}\\
\frac{d}{dz}\alpha_{2}(z) & =-ik_{2}\alpha_{2}(z)-ig_{21}\alpha_{1}(z). \label{eq:couple2}
\end{align}
Here, the subscript ``1'', ``2'' represent waveguide 1 (width $w_{1}$) and 2 (width $w_{2}$), $k_{1(2)}$ is the wavenumber of the mode in waveguide $1(2)$,
and $g_{12},\,g_{21}$ are the coupling coefficients between two modes. For lossless waveguides, coefficient $g$ must satisfy $g_{12}=g_{21}^{*}$ due to the power conservation \cite{Couple-1994}. Setting $g_{12}g_{21}=|g|^{2}$ and solving Eqs.$\,$(\ref{eq:couple1}) and (\ref{eq:couple2}), we obtain the wavenumber eigenvalues of
\begin{equation}
k_{\pm}=\frac{k_{1}+k_{2}}{2}\pm\sqrt{\left(\frac{k_{1}-k_{2}}{2}\right)^{2}+|g|^{2}}\label{eq:k+-}
\end{equation}
for hybrid modes in two waveguides. When $k_{1}=k_{2}=k$, the difference between $k_{\pm}$ reaches the minimum $2|g|$, and the eigenmodes are the $a_{1}\pm a_{2}$, as an equal superposition of two waveguide modes. If the phonon input at the first waveguide with $a_{1}(0)=1$ and $a_{2}(0)=0$, we arrive at 
\begin{align}
\alpha_{1}\left(z\right) & =\cos\left(|g|z\right)e^{-ikz},\\
\alpha_{2}\left(z\right) & =\sin\left(|g|z\right)e^{-i(kz+\frac{\pi}{2})}.
\end{align}
Thus, the phonons in the first waveguide $\alpha_{1}$ could be fully transported into the second waveguide $\alpha_{2}$ after a coupling distance of $\pi/2|g|$, or half of the phonons can be transmitted into mode $\alpha_{2}$ after a coupling distance $\pi/4|g|$. The larger the $|g|$, the quicker the energy exchanges.

\subsection{Coupling between two identical waveguides}

\begin{figure}
\includegraphics[width=4.3cm]{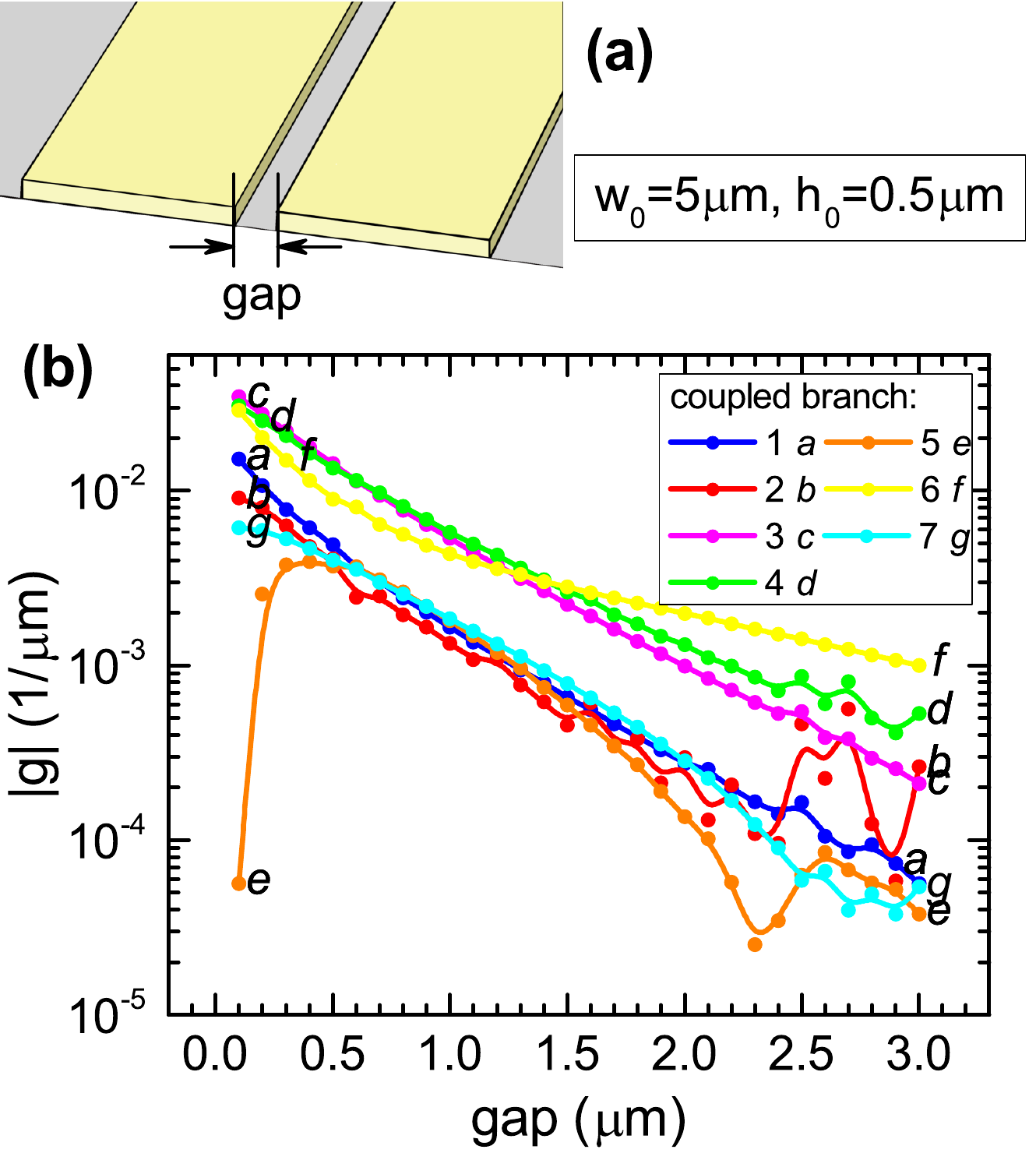}\hspace{1pt}\includegraphics[width=4.3cm]{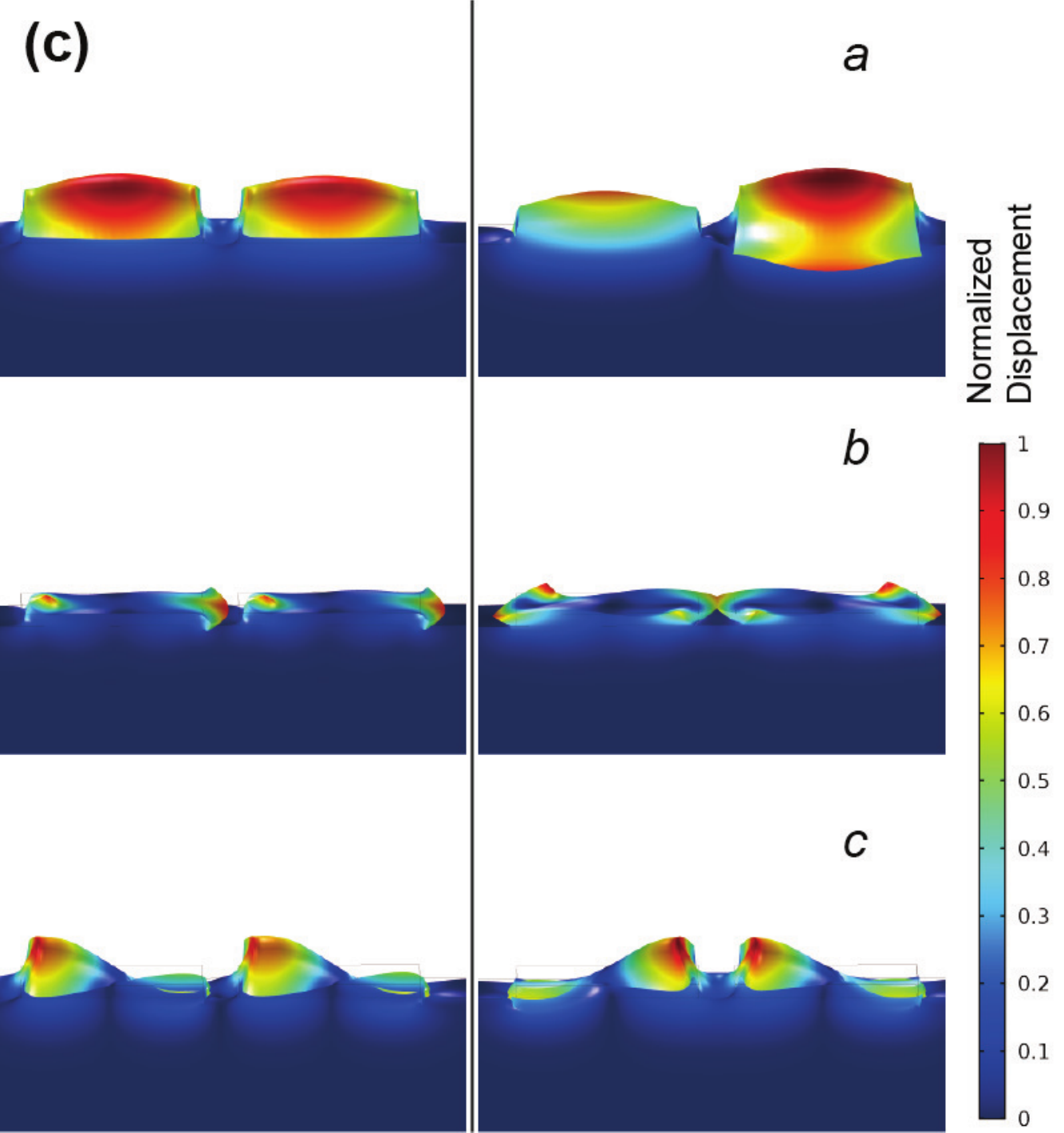}
\caption{(a) Schematic of the directional coupler. (b) The estimated coupling strengths $|g|$ as a function of $gap$. Set the $\lambda = 2\,\mathrm{\mu m} $ along guided direction. (c) The displacement field profiles for the in-phase and out-of-phase modes in the directional coupler, which is made by two identical waveguides with $gap=5\,\mathrm{\mu m}$. The color mappings show the strength of normalized displacement $|\vec{x}|/\max{(|\vec{x}|)}$ in each profile. The in-phase and out-of-phase figures share the same normalization, and all figures share the common color bar.}
\label{fig:identical-Directional-coupler}
\end{figure}

For two identical waveguides with $w_{1}=w_{2}$, the wavenumbers of the modes are the same $\left(k_{1}=k_{2}\right)$, thus all supported modes can couple between the two waveguides without phase mismatching. Here, we set the width $w_{0}=5\,\mathrm{\mu m}$ and height $h_{0}=0.5\,\mathrm{\mu m}$ for both waveguides, as shown in Fig.$\,$\ref{fig:identical-Directional-coupler}(a). From the analysis above, we can estimate the coupling strength $g$ between two waveguides by the splitting of $k_{\pm}$. 

Figure $\,$\ref{fig:identical-Directional-coupler}(b) shows the coupling strength of various waveguide modes with different gap, where the label \emph{a-g} corresponds to the modes introduced in Fig.$\,$\ref{fig:waveguide w0}. We set the $\lambda=2\,\mathrm{\mu m}$. Eigenmodes \emph{a, b} and \emph{c} in the coupled waveguides are shown in Fig.$\,$\ref{fig:identical-Directional-coupler}(c), featuring in-phase and out-of-phase hybridized modes in the coupled waveguides as predicted by the coupled-mode theory. Since the evanescent field exponentially decays with the distance to the waveguide, a reduction of the $g$ with increasing gap is expected. As shown by the numerical simulations, there is a clear exponential relation between the coupling strength and the gap, yet the trends of high order modes \emph{e} and \emph{f} slightly deviate from the exponential curve at small/large gaps. This can be attributed to the weaker confinement of the high
order modes, for which the perturbative approximation in coupled-mode theory is no longer accurate.

\subsection{Coupling between dissimilar waveguides}

\begin{figure}
\includegraphics[width=8.8cm]{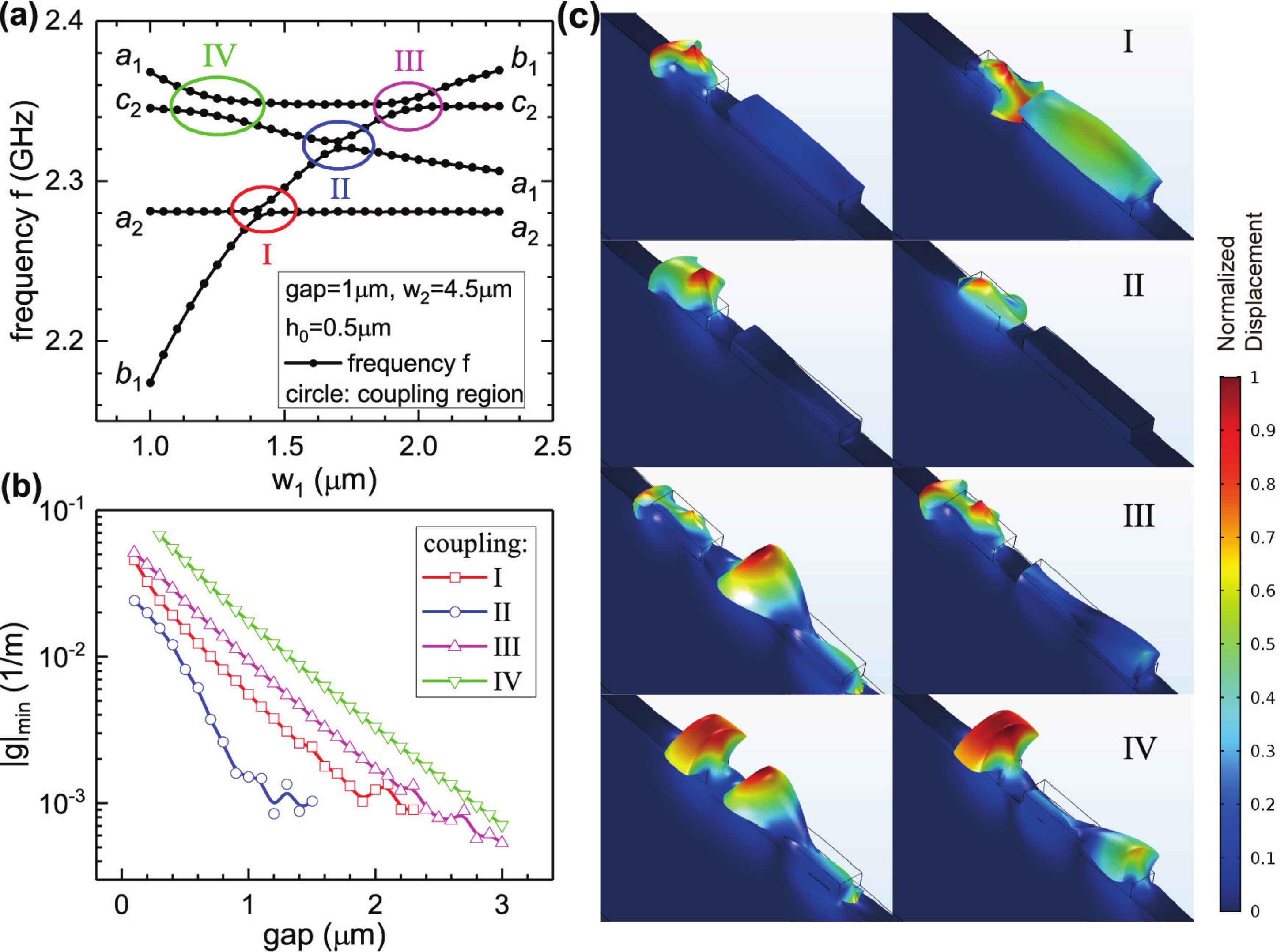}
\caption{(a) Modal frequencies in directional coupler of dissimilar waveguides, plotted against the waveguide
width $w_{1}$, with the other waveguide width and coupling gap fixed ($w_{2}=4.5\,\mathrm{\mu m}$, $gap=1\,\mathrm{\mu m}$). Due to the modal coupling, four avoided-crossing regions $\left\{ \mathrm{I,\,II,\,III,\,IV}\right\}$ are observed when the frequencies of modes in separate waveguides approach each other. The subscripts of waveguide mode labels $\left\{ 1,2\right\} $ denote the waveguide 1  and 2  respectively. (b) The modal coupling strength $|g|$ as a function of $gap$. (c) The mode profiles of the directional coupler in the four avoided crossing regions. The color mappings show the strength of normalized displacement $|\vec{x}|/\max{(|\vec{x}|)}$ in each profile. The in-phase and out-of-phase figures share the same normalization, and all figures share the common color bar.}
\label{fig:dissimilar-waveguide}
\end{figure}

The coupled-mode theory is also applicable to dissimilar modes in dissimilar waveguides, therefore opens the possibility for mode conversions with different polarization or mode orders as long as there is a finite mode overlap between them. From Eq.$\,$\ref{eq:couple2}, the maximum ratio of energy transmittance between the two modes is $\mathrm{sinc}^{2}[1+\left(k_{1}-k_{2}\right)^{2}/4g^{2}]^{1/2}$. Thus, for efficient energy transfer, we should adjust waveguide width to match the wavenumber of the two dissimilar modes, i.e. fulfill the phase-matching condition. In Fig.$\,$\ref{fig:dissimilar-waveguide}(a),
the width of one waveguide ($w_{1}$) is changed while the width of the other is fixed ($w_{2}=4.5\,\mathrm{\mu m}$) in order to meet with the phase-matching condition
\begin{equation}
k_{1}(w_{1})=k_{2}(w_{2})\label{eq:avoid crossing requirement}
\end{equation}
The modal frequencies of coupled waveguides are shown in Fig.$\,$\ref{fig:dissimilar-waveguide}(a) as the $w_{1}$ is varied from $1\,\mathrm{\mu m}$ to $2.5\,\mathrm{\mu m}$. From the plot, we observe four avoid crossings, corresponding to four
regions of modal coupling: I : $(b_{1},a_{2})$, II: $(a_{1},b_{1})$,
III : $(b_{1},c_{2})$, IV : $(a_{1},c_{2})$. Fig.$\,$\ref{fig:dissimilar-waveguide}(c) displays the displacement fields of the four coupling regions at the minimum frequency differences. Similar to the identical waveguides, $g$ values of different modes also exhibit an exponential dependency on $gap$ (Fig.$\,$\ref{fig:dissimilar-waveguide}(b)).
Among four cases, coupling II is particularly interesting because it represents a special mechanism of ``self coupling'' between mode \emph{a} and \emph{b} within waveguide 1, corresponding to a double tunneling process that mode \emph{a} and mode \emph{b} in waveguide 1 both couple to waveguide 2, which mediates the coupling of mode \emph{a} and \emph{b}. The double tunneling mechanism
could also explain the much faster decaying of the coupling strength for case II. With such a mechanism, we would expect the realization of a single-waveguide mode converter \cite{shen2019apl} in the future with the assistance of an ancillary waveguide. 

\section{Coupling to IDT}

\begin{figure}
\includegraphics[width=8.8cm]{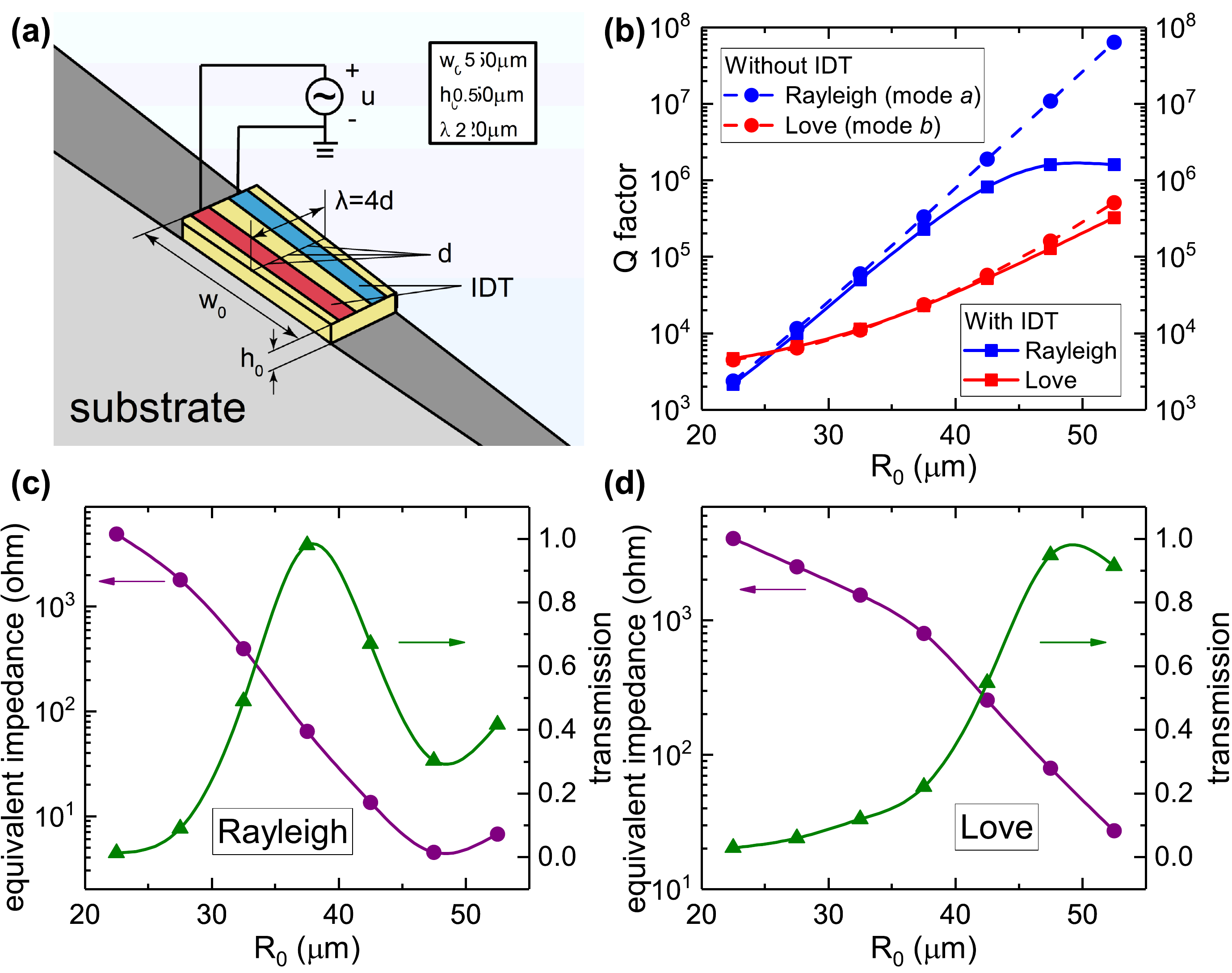}\caption{(a) Schematic illustration of the IDT integrated ring resonator, with
the electrodes of IDTs connected to external microwave cable (not shown) for signal input. (b) The influence
of the IDT on the mode $Q$ factors. (c) and (d) The equivalent impedance
and the energy transmission (conversion efficiency) from microwave to phonon as functions of
the ring radius for S-Rayleigh and A-Love modes, respectively.
}
\label{fig:couple with IDT}
\end{figure}

Aside from the vibrational properties of phononic waveguides and ring resonators, the excitation and detection of phonons are also of practical importance
and interests, for example, in phononic implementation of  microwave delay lines and filters. Based on the model of ring resonator presented in Sec.$\,$\mbox{III}, we add IDT electrodes on the top surface of the ring resonator (Fig.$\,$\ref{fig:couple with IDT}(a)), and numerically investigate the coupling between IDT and phononic ring resonators and its efficiency to excite phonons. In this model, only the electrical effect of the IDT electrodes to the resonator is considered, while the mechanical and other effects due to loading of the electrodes are ignored (such as mass loading).


We first evaluate the quality factor of the mechanical ring resonator under the presence of IDT electrodes. The width of each electrode is $1/4$ of the acoustic wavelength and the electrodes cover half of the ring resonator area. For the S-Rayleigh mode \emph{a }with wavelength of $2\,\mathrm{\mu m}$ in a $5\,\mathrm{\mu m}$-wide ring resonator without IDT, the quality factor increases exponentially with the radius of the resonator as the quality factor is limited by the radiation loss; with the IDT electrodes, the quality factor saturates when the radius is larger than $40\,\mathrm{\mu m}$, as shown in Fig.$\,$\ref{fig:couple with IDT}(b). The presence of the electrodes modifies the local phonon velocity, due to the $\delta v/v$ effect \cite{Hughes1972}. Hence, the IDTs break the cylindrical symmetry of the perfect ring, leading to extra coupling of the phonon mode in ring to the leaky phonon modes in the bulk substrate. The same effect also happens to the Love mode. When the radius is small and radiation loss is dominant, the quality factors with and without IDT electrodes are almost the same. At larger radius in our simulation, the IDT electrodes induced loss becomes comparable with radiation loss and the quality factor with IDT electrodes are slightly lower. 

To evaluate the external coupling to the phonon resonator, we also simulate the equivalent impedance of the IDT coupled mechanical ring resonator in order to achieve impedance matching to $Z_0 = 50\,\Omega$ transmission lines for maximal acoustic launching efficiency. In the COMSOL simulation, by assigning voltage $V$ on the electrodes and measuring the current flow $I$ on the electrodes at mechanical resonance frequency, the equivalent impedance, $Z = V/I$, can be extracted. The microwave reflectivity can be expressed by
\begin{equation}
r=\frac{Z-Z_{0}}{Z+Z_{0}}
\end{equation}
and the corresponding microwave to mechanical energy conversion efficiency $T= 1-\left|r^{2}\right|$. The simulated results are plotted in Fig.$\,$\ref{fig:couple with IDT}(c) and (d) for quasi-Rayleigh and quasi-Love mode, respectively. The results indicate that the effective impedance of the IDT coupled mechanical resonator decreases exponentially with the radius and the impedance matching condition $Z=Z_{0}$ can be satisfied with a certain radius, which gives $100\%$ microwave-to-phonon
conversion efficiency. To explain the dependence of the impedance on the radius, we consider the Butterworth-Van Dyke circuit model of the piezomechanical resonator which consists of a static capacitance $C_0$ in parallel with a motional series RLC circuit. The admittance of the circuit is \cite{Dahmani2020pral}
\begin{align}
    Y(\omega)=&-i\omega C_0+\frac{1}{1/(-i\omega C_m)-i\omega L_m+R_m}\nonumber\\
    =&-i\omega C_0+i\omega_s^2C_m\frac{\omega}{\omega^2+i\omega_s\omega/Q_i-\omega_s^2}\nonumber\\
    =&-i\omega C_0+\frac{iC_m\omega_s^2\omega(\omega^2-\omega_s^2)}{(\omega^2-\omega_s^2)^2+\omega_s^2\omega^2/Q_i^2}\nonumber \\
    &+\frac{\omega_s^3\omega^2C_m/Q_i}{(\omega^2-\omega_s^2)^2+\omega_s^2\omega^2/Q_i^2}
\end{align}
where $C_m$, $L_m$ and $R_m$ are the motional capacitance, inductance and resistance. The mechancial resonance $\omega_s=1/\sqrt{L_mC_m}$ and its intrinsic quality factor $Q_i=1/\omega_sR_mC_m$. Note that $\omega_s$ is the resonant frequency without the presence of the IDT electrodes, and suppose that with IDT on it, the resonant frequency is $\omega_f$. This resonant frequency shift originates from the $\delta v/v$ effect and the value $(\omega_f-\omega_s)/\omega_s$ equals $\frac{1}{2}\delta v/v$ (as the electrodes cover only half of the ring resonator). In the regime where the quality factor is limited by intrinsic radiation loss instead of the $\delta v/v$ scattering, we have $(\omega_f-\omega_s)/\omega_s=\delta v/v\ll1/Q_i$. In this case, the admittance can be simplified to
\begin{equation}
    Y(\omega_f) = -i\omega_f C_0 + \omega_f C_mQ_i
\end{equation}
with the real part being dominant, we get the relation between the equivalent impedance and its quality factor
\begin{equation}
    Z(\omega_f)=\frac{1}{\omega_f C_m Q_i}
\end{equation}
We see that the equivalent impedance is inversely proportional to its quality factor. The motional capacitance $C_m$ is proportional to the electrode area like $C_0$ \cite{Arnau2000rosi}, so it is linearly proportional to the radius as the surface area increases while the quality factor increases exponentially with radius. So in Fig.$\,$\ref{fig:couple with IDT}(c) and (d), the impedance's dependence with radius is mostly exponential, consistent with the quality factor simulated in Fig.$\,$\ref{fig:couple with IDT}(b). The microwave-to-phonon conversion efficiency can also be understood from coupling condition point of view where the external coupling between the electrodes and the resonator is weak, but by changing the resonator's intrinsic energy decay rate, critical coupling condition can be satisfied and all the microwave energy can be converted into mechanical energy.

Therefore, the IDT integrated mechanical resonator can enhance microwave-phonon conversion efficiency, boost the performances of applications that require high excitation and collection efficiency and provide coherent interface between phononic and superconducting circuits.

\section{Conclusion}

In conclusion, we have quantitatively studied phononic mode properties in unsuspended strip waveguides and ring resonators. Our numerical results demonstrate efficient confinement of phonon in the practical GaN-on-sapphire microstructures, which provide a scalable platform of phononic integrated circuits for acoustic signal processing and enhanced phonon-matter or phonon-light interactions. Basic components for a phononic circuit are thoroughly investigated, including the directional coupler made by two waveguides, quasi-Rayleigh mode to quasi-Love mode conversion in coupled non-identical waveguides, ring resonator, as well as microwave photon-to-phonon conversion for efficient input and output coupling. It is worth noting that the studies in this paper can be generalized to other frequencies by rescaling the geometry parameters. For example, $10\,$GHz phononic waveguide modes could be realized with a  thickness of $~125\,$nm and a width of $500\,$nm, whose geometry is compatible with the photonic integrated circuits and promises the applications in optical Brillouin scattering. 

Although not addressed in this paper, it is worth noting that there
are various other loss mechanisms \cite{Imboden14pr}. Some losses don't
exist or can be neglected in our strip waveguide SAW devices, for
example, our system doesn't use any suspended structures so it isn't
subject to clamping or support losses. For the circuit damping caused
by drive-and-detection circuit, we evaluate the equivalent impedance
and $Q$ factor in Sec. V. Scattering losses from crystal defects
or thermal phonons are ignored, though thermoelastic damping can be
one primary limits of the $Q$ factors at room temperature. Recently,
people has done phononic band structure engineering to reduce scattering
of SAW into bulk modes \cite{Xu2018apl, Shao19pra}. However, scattering
losses from surface roughness due to fabrication disorder, can be
critical for miniaturizing phononic systems and thus for high-frequency
application \cite{Safavi-Naeini19o}, since the scattering loss coefficient
has a quadratic dependency with repect to the guided wave frequency
\cite{Payne94oqe, Hughes05prl, Melati14aop}, it could be the main limitation
to efficient photo-phonon interaction \cite{Safavi-Naeini19o, Liu19o}.
In addition, scaling to smaller geometry increases surface-to-volume
ratios which may give rise to more surface loss, this loss mechanism
has been found in other micromechanical systems \cite{Yasumura00jms}.
Besides, we do not consider material losses due to any intrinsic microscopic
processes. The damping from surrounding medium contributes to additional
loss if the devices are operating under atmosphere or viscous medium.

We believe the phononic waveguides and ring resonators studied in this paper can be applied in future studies on phononic circuits \cite{OlssonIII2009,Maldovan2013,Hatanaka2014},
gyroscopic sensors \cite{Lao1980,Zou2016,Fu2019nc},
integrated acoustic-optics modulators \cite{Fuhrmann2011,Tadesse2014,Uan2015,Tadesse2015},
circulators \cite{Fu2015}, integrated delay line and data bus that
communicates the quantum bits \cite{Gustafsson2014}. 
\begin{acknowledgments}
This work was supported by DARPA/MTO's PRIGM: AIMS program through a grant from SPAWAR (N66001-16-1-4026).
H.X.T. acknowledges funding support from Army Research Office (W911NF-18-1-0020) and Packard Fellowship in Science and Engineering. C.L.Z. thanks Liang Jiang for helpful discussions.  
\end{acknowledgments}

\section*{Data Availability}
The data that support the findings of this study are available from the corresponding author upon reasonable request.

\bibliographystyle{Zou}
\phantomsection\addcontentsline{toc}{section}{\refname}\bibliography{SAWsimulation}

\begin{thebibliography}{111}%
\makeatletter
\providecommand \@ifxundefined [1]{%
 \@ifx{#1\undefined}
}%
\providecommand \@ifnum [1]{%
 \ifnum #1\expandafter \@firstoftwo
 \else \expandafter \@secondoftwo
 \fi
}%
\providecommand \@ifx [1]{%
 \ifx #1\expandafter \@firstoftwo
 \else \expandafter \@secondoftwo
 \fi
}%
\providecommand \natexlab [1]{#1}%
\providecommand \enquote  [1]{``#1''}%
\providecommand \bibnamefont  [1]{#1}%
\providecommand \bibfnamefont [1]{#1}%
\providecommand \citenamefont [1]{#1}%
\providecommand \href@noop [0]{\@secondoftwo}%
\providecommand \href [0]{\begingroup \@sanitize@url \@href}%
\providecommand \@href[1]{\@@startlink{#1}\@@href}%
\providecommand \@@href[1]{\endgroup#1\@@endlink}%
\providecommand \@sanitize@url [0]{\catcode `\\12\catcode `\$12\catcode
  `\&12\catcode `\#12\catcode `\^12\catcode `\_12\catcode `\%12\relax}%
\providecommand \@@startlink[1]{}%
\providecommand \@@endlink[0]{}%
\providecommand \url  [0]{\begingroup\@sanitize@url \@url }%
\providecommand \@url [1]{\endgroup\@href {#1}{\urlprefix }}%
\providecommand \urlprefix  [0]{URL }%
\providecommand \Eprint [0]{\href }%
\providecommand \doibase [0]{http://dx.doi.org/}%
\providecommand \selectlanguage [0]{\@gobble}%
\providecommand \bibinfo  [0]{\@secondoftwo}%
\providecommand \bibfield  [0]{\@secondoftwo}%
\providecommand \translation [1]{[#1]}%
\providecommand \BibitemOpen [0]{}%
\providecommand \bibitemStop [0]{}%
\providecommand \bibitemNoStop [0]{.\EOS\space}%
\providecommand \EOS [0]{\spacefactor3000\relax}%
\providecommand \BibitemShut  [1]{\csname bibitem#1\endcsname}%
\let\auto@bib@innerbib\@empty
\bibitem [{\citenamefont {Campbell}(1998)}]{campbell1998surface}%
  \BibitemOpen
  \bibfield  {author} {\bibinfo {author} {\bibfnamefont {C.}~\bibnamefont
  {Campbell}},\ }\href@noop {} {\emph {\bibinfo {title} {Surface acoustic wave
  devices for mobile and wireless communications}}}\ (\bibinfo  {publisher}
  {Academic press},\ \bibinfo {year} {1998})\BibitemShut {NoStop}%
\bibitem [{\citenamefont {Lao}(1980)}]{Lao1980}%
  \BibitemOpen
  \bibfield  {author} {\bibinfo {author} {\bibfnamefont {B.~Y.}\ \bibnamefont
  {Lao}},\ }\bibfield  {title} {\enquote {\bibinfo {title} {{Gyroscopic Effect
  in Surface Acoustic Waves}},}\ }\href {\doibase 10.1109/ULTSYM.1980.197487}
  {\bibfield  {journal} {\bibinfo  {journal} {1980 Ultrason. Symp.}\ ,\
  \bibinfo {pages} {687}} (\bibinfo {year} {1980})}\BibitemShut {NoStop}%
\bibitem [{\citenamefont {Friend}\ and\ \citenamefont
  {Yeo}(2011)}]{Friend2011}%
  \BibitemOpen
  \bibfield  {author} {\bibinfo {author} {\bibfnamefont {J.}~\bibnamefont
  {Friend}}\ and\ \bibinfo {author} {\bibfnamefont {L.~Y.}\ \bibnamefont
  {Yeo}},\ }\bibfield  {title} {\enquote {\bibinfo {title} {{Microscale
  acoustofluidics: Microfluidics driven via acoustics and ultrasonics}},}\
  }\href {\doibase 10.1103/RevModPhys.83.647} {\bibfield  {journal} {\bibinfo
  {journal} {Rev. Mod. Phys.}\ }\textbf {\bibinfo {volume} {83}},\ \bibinfo
  {pages} {647} (\bibinfo {year} {2011})}\BibitemShut {NoStop}%
\bibitem [{\citenamefont {Campbell}(1989)}]{Campbell1989}%
  \BibitemOpen
  \bibfield  {author} {\bibinfo {author} {\bibfnamefont {C.~K.}\ \bibnamefont
  {Campbell}},\ }\bibfield  {title} {\enquote {\bibinfo {title} {{Applications
  of surface acoustic and shallow bulk acoustic wave devices}},}\ }\href
  {\doibase 10.1109/5.40664} {\bibfield  {journal} {\bibinfo  {journal} {Proc.
  IEEE}\ }\textbf {\bibinfo {volume} {77}},\ \bibinfo {pages} {1453} (\bibinfo
  {year} {1989})}\BibitemShut {NoStop}%
\bibitem [{\citenamefont {Mcneil}\ \emph {et~al.}(2011)\citenamefont {Mcneil},
  \citenamefont {Kataoka}, \citenamefont {Ford}, \citenamefont {Barnes},
  \citenamefont {Anderson}, \citenamefont {Jones}, \citenamefont {Farrer},\
  and\ \citenamefont {Ritchie}}]{Mcneil2011}%
  \BibitemOpen
  \bibfield  {author} {\bibinfo {author} {\bibfnamefont {R.~P.~G.}\
  \bibnamefont {Mcneil}}, \bibinfo {author} {\bibfnamefont {M.}~\bibnamefont
  {Kataoka}}, \bibinfo {author} {\bibfnamefont {C.~J.~B.}\ \bibnamefont
  {Ford}}, \bibinfo {author} {\bibfnamefont {C.~H.~W.}\ \bibnamefont {Barnes}},
  \bibinfo {author} {\bibfnamefont {D.}~\bibnamefont {Anderson}}, \bibinfo
  {author} {\bibfnamefont {G.~A.~C.}\ \bibnamefont {Jones}}, \bibinfo {author}
  {\bibfnamefont {I.}~\bibnamefont {Farrer}}, \ and\ \bibinfo {author}
  {\bibfnamefont {D.~A.}\ \bibnamefont {Ritchie}},\ }\bibfield  {title}
  {\enquote {\bibinfo {title} {{On-demand single-electron transfer between
  distant quantum dots}},}\ }\href {\doibase 10.1038/nature10444} {\bibfield
  {journal} {\bibinfo  {journal} {Nature}\ }\textbf {\bibinfo {volume} {477}},\
  \bibinfo {pages} {439} (\bibinfo {year} {2011})}\BibitemShut {NoStop}%
\bibitem [{\citenamefont {Hermelin}\ \emph {et~al.}(2011)\citenamefont
  {Hermelin}, \citenamefont {Takada}, \citenamefont {Yamamoto}, \citenamefont
  {Tarucha}, \citenamefont {Wieck}, \citenamefont {Saminadayar}, \citenamefont
  {B{\"a}uerle},\ and\ \citenamefont {Meunier}}]{Hermelin2011}%
  \BibitemOpen
  \bibfield  {author} {\bibinfo {author} {\bibfnamefont {S.}~\bibnamefont
  {Hermelin}}, \bibinfo {author} {\bibfnamefont {S.}~\bibnamefont {Takada}},
  \bibinfo {author} {\bibfnamefont {M.}~\bibnamefont {Yamamoto}}, \bibinfo
  {author} {\bibfnamefont {S.}~\bibnamefont {Tarucha}}, \bibinfo {author}
  {\bibfnamefont {A.~D.}\ \bibnamefont {Wieck}}, \bibinfo {author}
  {\bibfnamefont {L.}~\bibnamefont {Saminadayar}}, \bibinfo {author}
  {\bibfnamefont {C.}~\bibnamefont {B{\"a}uerle}}, \ and\ \bibinfo {author}
  {\bibfnamefont {T.}~\bibnamefont {Meunier}},\ }\bibfield  {title} {\enquote
  {\bibinfo {title} {Electrons surfing on a sound wave as a platform for
  quantum optics with flying electrons},}\ }\href {\doibase
  10.1038/nature10416} {\bibfield  {journal} {\bibinfo  {journal} {Nature}\
  }\textbf {\bibinfo {volume} {477}},\ \bibinfo {pages} {435} (\bibinfo {year}
  {2011})}\BibitemShut {NoStop}%
\bibitem [{\citenamefont {Chen}\ \emph {et~al.}(2015)\citenamefont {Chen},
  \citenamefont {Sato}, \citenamefont {Kosaka}, \citenamefont {Hashisaka},
  \citenamefont {Muraki},\ and\ \citenamefont {Fujisawa}}]{Chen2015}%
  \BibitemOpen
  \bibfield  {author} {\bibinfo {author} {\bibfnamefont {J.~C.~H.}\
  \bibnamefont {Chen}}, \bibinfo {author} {\bibfnamefont {Y.}~\bibnamefont
  {Sato}}, \bibinfo {author} {\bibfnamefont {R.}~\bibnamefont {Kosaka}},
  \bibinfo {author} {\bibfnamefont {M.}~\bibnamefont {Hashisaka}}, \bibinfo
  {author} {\bibfnamefont {K.}~\bibnamefont {Muraki}}, \ and\ \bibinfo {author}
  {\bibfnamefont {T.}~\bibnamefont {Fujisawa}},\ }\bibfield  {title} {\enquote
  {\bibinfo {title} {{Enhanced electron-phonon coupling for a semiconductor
  charge qubit in a surface phonon cavity}},}\ }\href {\doibase
  10.1038/srep15176} {\bibfield  {journal} {\bibinfo  {journal} {Sci. Rep.}\
  }\textbf {\bibinfo {volume} {5}},\ \bibinfo {pages} {15176} (\bibinfo {year}
  {2015})}\BibitemShut {NoStop}%
\bibitem [{\citenamefont {Golter}\ \emph
  {et~al.}(2016{\natexlab{a}})\citenamefont {Golter}, \citenamefont {Oo},
  \citenamefont {Amezcua}, \citenamefont {Stewart},\ and\ \citenamefont
  {Wang}}]{Golter2016}%
  \BibitemOpen
  \bibfield  {author} {\bibinfo {author} {\bibfnamefont {D.~A.}\ \bibnamefont
  {Golter}}, \bibinfo {author} {\bibfnamefont {T.}~\bibnamefont {Oo}}, \bibinfo
  {author} {\bibfnamefont {M.}~\bibnamefont {Amezcua}}, \bibinfo {author}
  {\bibfnamefont {K.~A.}\ \bibnamefont {Stewart}}, \ and\ \bibinfo {author}
  {\bibfnamefont {H.}~\bibnamefont {Wang}},\ }\bibfield  {title} {\enquote
  {\bibinfo {title} {{Optomechanical Quantum Control of a Nitrogen-Vacancy
  Center in Diamond}},}\ }\href {\doibase 10.1103/PhysRevLett.116.143602}
  {\bibfield  {journal} {\bibinfo  {journal} {Phys. Rev. Lett.}\ }\textbf
  {\bibinfo {volume} {116}},\ \bibinfo {pages} {143602} (\bibinfo {year}
  {2016}{\natexlab{a}})}\BibitemShut {NoStop}%
\bibitem [{\citenamefont {Gustafsson}\ \emph {et~al.}(2014)\citenamefont
  {Gustafsson}, \citenamefont {Aref}, \citenamefont {Kockum}, \citenamefont
  {Ekstrom}, \citenamefont {Johansson},\ and\ \citenamefont
  {Delsing}}]{Gustafsson2014}%
  \BibitemOpen
  \bibfield  {author} {\bibinfo {author} {\bibfnamefont {M.~V.}\ \bibnamefont
  {Gustafsson}}, \bibinfo {author} {\bibfnamefont {T.}~\bibnamefont {Aref}},
  \bibinfo {author} {\bibfnamefont {A.~F.}\ \bibnamefont {Kockum}}, \bibinfo
  {author} {\bibfnamefont {M.~K.}\ \bibnamefont {Ekstrom}}, \bibinfo {author}
  {\bibfnamefont {G.}~\bibnamefont {Johansson}}, \ and\ \bibinfo {author}
  {\bibfnamefont {P.}~\bibnamefont {Delsing}},\ }\bibfield  {title} {\enquote
  {\bibinfo {title} {{Propagating phonons coupled to an artificial atom}},}\
  }\href {\doibase 10.1126/science.1257219} {\bibfield  {journal} {\bibinfo
  {journal} {Science}\ }\textbf {\bibinfo {volume} {346}},\ \bibinfo {pages}
  {207} (\bibinfo {year} {2014})}\BibitemShut {NoStop}%
\bibitem [{\citenamefont {Satzinger}\ \emph {et~al.}(2018)\citenamefont
  {Satzinger}, \citenamefont {Zhong}, \citenamefont {Chang}, \citenamefont
  {Peairs}, \citenamefont {Bienfait}, \citenamefont {Chou}, \citenamefont
  {Cleland}, \citenamefont {Conner}, \citenamefont {Dumur}, \citenamefont
  {Grebel} \emph {et~al.}}]{Satzinger2018n}%
  \BibitemOpen
  \bibfield  {author} {\bibinfo {author} {\bibfnamefont {K.~J.}\ \bibnamefont
  {Satzinger}}, \bibinfo {author} {\bibfnamefont {Y.}~\bibnamefont {Zhong}},
  \bibinfo {author} {\bibfnamefont {H.-S.}\ \bibnamefont {Chang}}, \bibinfo
  {author} {\bibfnamefont {G.~A.}\ \bibnamefont {Peairs}}, \bibinfo {author}
  {\bibfnamefont {A.}~\bibnamefont {Bienfait}}, \bibinfo {author}
  {\bibfnamefont {M.-H.}\ \bibnamefont {Chou}}, \bibinfo {author}
  {\bibfnamefont {A.}~\bibnamefont {Cleland}}, \bibinfo {author} {\bibfnamefont
  {C.~R.}\ \bibnamefont {Conner}}, \bibinfo {author} {\bibfnamefont
  {{\'E}.}~\bibnamefont {Dumur}}, \bibinfo {author} {\bibfnamefont
  {J.}~\bibnamefont {Grebel}},  \emph {et~al.},\ }\bibfield  {title} {\enquote
  {\bibinfo {title} {Quantum control of surface acoustic-wave phonons},}\
  }\href {\doibase 10.1038/s41586-018-0719-5} {\bibfield  {journal} {\bibinfo
  {journal} {Nature}\ }\textbf {\bibinfo {volume} {563}},\ \bibinfo {pages}
  {661} (\bibinfo {year} {2018})}\BibitemShut {NoStop}%
\bibitem [{\citenamefont {Bienfait}\ \emph {et~al.}(2019)\citenamefont
  {Bienfait}, \citenamefont {Satzinger}, \citenamefont {Zhong}, \citenamefont
  {Chang}, \citenamefont {Chou}, \citenamefont {Conner}, \citenamefont {Dumur},
  \citenamefont {Grebel}, \citenamefont {Peairs}, \citenamefont {Povey} \emph
  {et~al.}}]{Bienfait2019s}%
  \BibitemOpen
  \bibfield  {author} {\bibinfo {author} {\bibfnamefont {A.}~\bibnamefont
  {Bienfait}}, \bibinfo {author} {\bibfnamefont {K.~J.}\ \bibnamefont
  {Satzinger}}, \bibinfo {author} {\bibfnamefont {Y.}~\bibnamefont {Zhong}},
  \bibinfo {author} {\bibfnamefont {H.-S.}\ \bibnamefont {Chang}}, \bibinfo
  {author} {\bibfnamefont {M.-H.}\ \bibnamefont {Chou}}, \bibinfo {author}
  {\bibfnamefont {C.~R.}\ \bibnamefont {Conner}}, \bibinfo {author}
  {\bibfnamefont {{\'E}.}~\bibnamefont {Dumur}}, \bibinfo {author}
  {\bibfnamefont {J.}~\bibnamefont {Grebel}}, \bibinfo {author} {\bibfnamefont
  {G.~A.}\ \bibnamefont {Peairs}}, \bibinfo {author} {\bibfnamefont {R.~G.}\
  \bibnamefont {Povey}},  \emph {et~al.},\ }\bibfield  {title} {\enquote
  {\bibinfo {title} {Phonon-mediated quantum state transfer and remote qubit
  entanglement},}\ }\href {\doibase 10.1126/science.aaw8415} {\bibfield
  {journal} {\bibinfo  {journal} {Science}\ }\textbf {\bibinfo {volume}
  {364}},\ \bibinfo {pages} {368} (\bibinfo {year} {2019})}\BibitemShut
  {NoStop}%
\bibitem [{\citenamefont {Fuhrmann}\ \emph {et~al.}(2011)\citenamefont
  {Fuhrmann}, \citenamefont {Thon}, \citenamefont {Kim}, \citenamefont
  {Bouwmeester}, \citenamefont {Petroff}, \citenamefont {Wixforth},\ and\
  \citenamefont {Krenner}}]{Fuhrmann2011}%
  \BibitemOpen
  \bibfield  {author} {\bibinfo {author} {\bibfnamefont {D.~A.}\ \bibnamefont
  {Fuhrmann}}, \bibinfo {author} {\bibfnamefont {S.~M.}\ \bibnamefont {Thon}},
  \bibinfo {author} {\bibfnamefont {H.}~\bibnamefont {Kim}}, \bibinfo {author}
  {\bibfnamefont {D.}~\bibnamefont {Bouwmeester}}, \bibinfo {author}
  {\bibfnamefont {P.~M.}\ \bibnamefont {Petroff}}, \bibinfo {author}
  {\bibfnamefont {A.}~\bibnamefont {Wixforth}}, \ and\ \bibinfo {author}
  {\bibfnamefont {H.~J.}\ \bibnamefont {Krenner}},\ }\bibfield  {title}
  {\enquote {\bibinfo {title} {{Dynamic modulation of photonic crystal
  nanocavities using gigahertz acoustic phonons}},}\ }\href {\doibase
  10.1038/nphoton.2011.208} {\bibfield  {journal} {\bibinfo  {journal} {Nat.
  Photonics}\ }\textbf {\bibinfo {volume} {5}},\ \bibinfo {pages} {605}
  (\bibinfo {year} {2011})}\BibitemShut {NoStop}%
\bibitem [{\citenamefont {Tadesse}\ and\ \citenamefont
  {Li}(2014{\natexlab{a}})}]{Tadesse2014}%
  \BibitemOpen
  \bibfield  {author} {\bibinfo {author} {\bibfnamefont {S.~A.}\ \bibnamefont
  {Tadesse}}\ and\ \bibinfo {author} {\bibfnamefont {M.}~\bibnamefont {Li}},\
  }\bibfield  {title} {\enquote {\bibinfo {title} {{Sub-optical wavelength
  acoustic wave modulation of integrated photonic resonators at microwave
  frequencies.}}}\ }\href {\doibase 10.1038/ncomms6402} {\bibfield  {journal}
  {\bibinfo  {journal} {Nat. Commun.}\ }\textbf {\bibinfo {volume} {5}},\
  \bibinfo {pages} {5402} (\bibinfo {year} {2014}{\natexlab{a}})}\BibitemShut
  {NoStop}%
\bibitem [{\citenamefont {Uan}\ \emph {et~al.}(2015)\citenamefont {Uan},
  \citenamefont {Adesse}, \citenamefont {Iu}, \citenamefont {Li}, \citenamefont
  {Tadesse}, \citenamefont {Liu},\ and\ \citenamefont {Li}}]{Uan2015}%
  \BibitemOpen
  \bibfield  {author} {\bibinfo {author} {\bibfnamefont {H.~L.~I.}\
  \bibnamefont {Uan}}, \bibinfo {author} {\bibfnamefont {S.~E. A.~T.}\
  \bibnamefont {Adesse}}, \bibinfo {author} {\bibfnamefont {Q.~I. Y. U.~L.}\
  \bibnamefont {Iu}}, \bibinfo {author} {\bibfnamefont {H.}~\bibnamefont {Li}},
  \bibinfo {author} {\bibfnamefont {S.~A.}\ \bibnamefont {Tadesse}}, \bibinfo
  {author} {\bibfnamefont {Q.}~\bibnamefont {Liu}}, \ and\ \bibinfo {author}
  {\bibfnamefont {M.}~\bibnamefont {Li}},\ }\bibfield  {title} {\enquote
  {\bibinfo {title} {{Nanophotonic cavity optomechanics with propagating
  acoustic waves at frequencies up to 12 GHz}},}\ }\href@noop {} {\bibfield
  {journal} {\bibinfo  {journal} {Optica}\ }\textbf {\bibinfo {volume} {2}},\
  \bibinfo {pages} {826} (\bibinfo {year} {2015})}\BibitemShut {NoStop}%
\bibitem [{\citenamefont {Tadesse}\ \emph {et~al.}(2015)\citenamefont
  {Tadesse}, \citenamefont {Li}, \citenamefont {Liu},\ and\ \citenamefont
  {Li}}]{Tadesse2015}%
  \BibitemOpen
  \bibfield  {author} {\bibinfo {author} {\bibfnamefont {S.~A.}\ \bibnamefont
  {Tadesse}}, \bibinfo {author} {\bibfnamefont {H.}~\bibnamefont {Li}},
  \bibinfo {author} {\bibfnamefont {Q.}~\bibnamefont {Liu}}, \ and\ \bibinfo
  {author} {\bibfnamefont {M.}~\bibnamefont {Li}},\ }\bibfield  {title}
  {\enquote {\bibinfo {title} {{Acousto-optic modulation of a photonic crystal
  nanocavity with Lamb waves in microwave K band}},}\ }\href {\doibase
  10.1063/1.4935981} {\bibfield  {journal} {\bibinfo  {journal} {Appl. Phys.
  Lett.}\ }\textbf {\bibinfo {volume} {107}} (\bibinfo {year} {2015}),\
  10.1063/1.4935981}\BibitemShut {NoStop}%
\bibitem [{\citenamefont {Kurizki}\ \emph {et~al.}(2015)\citenamefont
  {Kurizki}, \citenamefont {Bertet}, \citenamefont {Kubo}, \citenamefont
  {M{\o}lmer}, \citenamefont {Petrosyan}, \citenamefont {Rabl},\ and\
  \citenamefont {Schmiedmayer}}]{Kurizki2015}%
  \BibitemOpen
  \bibfield  {author} {\bibinfo {author} {\bibfnamefont {G.}~\bibnamefont
  {Kurizki}}, \bibinfo {author} {\bibfnamefont {P.}~\bibnamefont {Bertet}},
  \bibinfo {author} {\bibfnamefont {Y.}~\bibnamefont {Kubo}}, \bibinfo {author}
  {\bibfnamefont {K.}~\bibnamefont {M{\o}lmer}}, \bibinfo {author}
  {\bibfnamefont {D.}~\bibnamefont {Petrosyan}}, \bibinfo {author}
  {\bibfnamefont {P.}~\bibnamefont {Rabl}}, \ and\ \bibinfo {author}
  {\bibfnamefont {J.}~\bibnamefont {Schmiedmayer}},\ }\bibfield  {title}
  {\enquote {\bibinfo {title} {{Quantum technologies with hybrid systems}},}\
  }\href {\doibase 10.1073/pnas.1419326112} {\bibfield  {journal} {\bibinfo
  {journal} {Proc. Natl. Acad. Sci.}\ }\textbf {\bibinfo {volume} {112}},\
  \bibinfo {pages} {3866} (\bibinfo {year} {2015})}\BibitemShut {NoStop}%
\bibitem [{\citenamefont {Schuetz}\ \emph {et~al.}(2015)\citenamefont
  {Schuetz}, \citenamefont {Kessler}, \citenamefont {Giedke}, \citenamefont
  {Vandersypen}, \citenamefont {Lukin},\ and\ \citenamefont
  {Cirac}}]{Schuetz2015}%
  \BibitemOpen
  \bibfield  {author} {\bibinfo {author} {\bibfnamefont {M.~J.~A.}\
  \bibnamefont {Schuetz}}, \bibinfo {author} {\bibfnamefont {E.~M.}\
  \bibnamefont {Kessler}}, \bibinfo {author} {\bibfnamefont {G.}~\bibnamefont
  {Giedke}}, \bibinfo {author} {\bibfnamefont {L.~M.~K.}\ \bibnamefont
  {Vandersypen}}, \bibinfo {author} {\bibfnamefont {M.~D.}\ \bibnamefont
  {Lukin}}, \ and\ \bibinfo {author} {\bibfnamefont {J.~I.}\ \bibnamefont
  {Cirac}},\ }\bibfield  {title} {\enquote {\bibinfo {title} {{Universal
  Quantum Transducers Based on Surface Acoustic Waves}},}\ }\href {\doibase
  10.1103/PhysRevX.5.031031} {\bibfield  {journal} {\bibinfo  {journal} {Phys.
  Rev. X}\ }\textbf {\bibinfo {volume} {5}},\ \bibinfo {pages} {031031}
  (\bibinfo {year} {2015})}\BibitemShut {NoStop}%
\bibitem [{\citenamefont {Shumeiko}(2016)}]{Shumeiko2016}%
  \BibitemOpen
  \bibfield  {author} {\bibinfo {author} {\bibfnamefont {V.~S.}\ \bibnamefont
  {Shumeiko}},\ }\bibfield  {title} {\enquote {\bibinfo {title} {{Quantum
  acousto-optic transducer for superconducting qubits}},}\ }\href {\doibase
  10.1103/PhysRevA.93.023838} {\bibfield  {journal} {\bibinfo  {journal} {Phys.
  Rev. A}\ }\textbf {\bibinfo {volume} {93}},\ \bibinfo {pages} {023838}
  (\bibinfo {year} {2016})}\BibitemShut {NoStop}%
\bibitem [{\citenamefont {Clerk}\ \emph {et~al.}(2020)\citenamefont {Clerk},
  \citenamefont {Lehnert}, \citenamefont {Bertet}, \citenamefont {Petta},\ and\
  \citenamefont {Nakamura}}]{Clerk2020np}%
  \BibitemOpen
  \bibfield  {author} {\bibinfo {author} {\bibfnamefont {A.}~\bibnamefont
  {Clerk}}, \bibinfo {author} {\bibfnamefont {K.}~\bibnamefont {Lehnert}},
  \bibinfo {author} {\bibfnamefont {P.}~\bibnamefont {Bertet}}, \bibinfo
  {author} {\bibfnamefont {J.}~\bibnamefont {Petta}}, \ and\ \bibinfo {author}
  {\bibfnamefont {Y.}~\bibnamefont {Nakamura}},\ }\bibfield  {title} {\enquote
  {\bibinfo {title} {Hybrid quantum systems with circuit quantum
  electrodynamics},}\ }\href {\doibase 10.1038/s41567-020-0797-9} {\bibfield
  {journal} {\bibinfo  {journal} {Nature Physics}\ ,\ \bibinfo {pages} {1}}
  (\bibinfo {year} {2020})}\BibitemShut {NoStop}%
\bibitem [{\citenamefont {Manenti}\ \emph {et~al.}(2017)\citenamefont
  {Manenti}, \citenamefont {Kockum}, \citenamefont {Patterson}, \citenamefont
  {Behrle}, \citenamefont {Rahamim}, \citenamefont {Tancredi}, \citenamefont
  {Nori},\ and\ \citenamefont {Leek}}]{Manenti2017nc}%
  \BibitemOpen
  \bibfield  {author} {\bibinfo {author} {\bibfnamefont {R.}~\bibnamefont
  {Manenti}}, \bibinfo {author} {\bibfnamefont {A.~F.}\ \bibnamefont {Kockum}},
  \bibinfo {author} {\bibfnamefont {A.}~\bibnamefont {Patterson}}, \bibinfo
  {author} {\bibfnamefont {T.}~\bibnamefont {Behrle}}, \bibinfo {author}
  {\bibfnamefont {J.}~\bibnamefont {Rahamim}}, \bibinfo {author} {\bibfnamefont
  {G.}~\bibnamefont {Tancredi}}, \bibinfo {author} {\bibfnamefont
  {F.}~\bibnamefont {Nori}}, \ and\ \bibinfo {author} {\bibfnamefont {P.~J.}\
  \bibnamefont {Leek}},\ }\bibfield  {title} {\enquote {\bibinfo {title}
  {Circuit quantum acoustodynamics with surface acoustic waves},}\ }\href
  {\doibase 10.1038/s41467-017-01063-9} {\bibfield  {journal} {\bibinfo
  {journal} {Nature communications}\ }\textbf {\bibinfo {volume} {8}},\
  \bibinfo {pages} {1} (\bibinfo {year} {2017})}\BibitemShut {NoStop}%
\bibitem [{\citenamefont {Moores}\ \emph {et~al.}(2018)\citenamefont {Moores},
  \citenamefont {Sletten}, \citenamefont {Viennot},\ and\ \citenamefont
  {Lehnert}}]{Moores2018prl}%
  \BibitemOpen
  \bibfield  {author} {\bibinfo {author} {\bibfnamefont {B.~A.}\ \bibnamefont
  {Moores}}, \bibinfo {author} {\bibfnamefont {L.~R.}\ \bibnamefont {Sletten}},
  \bibinfo {author} {\bibfnamefont {J.~J.}\ \bibnamefont {Viennot}}, \ and\
  \bibinfo {author} {\bibfnamefont {K.}~\bibnamefont {Lehnert}},\ }\bibfield
  {title} {\enquote {\bibinfo {title} {Cavity quantum acoustic device in the
  multimode strong coupling regime},}\ }\href {\doibase
  10.1103/PhysRevLett.120.227701} {\bibfield  {journal} {\bibinfo  {journal}
  {Physical review letters}\ }\textbf {\bibinfo {volume} {120}},\ \bibinfo
  {pages} {227701} (\bibinfo {year} {2018})}\BibitemShut {NoStop}%
\bibitem [{\citenamefont {Noguchi}\ \emph {et~al.}(2017)\citenamefont
  {Noguchi}, \citenamefont {Yamazaki}, \citenamefont {Tabuchi},\ and\
  \citenamefont {Nakamura}}]{Noguchi2017prl}%
  \BibitemOpen
  \bibfield  {author} {\bibinfo {author} {\bibfnamefont {A.}~\bibnamefont
  {Noguchi}}, \bibinfo {author} {\bibfnamefont {R.}~\bibnamefont {Yamazaki}},
  \bibinfo {author} {\bibfnamefont {Y.}~\bibnamefont {Tabuchi}}, \ and\
  \bibinfo {author} {\bibfnamefont {Y.}~\bibnamefont {Nakamura}},\ }\bibfield
  {title} {\enquote {\bibinfo {title} {Qubit-assisted transduction for a
  detection of surface acoustic waves near the quantum limit},}\ }\href
  {\doibase 10.1103/PhysRevLett.119.180505} {\bibfield  {journal} {\bibinfo
  {journal} {Physical review letters}\ }\textbf {\bibinfo {volume} {119}},\
  \bibinfo {pages} {180505} (\bibinfo {year} {2017})}\BibitemShut {NoStop}%
\bibitem [{\citenamefont {Yen}(1972)}]{Yen1972}%
  \BibitemOpen
  \bibfield  {author} {\bibinfo {author} {\bibfnamefont {K.~H.}\ \bibnamefont
  {Yen}},\ }\bibfield  {title} {\enquote {\bibinfo {title} {{Broadband
  Efficient Excitation of the Thin-Ribbon Waveguide for Surface Acoustic
  Waves}},}\ }\href {\doibase 10.1063/1.1654151} {\bibfield  {journal}
  {\bibinfo  {journal} {Appl. Phys. Lett.}\ }\textbf {\bibinfo {volume} {20}},\
  \bibinfo {pages} {284} (\bibinfo {year} {1972})}\BibitemShut {NoStop}%
\bibitem [{\citenamefont {Oliner}(1976)}]{Oliner1976}%
  \BibitemOpen
  \bibfield  {author} {\bibinfo {author} {\bibfnamefont {A.}~\bibnamefont
  {Oliner}},\ }\bibfield  {title} {\enquote {\bibinfo {title} {{Waveguides for
  acoustic surface waves: A review}},}\ }\href {\doibase
  10.1109/PROC.1976.10185} {\bibfield  {journal} {\bibinfo  {journal} {Proc.
  IEEE}\ }\textbf {\bibinfo {volume} {64}},\ \bibinfo {pages} {615} (\bibinfo
  {year} {1976})}\BibitemShut {NoStop}%
\bibitem [{\citenamefont {Oliner}(1978)}]{Oliner1978}%
  \BibitemOpen
  \bibfield  {author} {\bibinfo {author} {\bibfnamefont {A.~A.}\ \bibnamefont
  {Oliner}},\ }\bibfield  {title} {\enquote {\bibinfo {title} {{Waveguides for
  surface waves}},}\ }in\ \href {\doibase 10.1007/3-540-08575-0_12} {\emph
  {\bibinfo {booktitle} {Acoust. Surf. Waves}}}\ (\bibinfo  {publisher}
  {Springer Berlin Heidelberg},\ \bibinfo {year} {1978})\ pp.\ \bibinfo {pages}
  {187--223}\BibitemShut {NoStop}%
\bibitem [{\citenamefont {Knox}\ and\ \citenamefont {Owen}(1970)}]{Knox1970}%
  \BibitemOpen
  \bibfield  {author} {\bibinfo {author} {\bibfnamefont {R.}~\bibnamefont
  {Knox}}\ and\ \bibinfo {author} {\bibfnamefont {D.}~\bibnamefont {Owen}},\
  }\bibfield  {title} {\enquote {\bibinfo {title} {{Distributed Components in
  Microwave Elastic Surface Wave Circuits}},}\ }in\ \href {\doibase
  10.1109/GMTT.1970.1122850} {\emph {\bibinfo {booktitle} {G-MTT 1970 Int.
  Microw. Symp.}}},\ Vol.~\bibinfo {volume} {1}\ (\bibinfo  {publisher}
  {IEEE},\ \bibinfo {year} {1970})\ pp.\ \bibinfo {pages}
  {370--374}\BibitemShut {NoStop}%
\bibitem [{\citenamefont {Sandy}\ and\ \citenamefont
  {Parker}(1976)}]{Sandy1976}%
  \BibitemOpen
  \bibfield  {author} {\bibinfo {author} {\bibfnamefont {F.}~\bibnamefont
  {Sandy}}\ and\ \bibinfo {author} {\bibfnamefont {T.}~\bibnamefont {Parker}},\
  }\bibfield  {title} {\enquote {\bibinfo {title} {{Surface Acoustic Wave Ring
  Filter}},}\ }in\ \href {\doibase 10.1109/ULTSYM.1976.196704} {\emph {\bibinfo
  {booktitle} {1976 Ultrason. Symp.}}},\ Vol.~\bibinfo {volume} {79}\ (\bibinfo
   {publisher} {IEEE},\ \bibinfo {year} {1976})\ pp.\ \bibinfo {pages}
  {391--396}\BibitemShut {NoStop}%
\bibitem [{\citenamefont {Goryachev}\ \emph {et~al.}(2014)\citenamefont
  {Goryachev}, \citenamefont {Creedon}, \citenamefont {Ivanov}, \citenamefont
  {Galliou}, \citenamefont {Bourquin},\ and\ \citenamefont
  {Tobar}}]{Goryachev2014aipcp}%
  \BibitemOpen
  \bibfield  {author} {\bibinfo {author} {\bibfnamefont {M.}~\bibnamefont
  {Goryachev}}, \bibinfo {author} {\bibfnamefont {D.}~\bibnamefont {Creedon}},
  \bibinfo {author} {\bibfnamefont {E.}~\bibnamefont {Ivanov}}, \bibinfo
  {author} {\bibfnamefont {S.}~\bibnamefont {Galliou}}, \bibinfo {author}
  {\bibfnamefont {R.}~\bibnamefont {Bourquin}}, \ and\ \bibinfo {author}
  {\bibfnamefont {M.}~\bibnamefont {Tobar}},\ }\bibfield  {title} {\enquote
  {\bibinfo {title} {Extremely high q-factor mechanical modes in quartz bulk
  acoustic wave resonators at millikelvin temperature},}\ }in\ \href
  {https://aip.scitation.org/doi/abs/10.1063/1.4903104} {\emph {\bibinfo
  {booktitle} {AIP Conference Proceedings}}},\ Vol.\ \bibinfo {volume} {1633}\
  (\bibinfo {organization} {American Institute of Physics},\ \bibinfo {year}
  {2014})\ pp.\ \bibinfo {pages} {90--92}\BibitemShut {NoStop}%
\bibitem [{\citenamefont {Carvalho}\ \emph {et~al.}(2019)\citenamefont
  {Carvalho}, \citenamefont {Bourhill}, \citenamefont {Goryachev},
  \citenamefont {Galliou},\ and\ \citenamefont {Tobar}}]{Carvalho2019apl}%
  \BibitemOpen
  \bibfield  {author} {\bibinfo {author} {\bibfnamefont {N.}~\bibnamefont
  {Carvalho}}, \bibinfo {author} {\bibfnamefont {J.}~\bibnamefont {Bourhill}},
  \bibinfo {author} {\bibfnamefont {M.}~\bibnamefont {Goryachev}}, \bibinfo
  {author} {\bibfnamefont {S.}~\bibnamefont {Galliou}}, \ and\ \bibinfo
  {author} {\bibfnamefont {M.}~\bibnamefont {Tobar}},\ }\bibfield  {title}
  {\enquote {\bibinfo {title} {Piezo-optomechanical coupling of a 3d microwave
  resonator to a bulk acoustic wave crystalline resonator},}\ }\href
  {https://aip.scitation.org/doi/full/10.1063/1.5127997} {\bibfield  {journal}
  {\bibinfo  {journal} {Applied Physics Letters}\ }\textbf {\bibinfo {volume}
  {115}},\ \bibinfo {pages} {211102} (\bibinfo {year} {2019})}\BibitemShut
  {NoStop}%
\bibitem [{\citenamefont {Biryukov}\ \emph {et~al.}(2007)\citenamefont
  {Biryukov}, \citenamefont {Martin},\ and\ \citenamefont
  {Weihnacht}}]{Biryukov2007}%
  \BibitemOpen
  \bibfield  {author} {\bibinfo {author} {\bibfnamefont {S.~V.}\ \bibnamefont
  {Biryukov}}, \bibinfo {author} {\bibfnamefont {G.}~\bibnamefont {Martin}}, \
  and\ \bibinfo {author} {\bibfnamefont {M.}~\bibnamefont {Weihnacht}},\
  }\bibfield  {title} {\enquote {\bibinfo {title} {{Ring waveguide resonator on
  surface acoustic waves}},}\ }\href {\doibase 10.1063/1.2731683} {\bibfield
  {journal} {\bibinfo  {journal} {Appl. Phys. Lett.}\ }\textbf {\bibinfo
  {volume} {90}},\ \bibinfo {pages} {173503} (\bibinfo {year}
  {2007})}\BibitemShut {NoStop}%
\bibitem [{\citenamefont {Biryukov}\ \emph {et~al.}(2009)\citenamefont
  {Biryukov}, \citenamefont {Schmidt}, \citenamefont {Sotnikov}, \citenamefont
  {Weihnacht}, \citenamefont {Chemekova},\ and\ \citenamefont
  {Makarov}}]{Biryukov2009}%
  \BibitemOpen
  \bibfield  {author} {\bibinfo {author} {\bibfnamefont {S.~V.}\ \bibnamefont
  {Biryukov}}, \bibinfo {author} {\bibfnamefont {H.}~\bibnamefont {Schmidt}},
  \bibinfo {author} {\bibfnamefont {A.~V.}\ \bibnamefont {Sotnikov}}, \bibinfo
  {author} {\bibfnamefont {M.}~\bibnamefont {Weihnacht}}, \bibinfo {author}
  {\bibfnamefont {T.~Y.}\ \bibnamefont {Chemekova}}, \ and\ \bibinfo {author}
  {\bibfnamefont {Y.~N.}\ \bibnamefont {Makarov}},\ }\bibfield  {title}
  {\enquote {\bibinfo {title} {{Ring waveguide resonator on surface acoustic
  waves: First experiments}},}\ }\href {\doibase 10.1063/1.3272027} {\bibfield
  {journal} {\bibinfo  {journal} {J. Appl. Phys.}\ }\textbf {\bibinfo {volume}
  {106}},\ \bibinfo {pages} {130} (\bibinfo {year} {2009})}\BibitemShut
  {NoStop}%
\bibitem [{\citenamefont {Maznev}(2009)}]{Maznev2009}%
  \BibitemOpen
  \bibfield  {author} {\bibinfo {author} {\bibfnamefont {A.}~\bibnamefont
  {Maznev}},\ }\bibfield  {title} {\enquote {\bibinfo {title} {{Laser-generated
  surface acoustic waves in a ring-shaped waveguide resonator}},}\ }\href
  {\doibase 10.1016/j.ultras.2008.04.007} {\bibfield  {journal} {\bibinfo
  {journal} {Ultrasonics}\ }\textbf {\bibinfo {volume} {49}},\ \bibinfo {pages}
  {1} (\bibinfo {year} {2009})}\BibitemShut {NoStop}%
\bibitem [{\citenamefont {Manenti}\ \emph {et~al.}(2016)\citenamefont
  {Manenti}, \citenamefont {Peterer}, \citenamefont {Nersisyan}, \citenamefont
  {Magnusson}, \citenamefont {Patterson},\ and\ \citenamefont
  {Leek}}]{Manenti2016}%
  \BibitemOpen
  \bibfield  {author} {\bibinfo {author} {\bibfnamefont {R.}~\bibnamefont
  {Manenti}}, \bibinfo {author} {\bibfnamefont {M.~J.}\ \bibnamefont
  {Peterer}}, \bibinfo {author} {\bibfnamefont {A.}~\bibnamefont {Nersisyan}},
  \bibinfo {author} {\bibfnamefont {E.~B.}\ \bibnamefont {Magnusson}}, \bibinfo
  {author} {\bibfnamefont {A.}~\bibnamefont {Patterson}}, \ and\ \bibinfo
  {author} {\bibfnamefont {P.~J.}\ \bibnamefont {Leek}},\ }\bibfield  {title}
  {\enquote {\bibinfo {title} {{Surface acoustic wave resonators in the quantum
  regime}},}\ }\href {\doibase 10.1103/PhysRevB.93.041411} {\bibfield
  {journal} {\bibinfo  {journal} {Phys. Rev. B}\ }\textbf {\bibinfo {volume}
  {93}},\ \bibinfo {pages} {041411} (\bibinfo {year} {2016})}\BibitemShut
  {NoStop}%
\bibitem [{\citenamefont {Liu}\ \emph {et~al.}(2009)\citenamefont {Liu},
  \citenamefont {Peng}, \citenamefont {Jia}, \citenamefont {Ke},\ and\
  \citenamefont {Liu}}]{Liu2009}%
  \BibitemOpen
  \bibfield  {author} {\bibinfo {author} {\bibfnamefont {F.}~\bibnamefont
  {Liu}}, \bibinfo {author} {\bibfnamefont {S.}~\bibnamefont {Peng}}, \bibinfo
  {author} {\bibfnamefont {H.}~\bibnamefont {Jia}}, \bibinfo {author}
  {\bibfnamefont {M.}~\bibnamefont {Ke}}, \ and\ \bibinfo {author}
  {\bibfnamefont {Z.}~\bibnamefont {Liu}},\ }\bibfield  {title} {\enquote
  {\bibinfo {title} {{Strongly localized acoustic surface waves propagating
  along a V-groove}},}\ }\href {\doibase 10.1063/1.3072346} {\bibfield
  {journal} {\bibinfo  {journal} {Appl. Phys. Lett.}\ }\textbf {\bibinfo
  {volume} {94}},\ \bibinfo {pages} {2007} (\bibinfo {year}
  {2009})}\BibitemShut {NoStop}%
\bibitem [{\citenamefont {Boucher}\ \emph {et~al.}(2014)\citenamefont
  {Boucher}, \citenamefont {Rauwerdink}, \citenamefont {Tahraoui},
  \citenamefont {Wenger}, \citenamefont {Yamamoto},\ and\ \citenamefont
  {Santos}}]{Boucher2014}%
  \BibitemOpen
  \bibfield  {author} {\bibinfo {author} {\bibfnamefont {P.}~\bibnamefont
  {Boucher}}, \bibinfo {author} {\bibfnamefont {S.}~\bibnamefont {Rauwerdink}},
  \bibinfo {author} {\bibfnamefont {A.}~\bibnamefont {Tahraoui}}, \bibinfo
  {author} {\bibfnamefont {C.}~\bibnamefont {Wenger}}, \bibinfo {author}
  {\bibfnamefont {Y.}~\bibnamefont {Yamamoto}}, \ and\ \bibinfo {author}
  {\bibfnamefont {P.~V.}\ \bibnamefont {Santos}},\ }\bibfield  {title}
  {\enquote {\bibinfo {title} {{Ring waveguides for gigahertz acoustic waves on
  silicon}},}\ }\href {\doibase 10.1063/1.4898814} {\bibfield  {journal}
  {\bibinfo  {journal} {Appl. Phys. Lett.}\ }\textbf {\bibinfo {volume}
  {105}},\ \bibinfo {pages} {161904} (\bibinfo {year} {2014})}\BibitemShut
  {NoStop}%
\bibitem [{\citenamefont {Fan}\ \emph {et~al.}(2016)\citenamefont {Fan},
  \citenamefont {Zou}, \citenamefont {Poot}, \citenamefont {Cheng},
  \citenamefont {Guo}, \citenamefont {Han},\ and\ \citenamefont
  {Tang}}]{wg_Fan2016}%
  \BibitemOpen
  \bibfield  {author} {\bibinfo {author} {\bibfnamefont {L.}~\bibnamefont
  {Fan}}, \bibinfo {author} {\bibfnamefont {C.-L.}\ \bibnamefont {Zou}},
  \bibinfo {author} {\bibfnamefont {M.}~\bibnamefont {Poot}}, \bibinfo {author}
  {\bibfnamefont {R.}~\bibnamefont {Cheng}}, \bibinfo {author} {\bibfnamefont
  {X.}~\bibnamefont {Guo}}, \bibinfo {author} {\bibfnamefont {X.}~\bibnamefont
  {Han}}, \ and\ \bibinfo {author} {\bibfnamefont {H.~X.}\ \bibnamefont
  {Tang}},\ }\bibfield  {title} {\enquote {\bibinfo {title} {Noise-free quantum
  optical frequency shifting driven by mechanics},}\ }\href
  {https://arxiv.org/pdf/1607.01823.pdf} {\bibfield  {journal} {\bibinfo
  {journal} {arXiv preprint arXiv:1607.01823}\ } (\bibinfo {year}
  {2016})}\BibitemShut {NoStop}%
\bibitem [{\citenamefont {Mohammadi}\ and\ \citenamefont
  {Adibi}(2011)}]{wg_Mohammadi2011}%
  \BibitemOpen
  \bibfield  {author} {\bibinfo {author} {\bibfnamefont {S.}~\bibnamefont
  {Mohammadi}}\ and\ \bibinfo {author} {\bibfnamefont {A.}~\bibnamefont
  {Adibi}},\ }\bibfield  {title} {\enquote {\bibinfo {title} {On chip complex
  signal processing devices using coupled phononic crystal slab resonators and
  waveguides},}\ }\href {\doibase 10.1063/1.3676168} {\bibfield  {journal}
  {\bibinfo  {journal} {AIP Advances}\ }\textbf {\bibinfo {volume} {1}},\
  \bibinfo {pages} {041903} (\bibinfo {year} {2011})}\BibitemShut {NoStop}%
\bibitem [{\citenamefont {Patel}\ \emph {et~al.}(2018)\citenamefont {Patel},
  \citenamefont {Wang}, \citenamefont {Jiang}, \citenamefont {Sarabalis},
  \citenamefont {Hill},\ and\ \citenamefont
  {Safavi-Naeini}}]{wg_PhysRevLett.121.040501}%
  \BibitemOpen
  \bibfield  {author} {\bibinfo {author} {\bibfnamefont {R.~N.}\ \bibnamefont
  {Patel}}, \bibinfo {author} {\bibfnamefont {Z.}~\bibnamefont {Wang}},
  \bibinfo {author} {\bibfnamefont {W.}~\bibnamefont {Jiang}}, \bibinfo
  {author} {\bibfnamefont {C.~J.}\ \bibnamefont {Sarabalis}}, \bibinfo {author}
  {\bibfnamefont {J.~T.}\ \bibnamefont {Hill}}, \ and\ \bibinfo {author}
  {\bibfnamefont {A.~H.}\ \bibnamefont {Safavi-Naeini}},\ }\bibfield  {title}
  {\enquote {\bibinfo {title} {Single-mode phononic wire},}\ }\href {\doibase
  10.1103/PhysRevLett.121.040501} {\bibfield  {journal} {\bibinfo  {journal}
  {Phys. Rev. Lett.}\ }\textbf {\bibinfo {volume} {121}},\ \bibinfo {pages}
  {040501} (\bibinfo {year} {2018})}\BibitemShut {NoStop}%
\bibitem [{\citenamefont {Fang}\ \emph {et~al.}(2016)\citenamefont {Fang},
  \citenamefont {Matheny}, \citenamefont {Luan},\ and\ \citenamefont
  {Painter}}]{wg_PnC_fang2016optical}%
  \BibitemOpen
  \bibfield  {author} {\bibinfo {author} {\bibfnamefont {K.}~\bibnamefont
  {Fang}}, \bibinfo {author} {\bibfnamefont {M.~H.}\ \bibnamefont {Matheny}},
  \bibinfo {author} {\bibfnamefont {X.}~\bibnamefont {Luan}}, \ and\ \bibinfo
  {author} {\bibfnamefont {O.}~\bibnamefont {Painter}},\ }\bibfield  {title}
  {\enquote {\bibinfo {title} {Optical transduction and routing of microwave
  phonons in cavity-optomechanical circuits},}\ }\href
  {https://www.nature.com/articles/nphoton.2016.107} {\bibfield  {journal}
  {\bibinfo  {journal} {Nat. Photon.}\ }\textbf {\bibinfo {volume} {10}},\
  \bibinfo {pages} {489} (\bibinfo {year} {2016})}\BibitemShut {NoStop}%
\bibitem [{\citenamefont {Hatanaka}\ \emph
  {et~al.}(2014{\natexlab{a}})\citenamefont {Hatanaka}, \citenamefont
  {Mahboob}, \citenamefont {Onomitsu},\ and\ \citenamefont
  {Yamaguchi}}]{wg_suspended_wg_hatanaka2014phonon}%
  \BibitemOpen
  \bibfield  {author} {\bibinfo {author} {\bibfnamefont {D.}~\bibnamefont
  {Hatanaka}}, \bibinfo {author} {\bibfnamefont {I.}~\bibnamefont {Mahboob}},
  \bibinfo {author} {\bibfnamefont {K.}~\bibnamefont {Onomitsu}}, \ and\
  \bibinfo {author} {\bibfnamefont {H.}~\bibnamefont {Yamaguchi}},\ }\bibfield
  {title} {\enquote {\bibinfo {title} {Phonon waveguides for electromechanical
  circuits},}\ }\href {https://www.nature.com/articles/nnano.2014.107}
  {\bibfield  {journal} {\bibinfo  {journal} {Nat. Nanotech.}\ }\textbf
  {\bibinfo {volume} {9}},\ \bibinfo {pages} {520} (\bibinfo {year}
  {2014}{\natexlab{a}})}\BibitemShut {NoStop}%
\bibitem [{\citenamefont {Xu}\ \emph {et~al.}(2018)\citenamefont {Xu},
  \citenamefont {Fu}, \citenamefont {Zou}, \citenamefont {Shen},\ and\
  \citenamefont {Tang}}]{Xu2018apl}%
  \BibitemOpen
  \bibfield  {author} {\bibinfo {author} {\bibfnamefont {Y.}~\bibnamefont
  {Xu}}, \bibinfo {author} {\bibfnamefont {W.}~\bibnamefont {Fu}}, \bibinfo
  {author} {\bibfnamefont {C.-l.}\ \bibnamefont {Zou}}, \bibinfo {author}
  {\bibfnamefont {Z.}~\bibnamefont {Shen}}, \ and\ \bibinfo {author}
  {\bibfnamefont {H.~X.}\ \bibnamefont {Tang}},\ }\bibfield  {title} {\enquote
  {\bibinfo {title} {High quality factor surface fabry-perot cavity of acoustic
  waves},}\ }\href@noop {} {\bibfield  {journal} {\bibinfo  {journal} {Applied
  Physics Letters}\ }\textbf {\bibinfo {volume} {112}},\ \bibinfo {pages}
  {073505} (\bibinfo {year} {2018})}\BibitemShut {NoStop}%
\bibitem [{\citenamefont {Golter}\ \emph
  {et~al.}(2016{\natexlab{b}})\citenamefont {Golter}, \citenamefont {Oo},
  \citenamefont {Amezcua}, \citenamefont {Stewart},\ and\ \citenamefont
  {Wang}}]{Golter16prl}%
  \BibitemOpen
  \bibfield  {author} {\bibinfo {author} {\bibfnamefont {D.~A.}\ \bibnamefont
  {Golter}}, \bibinfo {author} {\bibfnamefont {T.}~\bibnamefont {Oo}}, \bibinfo
  {author} {\bibfnamefont {M.}~\bibnamefont {Amezcua}}, \bibinfo {author}
  {\bibfnamefont {K.~A.}\ \bibnamefont {Stewart}}, \ and\ \bibinfo {author}
  {\bibfnamefont {H.}~\bibnamefont {Wang}},\ }\bibfield  {title} {\enquote
  {\bibinfo {title} {Optomechanical {Quantum} {Control} of a
  {Nitrogen}-{Vacancy} {Center} in {Diamond}},}\ }\href {\doibase
  10.1103/PhysRevLett.116.143602} {\bibfield  {journal} {\bibinfo  {journal}
  {Physical Review Letters}\ }\textbf {\bibinfo {volume} {116}},\ \bibinfo
  {pages} {143602} (\bibinfo {year} {2016}{\natexlab{b}})}\BibitemShut
  {NoStop}%
\bibitem [{\citenamefont {Huang}\ \emph {et~al.}(2010)\citenamefont {Huang},
  \citenamefont {Sun},\ and\ \citenamefont {Wu}}]{Huang10apl}%
  \BibitemOpen
  \bibfield  {author} {\bibinfo {author} {\bibfnamefont {C.-Y.}\ \bibnamefont
  {Huang}}, \bibinfo {author} {\bibfnamefont {J.-H.}\ \bibnamefont {Sun}}, \
  and\ \bibinfo {author} {\bibfnamefont {T.-T.}\ \bibnamefont {Wu}},\
  }\bibfield  {title} {\enquote {\bibinfo {title} {A two-port {ZnO}/silicon
  {Lamb} wave resonator using phononic crystals},}\ }\href {\doibase
  10.1063/1.3467145} {\bibfield  {journal} {\bibinfo  {journal} {Applied
  Physics Letters}\ }\textbf {\bibinfo {volume} {97}},\ \bibinfo {pages}
  {031913} (\bibinfo {year} {2010})}\BibitemShut {NoStop}%
\bibitem [{\citenamefont {Magnusson}\ \emph {et~al.}(2015)\citenamefont
  {Magnusson}, \citenamefont {Williams}, \citenamefont {Manenti}, \citenamefont
  {Nam}, \citenamefont {Nersisyan}, \citenamefont {Peterer}, \citenamefont
  {Ardavan},\ and\ \citenamefont {Leek}}]{Magnusson15apl}%
  \BibitemOpen
  \bibfield  {author} {\bibinfo {author} {\bibfnamefont {E.~B.}\ \bibnamefont
  {Magnusson}}, \bibinfo {author} {\bibfnamefont {B.~H.}\ \bibnamefont
  {Williams}}, \bibinfo {author} {\bibfnamefont {R.}~\bibnamefont {Manenti}},
  \bibinfo {author} {\bibfnamefont {M.-S.}\ \bibnamefont {Nam}}, \bibinfo
  {author} {\bibfnamefont {A.}~\bibnamefont {Nersisyan}}, \bibinfo {author}
  {\bibfnamefont {M.~J.}\ \bibnamefont {Peterer}}, \bibinfo {author}
  {\bibfnamefont {A.}~\bibnamefont {Ardavan}}, \ and\ \bibinfo {author}
  {\bibfnamefont {P.~J.}\ \bibnamefont {Leek}},\ }\bibfield  {title} {\enquote
  {\bibinfo {title} {Surface acoustic wave devices on bulk {ZnO} crystals at
  low temperature},}\ }\href {\doibase 10.1063/1.4908248} {\bibfield  {journal}
  {\bibinfo  {journal} {Applied Physics Letters}\ }\textbf {\bibinfo {volume}
  {106}},\ \bibinfo {pages} {063509} (\bibinfo {year} {2015})}\BibitemShut
  {NoStop}%
\bibitem [{\citenamefont {Ung}\ \emph {et~al.}(2017)\citenamefont {Ung},
  \citenamefont {Mutafopulos}, \citenamefont {Spink}, \citenamefont {Rambach},
  \citenamefont {Franke},\ and\ \citenamefont {Weitz}}]{Ung17lc}%
  \BibitemOpen
  \bibfield  {author} {\bibinfo {author} {\bibfnamefont {W.~L.}\ \bibnamefont
  {Ung}}, \bibinfo {author} {\bibfnamefont {K.}~\bibnamefont {Mutafopulos}},
  \bibinfo {author} {\bibfnamefont {P.}~\bibnamefont {Spink}}, \bibinfo
  {author} {\bibfnamefont {R.~W.}\ \bibnamefont {Rambach}}, \bibinfo {author}
  {\bibfnamefont {T.}~\bibnamefont {Franke}}, \ and\ \bibinfo {author}
  {\bibfnamefont {D.~A.}\ \bibnamefont {Weitz}},\ }\bibfield  {title} {\enquote
  {\bibinfo {title} {Enhanced surface acoustic wave cell sorting by {3D}
  microfluidic-chip design},}\ }\href {\doibase 10.1039/C7LC00715A} {\bibfield
  {journal} {\bibinfo  {journal} {Lab on a Chip}\ }\textbf {\bibinfo {volume}
  {17}},\ \bibinfo {pages} {4059} (\bibinfo {year} {2017})}\BibitemShut
  {NoStop}%
\bibitem [{\citenamefont {Shao}\ \emph {et~al.}(2019)\citenamefont {Shao},
  \citenamefont {Maity}, \citenamefont {Zheng}, \citenamefont {Wu},
  \citenamefont {Shams-Ansari}, \citenamefont {Sohn}, \citenamefont {Puma},
  \citenamefont {Gadalla}, \citenamefont {Zhang}, \citenamefont {Wang},
  \citenamefont {Hu}, \citenamefont {Lai},\ and\ \citenamefont
  {Loncar}}]{Shao19pra}%
  \BibitemOpen
  \bibfield  {author} {\bibinfo {author} {\bibfnamefont {L.}~\bibnamefont
  {Shao}}, \bibinfo {author} {\bibfnamefont {S.}~\bibnamefont {Maity}},
  \bibinfo {author} {\bibfnamefont {L.}~\bibnamefont {Zheng}}, \bibinfo
  {author} {\bibfnamefont {L.}~\bibnamefont {Wu}}, \bibinfo {author}
  {\bibfnamefont {A.}~\bibnamefont {Shams-Ansari}}, \bibinfo {author}
  {\bibfnamefont {Y.-I.}\ \bibnamefont {Sohn}}, \bibinfo {author}
  {\bibfnamefont {E.}~\bibnamefont {Puma}}, \bibinfo {author} {\bibfnamefont
  {M.}~\bibnamefont {Gadalla}}, \bibinfo {author} {\bibfnamefont
  {M.}~\bibnamefont {Zhang}}, \bibinfo {author} {\bibfnamefont
  {C.}~\bibnamefont {Wang}}, \bibinfo {author} {\bibfnamefont {E.}~\bibnamefont
  {Hu}}, \bibinfo {author} {\bibfnamefont {K.}~\bibnamefont {Lai}}, \ and\
  \bibinfo {author} {\bibfnamefont {M.}~\bibnamefont {Loncar}},\ }\bibfield
  {title} {\enquote {\bibinfo {title} {Phononic {Band} {Structure}
  {Engineering} for {High}-{Q} {Gigahertz} {Surface} {Acoustic} {Wave}
  {Resonators} on {Lithium} {Niobate}},}\ }\href {\doibase
  10.1103/PhysRevApplied.12.014022} {\bibfield  {journal} {\bibinfo  {journal}
  {Physical Review Applied}\ }\textbf {\bibinfo {volume} {12}},\ \bibinfo
  {pages} {014022} (\bibinfo {year} {2019})}\BibitemShut {NoStop}%
\bibitem [{\citenamefont {Fu}\ \emph {et~al.}(2017{\natexlab{a}})\citenamefont
  {Fu}, \citenamefont {Quan}, \citenamefont {Luo}, \citenamefont {Pang},
  \citenamefont {Guo}, \citenamefont {Wu}, \citenamefont {Ng}, \citenamefont
  {Zu},\ and\ \citenamefont {Fu}}]{Fu17apl}%
  \BibitemOpen
  \bibfield  {author} {\bibinfo {author} {\bibfnamefont {C.}~\bibnamefont
  {Fu}}, \bibinfo {author} {\bibfnamefont {A.~J.}\ \bibnamefont {Quan}},
  \bibinfo {author} {\bibfnamefont {J.~T.}\ \bibnamefont {Luo}}, \bibinfo
  {author} {\bibfnamefont {H.~F.}\ \bibnamefont {Pang}}, \bibinfo {author}
  {\bibfnamefont {Y.~J.}\ \bibnamefont {Guo}}, \bibinfo {author} {\bibfnamefont
  {Q.}~\bibnamefont {Wu}}, \bibinfo {author} {\bibfnamefont {W.~P.}\
  \bibnamefont {Ng}}, \bibinfo {author} {\bibfnamefont {X.~T.}\ \bibnamefont
  {Zu}}, \ and\ \bibinfo {author} {\bibfnamefont {Y.~Q.}\ \bibnamefont {Fu}},\
  }\bibfield  {title} {\enquote {\bibinfo {title} {Vertical jetting induced by
  shear horizontal leaky surface acoustic wave on 36° {Y}-{X} {LiTaO3}},}\
  }\href {\doibase 10.1063/1.4982073} {\bibfield  {journal} {\bibinfo
  {journal} {Applied Physics Letters}\ }\textbf {\bibinfo {volume} {110}},\
  \bibinfo {pages} {173501} (\bibinfo {year} {2017}{\natexlab{a}})}\BibitemShut
  {NoStop}%
\bibitem [{\citenamefont {Chu}\ \emph {et~al.}(2017)\citenamefont {Chu},
  \citenamefont {Kharel}, \citenamefont {Renninger}, \citenamefont {Burkhart},
  \citenamefont {Frunzio}, \citenamefont {Rakich},\ and\ \citenamefont
  {Schoelkopf}}]{Chu17s}%
  \BibitemOpen
  \bibfield  {author} {\bibinfo {author} {\bibfnamefont {Y.}~\bibnamefont
  {Chu}}, \bibinfo {author} {\bibfnamefont {P.}~\bibnamefont {Kharel}},
  \bibinfo {author} {\bibfnamefont {W.~H.}\ \bibnamefont {Renninger}}, \bibinfo
  {author} {\bibfnamefont {L.~D.}\ \bibnamefont {Burkhart}}, \bibinfo {author}
  {\bibfnamefont {L.}~\bibnamefont {Frunzio}}, \bibinfo {author} {\bibfnamefont
  {P.~T.}\ \bibnamefont {Rakich}}, \ and\ \bibinfo {author} {\bibfnamefont
  {R.~J.}\ \bibnamefont {Schoelkopf}},\ }\bibfield  {title} {\enquote {\bibinfo
  {title} {Quantum acoustics with superconducting qubits},}\ }\href {\doibase
  10.1126/science.aao1511} {\bibfield  {journal} {\bibinfo  {journal}
  {Science}\ }\textbf {\bibinfo {volume} {358}},\ \bibinfo {pages} {199}
  (\bibinfo {year} {2017})}\BibitemShut {NoStop}%
\bibitem [{\citenamefont {Aubert}\ \emph {et~al.}(2010)\citenamefont {Aubert},
  \citenamefont {Elmazria}, \citenamefont {Assouar}, \citenamefont {Bouvot},\
  and\ \citenamefont {Oudich}}]{Aubert10apl}%
  \BibitemOpen
  \bibfield  {author} {\bibinfo {author} {\bibfnamefont {T.}~\bibnamefont
  {Aubert}}, \bibinfo {author} {\bibfnamefont {O.}~\bibnamefont {Elmazria}},
  \bibinfo {author} {\bibfnamefont {B.}~\bibnamefont {Assouar}}, \bibinfo
  {author} {\bibfnamefont {L.}~\bibnamefont {Bouvot}}, \ and\ \bibinfo {author}
  {\bibfnamefont {M.}~\bibnamefont {Oudich}},\ }\bibfield  {title} {\enquote
  {\bibinfo {title} {Surface acoustic wave devices based on {AlN}/sapphire
  structure for high temperature applications},}\ }\href {\doibase
  10.1063/1.3430042} {\bibfield  {journal} {\bibinfo  {journal} {Applied
  Physics Letters}\ }\textbf {\bibinfo {volume} {96}},\ \bibinfo {pages}
  {203503} (\bibinfo {year} {2010})}\BibitemShut {NoStop}%
\bibitem [{\citenamefont {Tadesse}\ and\ \citenamefont
  {Li}(2014{\natexlab{b}})}]{Tadesse14nc}%
  \BibitemOpen
  \bibfield  {author} {\bibinfo {author} {\bibfnamefont {S.~A.}\ \bibnamefont
  {Tadesse}}\ and\ \bibinfo {author} {\bibfnamefont {M.}~\bibnamefont {Li}},\
  }\bibfield  {title} {\enquote {\bibinfo {title} {Sub-optical wavelength
  acoustic wave modulation of integrated photonic resonators at microwave
  frequencies},}\ }\href {\doibase 10.1038/ncomms6402} {\bibfield  {journal}
  {\bibinfo  {journal} {Nature Communications}\ }\textbf {\bibinfo {volume}
  {5}},\ \bibinfo {pages} {5402} (\bibinfo {year}
  {2014}{\natexlab{b}})}\BibitemShut {NoStop}%
\bibitem [{\citenamefont {Fujii}\ \emph {et~al.}(2013)\citenamefont {Fujii},
  \citenamefont {Odawara}, \citenamefont {Yamada}, \citenamefont {Omori},
  \citenamefont {Hashimoto}, \citenamefont {Torii}, \citenamefont {Umezawa},\
  and\ \citenamefont {Shikata}}]{Fujii13ituffc}%
  \BibitemOpen
  \bibfield  {author} {\bibinfo {author} {\bibfnamefont {S.}~\bibnamefont
  {Fujii}}, \bibinfo {author} {\bibfnamefont {T.}~\bibnamefont {Odawara}},
  \bibinfo {author} {\bibfnamefont {H.}~\bibnamefont {Yamada}}, \bibinfo
  {author} {\bibfnamefont {T.}~\bibnamefont {Omori}}, \bibinfo {author}
  {\bibfnamefont {K.-y.}\ \bibnamefont {Hashimoto}}, \bibinfo {author}
  {\bibfnamefont {H.}~\bibnamefont {Torii}}, \bibinfo {author} {\bibfnamefont
  {H.}~\bibnamefont {Umezawa}}, \ and\ \bibinfo {author} {\bibfnamefont
  {S.}~\bibnamefont {Shikata}},\ }\bibfield  {title} {\enquote {\bibinfo
  {title} {Low propagation loss in a one-port {SAW} resonator fabricated on
  single-crystal diamond for super-high-frequency applications},}\ }\href
  {\doibase 10.1109/TUFFC.2013.2656} {\bibfield  {journal} {\bibinfo  {journal}
  {IEEE Transactions on Ultrasonics, Ferroelectrics, and Frequency Control}\
  }\textbf {\bibinfo {volume} {60}},\ \bibinfo {pages} {986} (\bibinfo {year}
  {2013})}\BibitemShut {NoStop}%
\bibitem [{\citenamefont {Okamoto}\ \emph {et~al.}(2013)\citenamefont
  {Okamoto}, \citenamefont {Gourgout}, \citenamefont {Chang}, \citenamefont
  {Onomitsu}, \citenamefont {Mahboob}, \citenamefont {Chang},\ and\
  \citenamefont {Yamaguchi}}]{Okamoto13np}%
  \BibitemOpen
  \bibfield  {author} {\bibinfo {author} {\bibfnamefont {H.}~\bibnamefont
  {Okamoto}}, \bibinfo {author} {\bibfnamefont {A.}~\bibnamefont {Gourgout}},
  \bibinfo {author} {\bibfnamefont {C.-Y.}\ \bibnamefont {Chang}}, \bibinfo
  {author} {\bibfnamefont {K.}~\bibnamefont {Onomitsu}}, \bibinfo {author}
  {\bibfnamefont {I.}~\bibnamefont {Mahboob}}, \bibinfo {author} {\bibfnamefont
  {E.~Y.}\ \bibnamefont {Chang}}, \ and\ \bibinfo {author} {\bibfnamefont
  {H.}~\bibnamefont {Yamaguchi}},\ }\bibfield  {title} {\enquote {\bibinfo
  {title} {Coherent phonon manipulation in coupled mechanical resonators},}\
  }\href {\doibase 10.1038/nphys2665} {\bibfield  {journal} {\bibinfo
  {journal} {Nature Physics}\ }\textbf {\bibinfo {volume} {9}},\ \bibinfo
  {pages} {480} (\bibinfo {year} {2013})}\BibitemShut {NoStop}%
\bibitem [{\citenamefont {Metcalfe}\ \emph {et~al.}(2010)\citenamefont
  {Metcalfe}, \citenamefont {Carr}, \citenamefont {Muller}, \citenamefont
  {Solomon},\ and\ \citenamefont {Lawall}}]{Metcalfe10prl}%
  \BibitemOpen
  \bibfield  {author} {\bibinfo {author} {\bibfnamefont {M.}~\bibnamefont
  {Metcalfe}}, \bibinfo {author} {\bibfnamefont {S.~M.}\ \bibnamefont {Carr}},
  \bibinfo {author} {\bibfnamefont {A.}~\bibnamefont {Muller}}, \bibinfo
  {author} {\bibfnamefont {G.~S.}\ \bibnamefont {Solomon}}, \ and\ \bibinfo
  {author} {\bibfnamefont {J.}~\bibnamefont {Lawall}},\ }\bibfield  {title}
  {\enquote {\bibinfo {title} {Resolved {Sideband} {Emission} of
  \${\textbackslash}mathrm\{{InAs}\}/{\textbackslash}mathrm\{{GaAs}\}\$
  {Quantum} {Dots} {Strained} by {Surface} {Acoustic} {Waves}},}\ }\href
  {\doibase 10.1103/PhysRevLett.105.037401} {\bibfield  {journal} {\bibinfo
  {journal} {Physical Review Letters}\ }\textbf {\bibinfo {volume} {105}},\
  \bibinfo {pages} {037401} (\bibinfo {year} {2010})}\BibitemShut {NoStop}%
\bibitem [{\citenamefont {Valle}\ \emph {et~al.}(2019)\citenamefont {Valle},
  \citenamefont {Singh}, \citenamefont {Cryan}, \citenamefont {Kuball},\ and\
  \citenamefont {Balram}}]{Valle19apl}%
  \BibitemOpen
  \bibfield  {author} {\bibinfo {author} {\bibfnamefont {S.}~\bibnamefont
  {Valle}}, \bibinfo {author} {\bibfnamefont {M.}~\bibnamefont {Singh}},
  \bibinfo {author} {\bibfnamefont {M.~J.}\ \bibnamefont {Cryan}}, \bibinfo
  {author} {\bibfnamefont {M.}~\bibnamefont {Kuball}}, \ and\ \bibinfo {author}
  {\bibfnamefont {K.~C.}\ \bibnamefont {Balram}},\ }\bibfield  {title}
  {\enquote {\bibinfo {title} {High frequency guided mode resonances in
  mass-loaded, thin film gallium nitride surface acoustic wave devices},}\
  }\href {\doibase 10.1063/1.5123718} {\bibfield  {journal} {\bibinfo
  {journal} {Applied Physics Letters}\ }\textbf {\bibinfo {volume} {115}},\
  \bibinfo {pages} {212104} (\bibinfo {year} {2019})}\BibitemShut {NoStop}%
\bibitem [{\citenamefont {Wang}\ \emph {et~al.}(2015)\citenamefont {Wang},
  \citenamefont {Popa},\ and\ \citenamefont {Weinstein}}]{Wang1522iicmemsm}%
  \BibitemOpen
  \bibfield  {author} {\bibinfo {author} {\bibfnamefont {S.}~\bibnamefont
  {Wang}}, \bibinfo {author} {\bibfnamefont {L.~C.}\ \bibnamefont {Popa}}, \
  and\ \bibinfo {author} {\bibfnamefont {D.}~\bibnamefont {Weinstein}},\
  }\bibfield  {title} {\enquote {\bibinfo {title} {Tapered {Phononic} {Crystal}
  sawresonator in {GaN}},}\ }in\ \href {\doibase 10.1109/MEMSYS.2015.7051137}
  {\emph {\bibinfo {booktitle} {2015 28th {IEEE} {International} {Conference}
  on {Micro} {Electro} {Mechanical} {Systems} ({MEMS})}}}\ (\bibinfo {year}
  {2015})\ pp.\ \bibinfo {pages} {1028--1031}\BibitemShut {NoStop}%
\bibitem [{\citenamefont {Fu}\ \emph {et~al.}(2019{\natexlab{a}})\citenamefont
  {Fu}, \citenamefont {Shen}, \citenamefont {Xu}, \citenamefont {Zou},
  \citenamefont {Cheng}, \citenamefont {Han},\ and\ \citenamefont
  {Tang}}]{Fu2019nc}%
  \BibitemOpen
  \bibfield  {author} {\bibinfo {author} {\bibfnamefont {W.}~\bibnamefont
  {Fu}}, \bibinfo {author} {\bibfnamefont {Z.}~\bibnamefont {Shen}}, \bibinfo
  {author} {\bibfnamefont {Y.}~\bibnamefont {Xu}}, \bibinfo {author}
  {\bibfnamefont {C.-L.}\ \bibnamefont {Zou}}, \bibinfo {author} {\bibfnamefont
  {R.}~\bibnamefont {Cheng}}, \bibinfo {author} {\bibfnamefont
  {X.}~\bibnamefont {Han}}, \ and\ \bibinfo {author} {\bibfnamefont {H.~X.}\
  \bibnamefont {Tang}},\ }\bibfield  {title} {\enquote {\bibinfo {title}
  {Phononic integrated circuitry and spin--orbit interaction of phonons},}\
  }\href {\doibase 10.1038/s41467-019-10852-3} {\bibfield  {journal} {\bibinfo
  {journal} {Nature communications}\ }\textbf {\bibinfo {volume} {10}},\
  \bibinfo {pages} {1} (\bibinfo {year} {2019}{\natexlab{a}})}\BibitemShut
  {NoStop}%
\bibitem [{\citenamefont {Morgan}\ and\ \citenamefont
  {Paige}(2007)}]{SAW-book}%
  \BibitemOpen
  \bibfield  {author} {\bibinfo {author} {\bibfnamefont {D.}~\bibnamefont
  {Morgan}}\ and\ \bibinfo {author} {\bibfnamefont {E.}~\bibnamefont {Paige}},\
  }\bibfield  {title} {\enquote {\bibinfo {title} {2 - acoustic waves in
  elastic solids},}\ }in\ \href {\doibase
  http://dx.doi.org/10.1016/B978-012372537-0/50004-2} {\emph {\bibinfo
  {booktitle} {Surface Acoustic Wave Filters (Second edition)}}},\ \bibinfo
  {series and number} {Studies in Electrical and Electronic Engineering},\
  \bibinfo {editor} {edited by\ \bibinfo {editor} {\bibfnamefont
  {D.}~\bibnamefont {Morgan}}\ and\ \bibinfo {editor} {\bibfnamefont
  {E.}~\bibnamefont {Paige}}}\ (\bibinfo  {publisher} {Academic Press},\
  \bibinfo {address} {Oxford},\ \bibinfo {year} {2007})\ \bibinfo {edition}
  {second edition}\ ed.,\ pp.\ \bibinfo {pages} {38 -- 67}\BibitemShut
  {NoStop}%
\bibitem [{\citenamefont {Pang}\ \emph {et~al.}(2013)\citenamefont {Pang},
  \citenamefont {Fu}, \citenamefont {Li}, \citenamefont {Li}, \citenamefont
  {Ma}, \citenamefont {Placido}, \citenamefont {Walton},\ and\ \citenamefont
  {Zu}}]{Pang13saap}%
  \BibitemOpen
  \bibfield  {author} {\bibinfo {author} {\bibfnamefont {H.-F.}\ \bibnamefont
  {Pang}}, \bibinfo {author} {\bibfnamefont {Y.-Q.}\ \bibnamefont {Fu}},
  \bibinfo {author} {\bibfnamefont {Z.-J.}\ \bibnamefont {Li}}, \bibinfo
  {author} {\bibfnamefont {Y.}~\bibnamefont {Li}}, \bibinfo {author}
  {\bibfnamefont {J.-Y.}\ \bibnamefont {Ma}}, \bibinfo {author} {\bibfnamefont
  {F.}~\bibnamefont {Placido}}, \bibinfo {author} {\bibfnamefont {A.~J.}\
  \bibnamefont {Walton}}, \ and\ \bibinfo {author} {\bibfnamefont {X.-T.}\
  \bibnamefont {Zu}},\ }\bibfield  {title} {\enquote {\bibinfo {title} {Love
  mode surface acoustic wave ultraviolet sensor using {ZnO} films deposited on
  36° {Y}-cut {LiTaO3}},}\ }\href {\doibase 10.1016/j.sna.2013.01.016}
  {\bibfield  {journal} {\bibinfo  {journal} {Sensors and Actuators A:
  Physical}\ }\textbf {\bibinfo {volume} {193}},\ \bibinfo {pages} {87}
  (\bibinfo {year} {2013})}\BibitemShut {NoStop}%
\bibitem [{\citenamefont {Fu}\ \emph {et~al.}(2017{\natexlab{b}})\citenamefont
  {Fu}, \citenamefont {Luo}, \citenamefont {Nguyen}, \citenamefont {Walton},
  \citenamefont {Flewitt}, \citenamefont {Zu}, \citenamefont {Li},
  \citenamefont {McHale}, \citenamefont {Matthews}, \citenamefont {Iborra},
  \citenamefont {Du},\ and\ \citenamefont {Milne}}]{Fu17pms}%
  \BibitemOpen
  \bibfield  {author} {\bibinfo {author} {\bibfnamefont {Y.~Q.}\ \bibnamefont
  {Fu}}, \bibinfo {author} {\bibfnamefont {J.~K.}\ \bibnamefont {Luo}},
  \bibinfo {author} {\bibfnamefont {N.~T.}\ \bibnamefont {Nguyen}}, \bibinfo
  {author} {\bibfnamefont {A.~J.}\ \bibnamefont {Walton}}, \bibinfo {author}
  {\bibfnamefont {A.~J.}\ \bibnamefont {Flewitt}}, \bibinfo {author}
  {\bibfnamefont {X.~T.}\ \bibnamefont {Zu}}, \bibinfo {author} {\bibfnamefont
  {Y.}~\bibnamefont {Li}}, \bibinfo {author} {\bibfnamefont {G.}~\bibnamefont
  {McHale}}, \bibinfo {author} {\bibfnamefont {A.}~\bibnamefont {Matthews}},
  \bibinfo {author} {\bibfnamefont {E.}~\bibnamefont {Iborra}}, \bibinfo
  {author} {\bibfnamefont {H.}~\bibnamefont {Du}}, \ and\ \bibinfo {author}
  {\bibfnamefont {W.~I.}\ \bibnamefont {Milne}},\ }\bibfield  {title} {\enquote
  {\bibinfo {title} {Advances in piezoelectric thin films for acoustic
  biosensors, acoustofluidics and lab-on-chip applications},}\ }\href {\doibase
  10.1016/j.pmatsci.2017.04.006} {\bibfield  {journal} {\bibinfo  {journal}
  {Progress in Materials Science}\ }\textbf {\bibinfo {volume} {89}},\ \bibinfo
  {pages} {31} (\bibinfo {year} {2017}{\natexlab{b}})}\BibitemShut {NoStop}%
\bibitem [{\citenamefont {Fu}\ \emph {et~al.}(2019{\natexlab{b}})\citenamefont
  {Fu}, \citenamefont {Wang}, \citenamefont {Li}, \citenamefont {Lu},
  \citenamefont {Chen}, \citenamefont {Luo}, \citenamefont {Shen},
  \citenamefont {Wang}, \citenamefont {Song}, \citenamefont {Zeng},\ and\
  \citenamefont {Pan}}]{Fu19apl}%
  \BibitemOpen
  \bibfield  {author} {\bibinfo {author} {\bibfnamefont {S.}~\bibnamefont
  {Fu}}, \bibinfo {author} {\bibfnamefont {W.}~\bibnamefont {Wang}}, \bibinfo
  {author} {\bibfnamefont {Q.}~\bibnamefont {Li}}, \bibinfo {author}
  {\bibfnamefont {Z.}~\bibnamefont {Lu}}, \bibinfo {author} {\bibfnamefont
  {Z.}~\bibnamefont {Chen}}, \bibinfo {author} {\bibfnamefont {J.}~\bibnamefont
  {Luo}}, \bibinfo {author} {\bibfnamefont {J.}~\bibnamefont {Shen}}, \bibinfo
  {author} {\bibfnamefont {R.}~\bibnamefont {Wang}}, \bibinfo {author}
  {\bibfnamefont {C.}~\bibnamefont {Song}}, \bibinfo {author} {\bibfnamefont
  {F.}~\bibnamefont {Zeng}}, \ and\ \bibinfo {author} {\bibfnamefont
  {F.}~\bibnamefont {Pan}},\ }\bibfield  {title} {\enquote {\bibinfo {title}
  {High-frequency {V}-doped {ZnO}/{SiC} surface acoustic wave devices with
  enhanced electromechanical coupling coefficient},}\ }\href {\doibase
  10.1063/1.5086445} {\bibfield  {journal} {\bibinfo  {journal} {Applied
  Physics Letters}\ }\textbf {\bibinfo {volume} {114}},\ \bibinfo {pages}
  {113504} (\bibinfo {year} {2019}{\natexlab{b}})}\BibitemShut {NoStop}%
\bibitem [{\citenamefont {Deger}\ \emph {et~al.}(1998)\citenamefont {Deger},
  \citenamefont {Born}, \citenamefont {Angerer}, \citenamefont {Ambacher},
  \citenamefont {Stutzmann}, \citenamefont {Hornsteiner}, \citenamefont
  {Riha},\ and\ \citenamefont {Fischerauer}}]{Deger98apl}%
  \BibitemOpen
  \bibfield  {author} {\bibinfo {author} {\bibfnamefont {C.}~\bibnamefont
  {Deger}}, \bibinfo {author} {\bibfnamefont {E.}~\bibnamefont {Born}},
  \bibinfo {author} {\bibfnamefont {H.}~\bibnamefont {Angerer}}, \bibinfo
  {author} {\bibfnamefont {O.}~\bibnamefont {Ambacher}}, \bibinfo {author}
  {\bibfnamefont {M.}~\bibnamefont {Stutzmann}}, \bibinfo {author}
  {\bibfnamefont {J.}~\bibnamefont {Hornsteiner}}, \bibinfo {author}
  {\bibfnamefont {E.}~\bibnamefont {Riha}}, \ and\ \bibinfo {author}
  {\bibfnamefont {G.}~\bibnamefont {Fischerauer}},\ }\bibfield  {title}
  {\enquote {\bibinfo {title} {Sound velocity of {AlxGa1}{xN} thin films
  obtained by surface acoustic-wave measurements},}\ }\href {\doibase
  10.1063/1.121368} {\bibfield  {journal} {\bibinfo  {journal} {Applied Physics
  Letters}\ }\textbf {\bibinfo {volume} {72}},\ \bibinfo {pages} {2400}
  (\bibinfo {year} {1998})}\BibitemShut {NoStop}%
\bibitem [{\citenamefont {Nakahata}\ \emph {et~al.}(2003)\citenamefont
  {Nakahata}, \citenamefont {Fujii}, \citenamefont {Higaki}, \citenamefont
  {Hachigo}, \citenamefont {Kitabayashi}, \citenamefont {Shikata},\ and\
  \citenamefont {Fujimori}}]{Nakahata03sst}%
  \BibitemOpen
  \bibfield  {author} {\bibinfo {author} {\bibfnamefont {H.}~\bibnamefont
  {Nakahata}}, \bibinfo {author} {\bibfnamefont {S.}~\bibnamefont {Fujii}},
  \bibinfo {author} {\bibfnamefont {K.}~\bibnamefont {Higaki}}, \bibinfo
  {author} {\bibfnamefont {A.}~\bibnamefont {Hachigo}}, \bibinfo {author}
  {\bibfnamefont {H.}~\bibnamefont {Kitabayashi}}, \bibinfo {author}
  {\bibfnamefont {S.}~\bibnamefont {Shikata}}, \ and\ \bibinfo {author}
  {\bibfnamefont {N.}~\bibnamefont {Fujimori}},\ }\bibfield  {title} {\enquote
  {\bibinfo {title} {Diamond-based surface acoustic wave devices},}\ }\href
  {\doibase 10.1088/0268-1242/18/3/314} {\bibfield  {journal} {\bibinfo
  {journal} {Semiconductor Science and Technology}\ }\textbf {\bibinfo {volume}
  {18}},\ \bibinfo {pages} {S96} (\bibinfo {year} {2003})}\BibitemShut
  {NoStop}%
\bibitem [{\citenamefont {Pohl}\ \emph {et~al.}(2002)\citenamefont {Pohl},
  \citenamefont {Liu},\ and\ \citenamefont {Thompson}}]{Pohl02rmp}%
  \BibitemOpen
  \bibfield  {author} {\bibinfo {author} {\bibfnamefont {R.~O.}\ \bibnamefont
  {Pohl}}, \bibinfo {author} {\bibfnamefont {X.}~\bibnamefont {Liu}}, \ and\
  \bibinfo {author} {\bibfnamefont {E.}~\bibnamefont {Thompson}},\ }\bibfield
  {title} {\enquote {\bibinfo {title} {Low-temperature thermal conductivity and
  acoustic attenuation in amorphous solids},}\ }\href {\doibase
  10.1103/RevModPhys.74.991} {\bibfield  {journal} {\bibinfo  {journal}
  {Reviews of Modern Physics}\ }\textbf {\bibinfo {volume} {74}},\ \bibinfo
  {pages} {991} (\bibinfo {year} {2002})}\BibitemShut {NoStop}%
\bibitem [{\citenamefont {Azuhata}\ \emph {et~al.}(2003)\citenamefont
  {Azuhata}, \citenamefont {Takesada}, \citenamefont {Yagi}, \citenamefont
  {Shikanai}, \citenamefont {Chichibu}, \citenamefont {Torii}, \citenamefont
  {Nakamura}, \citenamefont {Sota}, \citenamefont {Cantwell}, \citenamefont
  {Eason},\ and\ \citenamefont {Litton}}]{Azuhata03jap}%
  \BibitemOpen
  \bibfield  {author} {\bibinfo {author} {\bibfnamefont {T.}~\bibnamefont
  {Azuhata}}, \bibinfo {author} {\bibfnamefont {M.}~\bibnamefont {Takesada}},
  \bibinfo {author} {\bibfnamefont {T.}~\bibnamefont {Yagi}}, \bibinfo {author}
  {\bibfnamefont {A.}~\bibnamefont {Shikanai}}, \bibinfo {author}
  {\bibfnamefont {S.}~\bibnamefont {Chichibu}}, \bibinfo {author}
  {\bibfnamefont {K.}~\bibnamefont {Torii}}, \bibinfo {author} {\bibfnamefont
  {A.}~\bibnamefont {Nakamura}}, \bibinfo {author} {\bibfnamefont
  {T.}~\bibnamefont {Sota}}, \bibinfo {author} {\bibfnamefont {G.}~\bibnamefont
  {Cantwell}}, \bibinfo {author} {\bibfnamefont {D.~B.}\ \bibnamefont {Eason}},
  \ and\ \bibinfo {author} {\bibfnamefont {C.~W.}\ \bibnamefont {Litton}},\
  }\bibfield  {title} {\enquote {\bibinfo {title} {Brillouin scattering study
  of {ZnO}},}\ }\href {\doibase 10.1063/1.1586466} {\bibfield  {journal}
  {\bibinfo  {journal} {Journal of Applied Physics}\ }\textbf {\bibinfo
  {volume} {94}},\ \bibinfo {pages} {968} (\bibinfo {year} {2003})}\BibitemShut
  {NoStop}%
\bibitem [{\citenamefont {Warner}\ \emph {et~al.}(1967)\citenamefont {Warner},
  \citenamefont {Onoe},\ and\ \citenamefont {Coquin}}]{Warner67jasa}%
  \BibitemOpen
  \bibfield  {author} {\bibinfo {author} {\bibfnamefont {A.~W.}\ \bibnamefont
  {Warner}}, \bibinfo {author} {\bibfnamefont {M.}~\bibnamefont {Onoe}}, \ and\
  \bibinfo {author} {\bibfnamefont {G.~A.}\ \bibnamefont {Coquin}},\ }\bibfield
   {title} {\enquote {\bibinfo {title} {Determination of {Elastic} and
  {Piezoelectric} {Constants} for {Crystals} in {Class} (3m)},}\ }\href
  {\doibase 10.1121/1.1910709} {\bibfield  {journal} {\bibinfo  {journal} {The
  Journal of the Acoustical Society of America}\ }\textbf {\bibinfo {volume}
  {42}},\ \bibinfo {pages} {1223} (\bibinfo {year} {1967})}\BibitemShut
  {NoStop}%
\bibitem [{\citenamefont {Strauch}(2011)}]{GaN-Poisson+Youngs}%
  \BibitemOpen
  \bibfield  {author} {\bibinfo {author} {\bibfnamefont {D.}~\bibnamefont
  {Strauch}},\ }\enquote {\bibinfo {title} {Gan: Poisson ratio, young's
  modulus, bulk modulus},}\ in\ \href {\doibase 10.1007/978-3-642-14148-5_226}
  {\emph {\bibinfo {booktitle} {New Data and Updates for IV-IV, III-V, II-VI
  and I-VII Compounds, their Mixed Crystals and Diluted Magnetic
  Semiconductors}}},\ \bibinfo {editor} {edited by\ \bibinfo {editor}
  {\bibfnamefont {U.}~\bibnamefont {R{\"o}ssler}}}\ (\bibinfo  {publisher}
  {Springer Berlin Heidelberg},\ \bibinfo {address} {Berlin, Heidelberg},\
  \bibinfo {year} {2011})\ pp.\ \bibinfo {pages} {409--412}\BibitemShut
  {NoStop}%
\bibitem [{\citenamefont {Flannery}\ \emph {et~al.}(2003)\citenamefont
  {Flannery}, \citenamefont {Whitfield},\ and\ \citenamefont
  {Jackman}}]{Flannery03sst}%
  \BibitemOpen
  \bibfield  {author} {\bibinfo {author} {\bibfnamefont {C.~M.}\ \bibnamefont
  {Flannery}}, \bibinfo {author} {\bibfnamefont {M.~D.}\ \bibnamefont
  {Whitfield}}, \ and\ \bibinfo {author} {\bibfnamefont {R.~B.}\ \bibnamefont
  {Jackman}},\ }\bibfield  {title} {\enquote {\bibinfo {title} {Acoustic wave
  properties of {CVD} diamond},}\ }\href {\doibase 10.1088/0268-1242/18/3/313}
  {\bibfield  {journal} {\bibinfo  {journal} {Semiconductor Science and
  Technology}\ }\textbf {\bibinfo {volume} {18}},\ \bibinfo {pages} {S86}
  (\bibinfo {year} {2003})}\BibitemShut {NoStop}%
\bibitem [{\citenamefont {Auld}(1990)}]{Auld1973}%
  \BibitemOpen
  \bibfield  {author} {\bibinfo {author} {\bibfnamefont {B.~A.}\ \bibnamefont
  {Auld}},\ }\href@noop {} {\emph {\bibinfo {title} {Acoustic fields and waves
  in solids}}},\ Vol.~\bibinfo {volume} {2}\ (\bibinfo  {publisher} {Krieger},\
  \bibinfo {year} {1990})\BibitemShut {NoStop}%
\bibitem [{\citenamefont {Hopcroft}\ \emph {et~al.}(2010)\citenamefont
  {Hopcroft}, \citenamefont {Nix},\ and\ \citenamefont
  {Kenny}}]{Hopcroft10jms}%
  \BibitemOpen
  \bibfield  {author} {\bibinfo {author} {\bibfnamefont {M.~A.}\ \bibnamefont
  {Hopcroft}}, \bibinfo {author} {\bibfnamefont {W.~D.}\ \bibnamefont {Nix}}, \
  and\ \bibinfo {author} {\bibfnamefont {T.~W.}\ \bibnamefont {Kenny}},\
  }\bibfield  {title} {\enquote {\bibinfo {title} {What is the {Young}'s
  {Modulus} of {Silicon}?}}\ }\href {\doibase 10.1109/JMEMS.2009.2039697}
  {\bibfield  {journal} {\bibinfo  {journal} {Journal of Microelectromechanical
  Systems}\ }\textbf {\bibinfo {volume} {19}},\ \bibinfo {pages} {229}
  (\bibinfo {year} {2010})}\BibitemShut {NoStop}%
\bibitem [{\citenamefont {Hughes}(1972)}]{Hughes1972}%
  \BibitemOpen
  \bibfield  {author} {\bibinfo {author} {\bibfnamefont {A.~J.}\ \bibnamefont
  {Hughes}},\ }\bibfield  {title} {\enquote {\bibinfo {title} {{Elastic Surface
  Wave Guidance by ($\Delta v/v$) Effect Guidance Structures}},}\ }\href
  {\doibase 10.1063/1.1661562} {\bibfield  {journal} {\bibinfo  {journal} {J.
  Appl. Phys.}\ }\textbf {\bibinfo {volume} {43}},\ \bibinfo {pages} {2569}
  (\bibinfo {year} {1972})}\BibitemShut {NoStop}%
\bibitem [{\citenamefont {Shen}\ \emph {et~al.}(2017)\citenamefont {Shen},
  \citenamefont {Han}, \citenamefont {Zou},\ and\ \citenamefont
  {Tang}}]{shen2017rosi}%
  \BibitemOpen
  \bibfield  {author} {\bibinfo {author} {\bibfnamefont {Z.}~\bibnamefont
  {Shen}}, \bibinfo {author} {\bibfnamefont {X.}~\bibnamefont {Han}}, \bibinfo
  {author} {\bibfnamefont {C.-L.}\ \bibnamefont {Zou}}, \ and\ \bibinfo
  {author} {\bibfnamefont {H.~X.}\ \bibnamefont {Tang}},\ }\bibfield  {title}
  {\enquote {\bibinfo {title} {Phase sensitive imaging of 10 ghz vibrations in
  an aln microdisk resonator},}\ }\href {\doibase 10.1063/1.4995008} {\bibfield
   {journal} {\bibinfo  {journal} {Review of Scientific Instruments}\ }\textbf
  {\bibinfo {volume} {88}},\ \bibinfo {pages} {123709} (\bibinfo {year}
  {2017})}\BibitemShut {NoStop}%
\bibitem [{\citenamefont {Tiersten}(1969)}]{Tiersten1969}%
  \BibitemOpen
  \bibfield  {author} {\bibinfo {author} {\bibfnamefont {H.}~\bibnamefont
  {Tiersten}},\ }\bibfield  {title} {\enquote {\bibinfo {title} {Elastic
  surface waves guided by thin films},}\ }\href {\doibase 10.1063/1.1657463}
  {\bibfield  {journal} {\bibinfo  {journal} {Journal of Applied Physics}\
  }\textbf {\bibinfo {volume} {40}},\ \bibinfo {pages} {770} (\bibinfo {year}
  {1969})}\BibitemShut {NoStop}%
\bibitem [{\citenamefont {Li}\ \emph {et~al.}(1977)\citenamefont {Li},
  \citenamefont {Oliner},\ and\ \citenamefont {Bertoni}}]{Microwave-I}%
  \BibitemOpen
  \bibfield  {author} {\bibinfo {author} {\bibfnamefont {R.}~\bibnamefont
  {Li}}, \bibinfo {author} {\bibfnamefont {A.}~\bibnamefont {Oliner}}, \ and\
  \bibinfo {author} {\bibfnamefont {H.}~\bibnamefont {Bertoni}},\ }\bibfield
  {title} {\enquote {\bibinfo {title} {Microwave network analyses of surface
  acoustic waveguides. i-flat overlay guides},}\ }\href {\doibase
  10.1109/T-SU.1977.30915} {\bibfield  {journal} {\bibinfo  {journal} {IEEE
  Transactions on Sonics Ultrasonics}\ }\textbf {\bibinfo {volume} {24}},\
  \bibinfo {pages} {66} (\bibinfo {year} {1977})}\BibitemShut {NoStop}%
\bibitem [{\citenamefont {Markman}\ \emph {et~al.}(1977)\citenamefont
  {Markman}, \citenamefont {Li}, \citenamefont {Oliner},\ and\ \citenamefont
  {Bertoni}}]{Microwave-II}%
  \BibitemOpen
  \bibfield  {author} {\bibinfo {author} {\bibfnamefont {S.}~\bibnamefont
  {Markman}}, \bibinfo {author} {\bibfnamefont {R.~C.}\ \bibnamefont {Li}},
  \bibinfo {author} {\bibfnamefont {A.}~\bibnamefont {Oliner}}, \ and\ \bibinfo
  {author} {\bibfnamefont {H.}~\bibnamefont {Bertoni}},\ }\bibfield  {title}
  {\enquote {\bibinfo {title} {Microwave network analyses of surface acoustic
  waveguides. ii-rectangular ridge guides},}\ }\href {\doibase
  10.1109/T-SU.1977.30916} {\bibfield  {journal} {\bibinfo  {journal} {IEEE
  Transactions on Sonics Ultrasonics}\ }\textbf {\bibinfo {volume} {24}},\
  \bibinfo {pages} {79} (\bibinfo {year} {1977})}\BibitemShut {NoStop}%
\bibitem [{\citenamefont {Love}(1911)}]{Love1911}%
  \BibitemOpen
  \bibfield  {author} {\bibinfo {author} {\bibfnamefont {A.}~\bibnamefont
  {Love}},\ }\bibfield  {title} {\enquote {\bibinfo {title} {Some problems of
  geodynamics},}\ }\href {http://adsabs.harvard.edu/abs/1911spge.book.....L}
  {\bibfield  {journal} {\bibinfo  {journal} {Some Problems of Geodynamics
  Publisher: Cambridge University Press, Cambridge}\ } (\bibinfo {year}
  {1911})}\BibitemShut {NoStop}%
\bibitem [{\citenamefont {Bougrov}\ \emph {et~al.}(2001)\citenamefont
  {Bougrov}, \citenamefont {Levinshtein}, \citenamefont {Rumyantsev},\ and\
  \citenamefont {Zubrilov}}]{GaN-density}%
  \BibitemOpen
  \bibfield  {author} {\bibinfo {author} {\bibfnamefont {V.}~\bibnamefont
  {Bougrov}}, \bibinfo {author} {\bibfnamefont {M.}~\bibnamefont
  {Levinshtein}}, \bibinfo {author} {\bibfnamefont {S.}~\bibnamefont
  {Rumyantsev}}, \ and\ \bibinfo {author} {\bibfnamefont {A.}~\bibnamefont
  {Zubrilov}},\ }\href@noop {} {\emph {\bibinfo {title} {Properties of Advanced
  Semiconductor Materials GaN, AlN, InN, BN, SiC, SiGe}}}\ (\bibinfo
  {publisher} {John Wiley \& Sons, Inc., New York},\ \bibinfo {year} {2001})\
  pp.\ \bibinfo {pages} {1--30}\BibitemShut {NoStop}%
\bibitem [{Sap()}]{Sapphire}%
  \BibitemOpen
  \href@noop {} {\enquote {\bibinfo {title} {{MolTech GmbH website} sapphire
  al2o3},}\ }\bibinfo {howpublished}
  {\url{http://www.mt-berlin.com/frames_cryst/descriptions/sapphire.htm}},\
  \bibinfo {note} {accessed: 2016-07-17}\BibitemShut {NoStop}%
\bibitem [{\citenamefont {Rayleigh}(1885)}]{Rayleigh1885}%
  \BibitemOpen
  \bibfield  {author} {\bibinfo {author} {\bibfnamefont {L.}~\bibnamefont
  {Rayleigh}},\ }\bibfield  {title} {\enquote {\bibinfo {title} {On waves
  propagated along the plane surface of an elastic solid},}\ }\href
  {http://onlinelibrary.wiley.com/doi/10.1112/plms/s1-17.1.4/full} {\bibfield
  {journal} {\bibinfo  {journal} {Proceedings of the London Mathematical
  Society}\ }\textbf {\bibinfo {volume} {1}},\ \bibinfo {pages} {4} (\bibinfo
  {year} {1885})}\BibitemShut {NoStop}%
\bibitem [{\citenamefont {Hill}\ \emph {et~al.}(2007)\citenamefont {Hill},
  \citenamefont {Oei}, \citenamefont {Smalbrugge}, \citenamefont {Zhu},
  \citenamefont {De~Vries}, \citenamefont {Van~Veldhoven}, \citenamefont
  {Van~Otten}, \citenamefont {Eijkemans}, \citenamefont {Turkiewicz},
  \citenamefont {De~Waardt} \emph {et~al.}}]{Mode-area-optics}%
  \BibitemOpen
  \bibfield  {author} {\bibinfo {author} {\bibfnamefont {M.~T.}\ \bibnamefont
  {Hill}}, \bibinfo {author} {\bibfnamefont {Y.-S.}\ \bibnamefont {Oei}},
  \bibinfo {author} {\bibfnamefont {B.}~\bibnamefont {Smalbrugge}}, \bibinfo
  {author} {\bibfnamefont {Y.}~\bibnamefont {Zhu}}, \bibinfo {author}
  {\bibfnamefont {T.}~\bibnamefont {De~Vries}}, \bibinfo {author}
  {\bibfnamefont {P.~J.}\ \bibnamefont {Van~Veldhoven}}, \bibinfo {author}
  {\bibfnamefont {F.~W.}\ \bibnamefont {Van~Otten}}, \bibinfo {author}
  {\bibfnamefont {T.~J.}\ \bibnamefont {Eijkemans}}, \bibinfo {author}
  {\bibfnamefont {J.~P.}\ \bibnamefont {Turkiewicz}}, \bibinfo {author}
  {\bibfnamefont {H.}~\bibnamefont {De~Waardt}},  \emph {et~al.},\ }\bibfield
  {title} {\enquote {\bibinfo {title} {Lasing in metallic-coated
  nanocavities},}\ }\href {\doibase 10.1038/nphoton.2007.171} {\bibfield
  {journal} {\bibinfo  {journal} {Nature Photonics}\ }\textbf {\bibinfo
  {volume} {1}},\ \bibinfo {pages} {589} (\bibinfo {year} {2007})}\BibitemShut
  {NoStop}%
\bibitem [{\citenamefont {Oulton}\ \emph {et~al.}(2008)\citenamefont {Oulton},
  \citenamefont {Bartal}, \citenamefont {Pile},\ and\ \citenamefont
  {Zhang}}]{Mode-area-2008}%
  \BibitemOpen
  \bibfield  {author} {\bibinfo {author} {\bibfnamefont {R.}~\bibnamefont
  {Oulton}}, \bibinfo {author} {\bibfnamefont {G.}~\bibnamefont {Bartal}},
  \bibinfo {author} {\bibfnamefont {D.}~\bibnamefont {Pile}}, \ and\ \bibinfo
  {author} {\bibfnamefont {X.}~\bibnamefont {Zhang}},\ }\bibfield  {title}
  {\enquote {\bibinfo {title} {Confinement and propagation characteristics of
  subwavelength plasmonic modes},}\ }\href
  {http://iopscience.iop.org/article/10.1088/1367-2630/10/10/105018/meta#}
  {\bibfield  {journal} {\bibinfo  {journal} {New Journal of Physics}\ }\textbf
  {\bibinfo {volume} {10}},\ \bibinfo {pages} {105018} (\bibinfo {year}
  {2008})}\BibitemShut {NoStop}%
\bibitem [{\citenamefont {Lagasse}\ \emph
  {et~al.}(1973{\natexlab{a}})\citenamefont {Lagasse}, \citenamefont {Mason},\
  and\ \citenamefont {Ash}}]{Lagasse1973}%
  \BibitemOpen
  \bibfield  {author} {\bibinfo {author} {\bibfnamefont {P.~E.}\ \bibnamefont
  {Lagasse}}, \bibinfo {author} {\bibfnamefont {I.~M.}\ \bibnamefont {Mason}},
  \ and\ \bibinfo {author} {\bibfnamefont {E.~A.}\ \bibnamefont {Ash}},\
  }\bibfield  {title} {\enquote {\bibinfo {title} {Acoustic surface
  waveguides---analysis and assessment},}\ }\href {\doibase
  10.1109/TMTT.1973.1127973} {\bibfield  {journal} {\bibinfo  {journal} {IEEE
  Transactions on Microwave Theory and Techniques}\ }\textbf {\bibinfo {volume}
  {21}},\ \bibinfo {pages} {225} (\bibinfo {year}
  {1973}{\natexlab{a}})}\BibitemShut {NoStop}%
\bibitem [{\citenamefont {Reddy}(2006)}]{Reddy2006}%
  \BibitemOpen
  \bibfield  {author} {\bibinfo {author} {\bibfnamefont {J.~N.}\ \bibnamefont
  {Reddy}},\ }\href@noop {} {\emph {\bibinfo {title} {Theory and analysis of
  elastic plates and shells}}}\ (\bibinfo  {publisher} {CRC press},\ \bibinfo
  {year} {2006})\BibitemShut {NoStop}%
\bibitem [{\citenamefont {Sinha}\ and\ \citenamefont
  {Tiersten}(1973)}]{Tiersten1973}%
  \BibitemOpen
  \bibfield  {author} {\bibinfo {author} {\bibfnamefont {B.}~\bibnamefont
  {Sinha}}\ and\ \bibinfo {author} {\bibfnamefont {H.}~\bibnamefont
  {Tiersten}},\ }\bibfield  {title} {\enquote {\bibinfo {title} {Elastic and
  piezoelectric surface waves guided by thin films},}\ }\href {\doibase
  10.1063/1.1662052} {\bibfield  {journal} {\bibinfo  {journal} {Journal of
  Applied Physics}\ }\textbf {\bibinfo {volume} {44}},\ \bibinfo {pages} {4831}
  (\bibinfo {year} {1973})}\BibitemShut {NoStop}%
\bibitem [{\citenamefont {Lagasse}\ \emph
  {et~al.}(1973{\natexlab{b}})\citenamefont {Lagasse}, \citenamefont {Mason},\
  and\ \citenamefont {Ash}}]{Paul1973}%
  \BibitemOpen
  \bibfield  {author} {\bibinfo {author} {\bibfnamefont {P.~E.}\ \bibnamefont
  {Lagasse}}, \bibinfo {author} {\bibfnamefont {I.~M.}\ \bibnamefont {Mason}},
  \ and\ \bibinfo {author} {\bibfnamefont {E.~A.}\ \bibnamefont {Ash}},\
  }\bibfield  {title} {\enquote {\bibinfo {title} {Acoustic surface
  waveguides---analysis and assessment},}\ }\href {\doibase
  10.1109/TMTT.1973.1127973} {\bibfield  {journal} {\bibinfo  {journal} {IEEE
  Transactions on Microwave Theory and Techniques}\ }\textbf {\bibinfo {volume}
  {21}},\ \bibinfo {pages} {225} (\bibinfo {year}
  {1973}{\natexlab{b}})}\BibitemShut {NoStop}%
\bibitem [{\citenamefont {Rayleigh}(1910)}]{WGM-1-Rayleigh}%
  \BibitemOpen
  \bibfield  {author} {\bibinfo {author} {\bibfnamefont {L.}~\bibnamefont
  {Rayleigh}},\ }\bibfield  {title} {\enquote {\bibinfo {title} {Cxii. the
  problem of the whispering gallery},}\ }\href
  {http://www.tandfonline.com/doi/pdf/10.1080/14786441008636993} {\bibfield
  {journal} {\bibinfo  {journal} {The London, Edinburgh, and Dublin
  Philosophical Magazine and Journal of Science}\ }\textbf {\bibinfo {volume}
  {20}},\ \bibinfo {pages} {1001} (\bibinfo {year} {1910})}\BibitemShut
  {NoStop}%
\bibitem [{\citenamefont {Rayleigh}(1914)}]{WGM-2-Rayleigh}%
  \BibitemOpen
  \bibfield  {author} {\bibinfo {author} {\bibfnamefont {L.}~\bibnamefont
  {Rayleigh}},\ }\bibfield  {title} {\enquote {\bibinfo {title} {Ix. further
  applications of bessel's functions of high order to the whispering gallery
  and allied problems},}\ }\href
  {http://www.tandfonline.com/doi/pdf/10.1080/14786440108635067} {\bibfield
  {journal} {\bibinfo  {journal} {The London, Edinburgh, and Dublin
  Philosophical Magazine and Journal of Science}\ }\textbf {\bibinfo {volume}
  {27}},\ \bibinfo {pages} {100} (\bibinfo {year} {1914})}\BibitemShut
  {NoStop}%
\bibitem [{\citenamefont {Oraevsky}(2002)}]{WGM-in-EM-waves}%
  \BibitemOpen
  \bibfield  {author} {\bibinfo {author} {\bibfnamefont {A.~N.}\ \bibnamefont
  {Oraevsky}},\ }\bibfield  {title} {\enquote {\bibinfo {title}
  {Whispering-gallery waves},}\ }\href {\doibase 10.1109/JSTQE.2005.862952}
  {\bibfield  {journal} {\bibinfo  {journal} {Quantum Electronics}\ }\textbf
  {\bibinfo {volume} {32}},\ \bibinfo {pages} {377} (\bibinfo {year}
  {2002})}\BibitemShut {NoStop}%
\bibitem [{\citenamefont {Matsko}\ and\ \citenamefont
  {Ilchenko}(2006)}]{OWGM-1}%
  \BibitemOpen
  \bibfield  {author} {\bibinfo {author} {\bibfnamefont {A.~B.}\ \bibnamefont
  {Matsko}}\ and\ \bibinfo {author} {\bibfnamefont {V.~S.}\ \bibnamefont
  {Ilchenko}},\ }\bibfield  {title} {\enquote {\bibinfo {title} {Optical
  resonators with whispering gallery modes i: basics},}\ }\href
  {https://www.researchgate.net/profile/Andrey_Matsko/publication/3410055_Optical_Resonators_With_Whispering-Gallery_Mode-Part_I/links/02e7e5283baeb47dc2000000/Optical-Resonators-With-Whispering-Gallery-Mode-Part-I.pdf}
  {\bibfield  {journal} {\bibinfo  {journal} {IEEE J. Sel. Top. Quantum
  Electron}\ }\textbf {\bibinfo {volume} {12}},\ \bibinfo {pages} {3} (\bibinfo
  {year} {2006})}\BibitemShut {NoStop}%
\bibitem [{\citenamefont {Ilchenko}\ and\ \citenamefont
  {Matsko}(2006)}]{OWGM-2}%
  \BibitemOpen
  \bibfield  {author} {\bibinfo {author} {\bibfnamefont {V.~S.}\ \bibnamefont
  {Ilchenko}}\ and\ \bibinfo {author} {\bibfnamefont {A.~B.}\ \bibnamefont
  {Matsko}},\ }\bibfield  {title} {\enquote {\bibinfo {title} {Optical
  resonators with whispering-gallery modes-part ii: applications},}\ }\href
  {\doibase 10.1109/JSTQE.2005.862943} {\bibfield  {journal} {\bibinfo
  {journal} {IEEE Journal of selected topics in quantum electronics}\ }\textbf
  {\bibinfo {volume} {12}},\ \bibinfo {pages} {15} (\bibinfo {year}
  {2006})}\BibitemShut {NoStop}%
\bibitem [{\citenamefont {Foreman}\ \emph {et~al.}(2015)\citenamefont
  {Foreman}, \citenamefont {Swaim},\ and\ \citenamefont
  {Vollmer}}]{WGM-applications}%
  \BibitemOpen
  \bibfield  {author} {\bibinfo {author} {\bibfnamefont {M.~R.}\ \bibnamefont
  {Foreman}}, \bibinfo {author} {\bibfnamefont {J.~D.}\ \bibnamefont {Swaim}},
  \ and\ \bibinfo {author} {\bibfnamefont {F.}~\bibnamefont {Vollmer}},\
  }\bibfield  {title} {\enquote {\bibinfo {title} {Whispering gallery mode
  sensors},}\ }\href {\doibase 10.1364/AOP.7.000168} {\bibfield  {journal}
  {\bibinfo  {journal} {Advances in optics and photonics}\ }\textbf {\bibinfo
  {volume} {7}},\ \bibinfo {pages} {168} (\bibinfo {year} {2015})}\BibitemShut
  {NoStop}%
\bibitem [{\citenamefont {Kippenberg}(2004)}]{Mode-Volume}%
  \BibitemOpen
  \bibfield  {author} {\bibinfo {author} {\bibfnamefont {T.~J.~A.}\
  \bibnamefont {Kippenberg}},\ }\emph {\bibinfo {title} {Nonlinear optics in
  ultra-high-Q whispering-gallery optical microcavities}},\ \href
  {http://k-lab.epfl.ch/files/content/sites/klab/files/publications/TJKippenbergThesis.pdf}
  {Ph.D. thesis},\ \bibinfo  {school} {California Institute of Technology}
  (\bibinfo {year} {2004})\BibitemShut {NoStop}%
\bibitem [{\citenamefont {Hsu}\ \emph {et~al.}(2016)\citenamefont {Hsu},
  \citenamefont {Zhen}, \citenamefont {Stone}, \citenamefont {Joannopoulos},\
  and\ \citenamefont {Solja{\v{c}}i{\'c}}}]{Hsu16nrm}%
  \BibitemOpen
  \bibfield  {author} {\bibinfo {author} {\bibfnamefont {C.~W.}\ \bibnamefont
  {Hsu}}, \bibinfo {author} {\bibfnamefont {B.}~\bibnamefont {Zhen}}, \bibinfo
  {author} {\bibfnamefont {A.~D.}\ \bibnamefont {Stone}}, \bibinfo {author}
  {\bibfnamefont {J.~D.}\ \bibnamefont {Joannopoulos}}, \ and\ \bibinfo
  {author} {\bibfnamefont {M.}~\bibnamefont {Solja{\v{c}}i{\'c}}},\ }\bibfield
  {title} {\enquote {\bibinfo {title} {Bound states in the continuum},}\ }\href
  {\doibase 10.1063/1.3467145} {\bibfield  {journal} {\bibinfo  {journal}
  {Nature Reviews Materials}\ }\textbf {\bibinfo {volume} {1}},\ \bibinfo
  {pages} {1} (\bibinfo {year} {2016})}\BibitemShut {NoStop}%
\bibitem [{\citenamefont {Chen}\ \emph {et~al.}(2016)\citenamefont {Chen},
  \citenamefont {Shen}, \citenamefont {Xiong}, \citenamefont {Dong},
  \citenamefont {Zou},\ and\ \citenamefont {Guo}}]{Chen16njp}%
  \BibitemOpen
  \bibfield  {author} {\bibinfo {author} {\bibfnamefont {Y.}~\bibnamefont
  {Chen}}, \bibinfo {author} {\bibfnamefont {Z.}~\bibnamefont {Shen}}, \bibinfo
  {author} {\bibfnamefont {X.}~\bibnamefont {Xiong}}, \bibinfo {author}
  {\bibfnamefont {C.-H.}\ \bibnamefont {Dong}}, \bibinfo {author}
  {\bibfnamefont {C.-L.}\ \bibnamefont {Zou}}, \ and\ \bibinfo {author}
  {\bibfnamefont {G.-C.}\ \bibnamefont {Guo}},\ }\bibfield  {title} {\enquote
  {\bibinfo {title} {Mechanical bound state in the continuum for optomechanical
  microresonators},}\ }\href
  {https://iopscience.iop.org/article/10.1088/1367-2630/18/6/063031} {\bibfield
   {journal} {\bibinfo  {journal} {New Journal of Physics}\ }\textbf {\bibinfo
  {volume} {18}},\ \bibinfo {pages} {063031} (\bibinfo {year}
  {2016})}\BibitemShut {NoStop}%
\bibitem [{\citenamefont {Yariv}(1973)}]{Couple-1973}%
  \BibitemOpen
  \bibfield  {author} {\bibinfo {author} {\bibfnamefont {A.}~\bibnamefont
  {Yariv}},\ }\bibfield  {title} {\enquote {\bibinfo {title} {Coupled-mode
  theory for guided-wave optics},}\ }\href {\doibase 10.1109/JQE.1973.1077767}
  {\bibfield  {journal} {\bibinfo  {journal} {IEEE Journal of Quantum
  Electronics}\ }\textbf {\bibinfo {volume} {9}},\ \bibinfo {pages} {919}
  (\bibinfo {year} {1973})}\BibitemShut {NoStop}%
\bibitem [{\citenamefont {Hardy}\ and\ \citenamefont
  {Streifer}(1985)}]{Couple-1985}%
  \BibitemOpen
  \bibfield  {author} {\bibinfo {author} {\bibfnamefont {A.}~\bibnamefont
  {Hardy}}\ and\ \bibinfo {author} {\bibfnamefont {W.}~\bibnamefont
  {Streifer}},\ }\bibfield  {title} {\enquote {\bibinfo {title} {Coupled mode
  theory of parallel waveguides},}\ }\href {\doibase 10.1109/JLT.1985.1074291}
  {\bibfield  {journal} {\bibinfo  {journal} {Journal of lightwave technology}\
  }\textbf {\bibinfo {volume} {3}},\ \bibinfo {pages} {1135} (\bibinfo {year}
  {1985})}\BibitemShut {NoStop}%
\bibitem [{\citenamefont {Haus}\ \emph {et~al.}(1987)\citenamefont {Haus},
  \citenamefont {Huang}, \citenamefont {Kawakami},\ and\ \citenamefont
  {Whitaker}}]{Couple-1994}%
  \BibitemOpen
  \bibfield  {author} {\bibinfo {author} {\bibfnamefont {H.}~\bibnamefont
  {Haus}}, \bibinfo {author} {\bibfnamefont {W.}~\bibnamefont {Huang}},
  \bibinfo {author} {\bibfnamefont {S.}~\bibnamefont {Kawakami}}, \ and\
  \bibinfo {author} {\bibfnamefont {N.}~\bibnamefont {Whitaker}},\ }\bibfield
  {title} {\enquote {\bibinfo {title} {Coupled-mode theory of optical
  waveguides},}\ }\href {\doibase 10.1109/JLT.1987.1075416} {\bibfield
  {journal} {\bibinfo  {journal} {Journal of Lightwave Technology}\ }\textbf
  {\bibinfo {volume} {5}},\ \bibinfo {pages} {16} (\bibinfo {year}
  {1987})}\BibitemShut {NoStop}%
\bibitem [{\citenamefont {Shen}\ \emph {et~al.}(2019)\citenamefont {Shen},
  \citenamefont {Fu}, \citenamefont {Cheng}, \citenamefont {Townley},
  \citenamefont {Zou},\ and\ \citenamefont {Tang}}]{shen2019apl}%
  \BibitemOpen
  \bibfield  {author} {\bibinfo {author} {\bibfnamefont {Z.}~\bibnamefont
  {Shen}}, \bibinfo {author} {\bibfnamefont {W.}~\bibnamefont {Fu}}, \bibinfo
  {author} {\bibfnamefont {R.}~\bibnamefont {Cheng}}, \bibinfo {author}
  {\bibfnamefont {H.}~\bibnamefont {Townley}}, \bibinfo {author} {\bibfnamefont
  {C.-L.}\ \bibnamefont {Zou}}, \ and\ \bibinfo {author} {\bibfnamefont
  {H.~X.}\ \bibnamefont {Tang}},\ }\bibfield  {title} {\enquote {\bibinfo
  {title} {Polarization mode hybridization and conversion in phononic wire
  waveguides},}\ }\href {\doibase 10.1063/1.5120844} {\bibfield  {journal}
  {\bibinfo  {journal} {Applied Physics Letters}\ }\textbf {\bibinfo {volume}
  {115}},\ \bibinfo {pages} {201901} (\bibinfo {year} {2019})}\BibitemShut
  {NoStop}%
\bibitem [{\citenamefont {Dahmani}\ \emph {et~al.}(2020)\citenamefont
  {Dahmani}, \citenamefont {Sarabalis}, \citenamefont {Jiang}, \citenamefont
  {Mayor},\ and\ \citenamefont {Safavi-Naeini}}]{Dahmani2020pral}%
  \BibitemOpen
  \bibfield  {author} {\bibinfo {author} {\bibfnamefont {Y.~D.}\ \bibnamefont
  {Dahmani}}, \bibinfo {author} {\bibfnamefont {C.~J.}\ \bibnamefont
  {Sarabalis}}, \bibinfo {author} {\bibfnamefont {W.}~\bibnamefont {Jiang}},
  \bibinfo {author} {\bibfnamefont {F.~M.}\ \bibnamefont {Mayor}}, \ and\
  \bibinfo {author} {\bibfnamefont {A.~H.}\ \bibnamefont {Safavi-Naeini}},\
  }\bibfield  {title} {\enquote {\bibinfo {title} {Piezoelectric transduction
  of a wavelength-scale mechanical waveguide},}\ }\href {\doibase
  10.1103/PhysRevApplied.13.024069} {\bibfield  {journal} {\bibinfo  {journal}
  {Physical Review Applied}\ }\textbf {\bibinfo {volume} {13}},\ \bibinfo
  {pages} {024069} (\bibinfo {year} {2020})}\BibitemShut {NoStop}%
\bibitem [{\citenamefont {Arnau}\ \emph {et~al.}(2000)\citenamefont {Arnau},
  \citenamefont {Sogorb},\ and\ \citenamefont {Jim{\'e}nez}}]{Arnau2000rosi}%
  \BibitemOpen
  \bibfield  {author} {\bibinfo {author} {\bibfnamefont {A.}~\bibnamefont
  {Arnau}}, \bibinfo {author} {\bibfnamefont {T.}~\bibnamefont {Sogorb}}, \
  and\ \bibinfo {author} {\bibfnamefont {Y.}~\bibnamefont {Jim{\'e}nez}},\
  }\bibfield  {title} {\enquote {\bibinfo {title} {A continuous motional series
  resonant frequency monitoring circuit and a new method of determining
  butterworth--van dyke parameters of a quartz crystal microbalance in fluid
  media},}\ }\href {\doibase 10.1063/1.1150649} {\bibfield  {journal} {\bibinfo
   {journal} {Review of scientific instruments}\ }\textbf {\bibinfo {volume}
  {71}},\ \bibinfo {pages} {2563} (\bibinfo {year} {2000})}\BibitemShut
  {NoStop}%
\bibitem [{\citenamefont {Imboden}\ and\ \citenamefont
  {Mohanty}(2014)}]{Imboden14pr}%
  \BibitemOpen
  \bibfield  {author} {\bibinfo {author} {\bibfnamefont {M.}~\bibnamefont
  {Imboden}}\ and\ \bibinfo {author} {\bibfnamefont {P.}~\bibnamefont
  {Mohanty}},\ }\bibfield  {title} {\enquote {\bibinfo {title} {Dissipation in
  nanoelectromechanical systems},}\ }\href {\doibase
  10.1016/j.physrep.2013.09.003} {\bibfield  {journal} {\bibinfo  {journal}
  {Physics Reports}\ }\bibinfo {series} {Dissipation in nano-electromechanical
  systems},\ \textbf {\bibinfo {volume} {534}},\ \bibinfo {pages} {89}
  (\bibinfo {year} {2014})}\BibitemShut {NoStop}%
\bibitem [{\citenamefont {Safavi-Naeini}\ \emph {et~al.}(2019)\citenamefont
  {Safavi-Naeini}, \citenamefont {Thourhout}, \citenamefont {Baets},\ and\
  \citenamefont {Laer}}]{Safavi-Naeini19o}%
  \BibitemOpen
  \bibfield  {author} {\bibinfo {author} {\bibfnamefont {A.~H.}\ \bibnamefont
  {Safavi-Naeini}}, \bibinfo {author} {\bibfnamefont {D.~V.}\ \bibnamefont
  {Thourhout}}, \bibinfo {author} {\bibfnamefont {R.}~\bibnamefont {Baets}}, \
  and\ \bibinfo {author} {\bibfnamefont {R.~V.}\ \bibnamefont {Laer}},\
  }\bibfield  {title} {\enquote {\bibinfo {title} {Controlling phonons and
  photons at the wavelength scale: integrated photonics meets integrated
  phononics},}\ }\href {\doibase 10.1364/OPTICA.6.000213} {\bibfield  {journal}
  {\bibinfo  {journal} {Optica}\ }\textbf {\bibinfo {volume} {6}},\ \bibinfo
  {pages} {213} (\bibinfo {year} {2019})}\BibitemShut {NoStop}%
\bibitem [{\citenamefont {Payne}\ and\ \citenamefont
  {Lacey}(1994)}]{Payne94oqe}%
  \BibitemOpen
  \bibfield  {author} {\bibinfo {author} {\bibfnamefont {F.~P.}\ \bibnamefont
  {Payne}}\ and\ \bibinfo {author} {\bibfnamefont {J.~P.~R.}\ \bibnamefont
  {Lacey}},\ }\bibfield  {title} {\enquote {\bibinfo {title} {A theoretical
  analysis of scattering loss from planar optical waveguides},}\ }\href
  {\doibase 10.1007/BF00708339} {\bibfield  {journal} {\bibinfo  {journal}
  {Optical and Quantum Electronics}\ }\textbf {\bibinfo {volume} {26}},\
  \bibinfo {pages} {977} (\bibinfo {year} {1994})}\BibitemShut {NoStop}%
\bibitem [{\citenamefont {Hughes}\ \emph {et~al.}(2005)\citenamefont {Hughes},
  \citenamefont {Ramunno}, \citenamefont {Young},\ and\ \citenamefont
  {Sipe}}]{Hughes05prl}%
  \BibitemOpen
  \bibfield  {author} {\bibinfo {author} {\bibfnamefont {S.}~\bibnamefont
  {Hughes}}, \bibinfo {author} {\bibfnamefont {L.}~\bibnamefont {Ramunno}},
  \bibinfo {author} {\bibfnamefont {J.~F.}\ \bibnamefont {Young}}, \ and\
  \bibinfo {author} {\bibfnamefont {J.~E.}\ \bibnamefont {Sipe}},\ }\bibfield
  {title} {\enquote {\bibinfo {title} {Extrinsic {Optical} {Scattering} {Loss}
  in {Photonic} {Crystal} {Waveguides}: {Role} of {Fabrication} {Disorder} and
  {Photon} {Group} {Velocity}},}\ }\href {\doibase
  10.1103/PhysRevLett.94.033903} {\bibfield  {journal} {\bibinfo  {journal}
  {Physical Review Letters}\ }\textbf {\bibinfo {volume} {94}},\ \bibinfo
  {pages} {033903} (\bibinfo {year} {2005})}\BibitemShut {NoStop}%
\bibitem [{\citenamefont {Melati}\ \emph {et~al.}(2014)\citenamefont {Melati},
  \citenamefont {Melloni},\ and\ \citenamefont {Morichetti}}]{Melati14aop}%
  \BibitemOpen
  \bibfield  {author} {\bibinfo {author} {\bibfnamefont {D.}~\bibnamefont
  {Melati}}, \bibinfo {author} {\bibfnamefont {A.}~\bibnamefont {Melloni}}, \
  and\ \bibinfo {author} {\bibfnamefont {F.}~\bibnamefont {Morichetti}},\
  }\bibfield  {title} {\enquote {\bibinfo {title} {Real photonic waveguides:
  guiding light through imperfections},}\ }\href {\doibase
  10.1364/AOP.6.000156} {\bibfield  {journal} {\bibinfo  {journal} {Advances in
  Optics and Photonics}\ }\textbf {\bibinfo {volume} {6}},\ \bibinfo {pages}
  {156} (\bibinfo {year} {2014})}\BibitemShut {NoStop}%
\bibitem [{\citenamefont {Liu}\ \emph {et~al.}(2019)\citenamefont {Liu},
  \citenamefont {Li},\ and\ \citenamefont {Li}}]{Liu19o}%
  \BibitemOpen
  \bibfield  {author} {\bibinfo {author} {\bibfnamefont {Q.}~\bibnamefont
  {Liu}}, \bibinfo {author} {\bibfnamefont {H.}~\bibnamefont {Li}}, \ and\
  \bibinfo {author} {\bibfnamefont {M.}~\bibnamefont {Li}},\ }\bibfield
  {title} {\enquote {\bibinfo {title} {Electromechanical {Brillouin} scattering
  in integrated optomechanical waveguides},}\ }\href {\doibase
  10.1364/OPTICA.6.000778} {\bibfield  {journal} {\bibinfo  {journal} {Optica}\
  }\textbf {\bibinfo {volume} {6}},\ \bibinfo {pages} {778} (\bibinfo {year}
  {2019})}\BibitemShut {NoStop}%
\bibitem [{\citenamefont {Yasumura}\ \emph {et~al.}(2000)\citenamefont
  {Yasumura}, \citenamefont {Stowe}, \citenamefont {Chow}, \citenamefont
  {Pfafman}, \citenamefont {Kenny}, \citenamefont {Stipe},\ and\ \citenamefont
  {Rugar}}]{Yasumura00jms}%
  \BibitemOpen
  \bibfield  {author} {\bibinfo {author} {\bibfnamefont {K.}~\bibnamefont
  {Yasumura}}, \bibinfo {author} {\bibfnamefont {T.}~\bibnamefont {Stowe}},
  \bibinfo {author} {\bibfnamefont {E.}~\bibnamefont {Chow}}, \bibinfo {author}
  {\bibfnamefont {T.}~\bibnamefont {Pfafman}}, \bibinfo {author} {\bibfnamefont
  {T.}~\bibnamefont {Kenny}}, \bibinfo {author} {\bibfnamefont
  {B.}~\bibnamefont {Stipe}}, \ and\ \bibinfo {author} {\bibfnamefont
  {D.}~\bibnamefont {Rugar}},\ }\bibfield  {title} {\enquote {\bibinfo {title}
  {Quality factors in micron- and submicron-thick cantilevers},}\ }\href
  {\doibase 10.1109/84.825786} {\bibfield  {journal} {\bibinfo  {journal}
  {Journal of Microelectromechanical Systems}\ }\textbf {\bibinfo {volume}
  {9}},\ \bibinfo {pages} {117} (\bibinfo {year} {2000})}\BibitemShut {NoStop}%
\bibitem [{\citenamefont {{Olsson III}}\ and\ \citenamefont
  {El-Kady}(2009)}]{OlssonIII2009}%
  \BibitemOpen
  \bibfield  {author} {\bibinfo {author} {\bibfnamefont {R.~H.}\ \bibnamefont
  {{Olsson III}}}\ and\ \bibinfo {author} {\bibfnamefont {I.}~\bibnamefont
  {El-Kady}},\ }\bibfield  {title} {\enquote {\bibinfo {title}
  {{Microfabricated phononic crystal devices and applications}},}\ }\href
  {\doibase 10.1088/0957-0233/20/1/012002} {\bibfield  {journal} {\bibinfo
  {journal} {Meas. Sci. Technol.}\ }\textbf {\bibinfo {volume} {20}},\ \bibinfo
  {pages} {012002} (\bibinfo {year} {2009})}\BibitemShut {NoStop}%
\bibitem [{\citenamefont {Maldovan}(2013)}]{Maldovan2013}%
  \BibitemOpen
  \bibfield  {author} {\bibinfo {author} {\bibfnamefont {M.}~\bibnamefont
  {Maldovan}},\ }\bibfield  {title} {\enquote {\bibinfo {title} {{Sound and
  heat revolutions in phononics}},}\ }\href {\doibase 10.1038/nature12608}
  {\bibfield  {journal} {\bibinfo  {journal} {Nature}\ }\textbf {\bibinfo
  {volume} {503}},\ \bibinfo {pages} {209} (\bibinfo {year}
  {2013})}\BibitemShut {NoStop}%
\bibitem [{\citenamefont {Hatanaka}\ \emph
  {et~al.}(2014{\natexlab{b}})\citenamefont {Hatanaka}, \citenamefont
  {Mahboob}, \citenamefont {Onomitsu},\ and\ \citenamefont
  {Yamaguchi}}]{Hatanaka2014}%
  \BibitemOpen
  \bibfield  {author} {\bibinfo {author} {\bibfnamefont {D.}~\bibnamefont
  {Hatanaka}}, \bibinfo {author} {\bibfnamefont {I.}~\bibnamefont {Mahboob}},
  \bibinfo {author} {\bibfnamefont {K.}~\bibnamefont {Onomitsu}}, \ and\
  \bibinfo {author} {\bibfnamefont {H.}~\bibnamefont {Yamaguchi}},\ }\bibfield
  {title} {\enquote {\bibinfo {title} {{Phonon waveguides for electromechanical
  circuits}},}\ }\href {\doibase 10.1038/nnano.2014.107} {\bibfield  {journal}
  {\bibinfo  {journal} {Nat. Nanotechnol.}\ }\textbf {\bibinfo {volume} {9}},\
  \bibinfo {pages} {520} (\bibinfo {year} {2014}{\natexlab{b}})}\BibitemShut
  {NoStop}%
\bibitem [{\citenamefont {Zou}\ \emph {et~al.}(2016)\citenamefont {Zou},
  \citenamefont {Fu},\ and\ \citenamefont {Tang}}]{Zou2016}%
  \BibitemOpen
  \bibfield  {author} {\bibinfo {author} {\bibfnamefont {C.-L.}\ \bibnamefont
  {Zou}}, \bibinfo {author} {\bibfnamefont {W.}~\bibnamefont {Fu}}, \ and\
  \bibinfo {author} {\bibfnamefont {H.~X.}\ \bibnamefont {Tang}},\ }\bibfield
  {title} {\enquote {\bibinfo {title} {Surface acoustic gyroscope},}\
  }\href@noop {} {\bibfield  {journal} {\bibinfo  {journal} {In Preparation}\ }
  (\bibinfo {year} {2016})}\BibitemShut {NoStop}%
\bibitem [{\citenamefont {Fu}\ \emph {et~al.}(2015)\citenamefont {Fu},
  \citenamefont {Shu}, \citenamefont {Zhang}, \citenamefont {Dong},
  \citenamefont {Zou},\ and\ \citenamefont {Guo}}]{Fu2015}%
  \BibitemOpen
  \bibfield  {author} {\bibinfo {author} {\bibfnamefont {W.}~\bibnamefont
  {Fu}}, \bibinfo {author} {\bibfnamefont {F.-J.}\ \bibnamefont {Shu}},
  \bibinfo {author} {\bibfnamefont {Y.-L.}\ \bibnamefont {Zhang}}, \bibinfo
  {author} {\bibfnamefont {C.-H.}\ \bibnamefont {Dong}}, \bibinfo {author}
  {\bibfnamefont {C.-L.}\ \bibnamefont {Zou}}, \ and\ \bibinfo {author}
  {\bibfnamefont {G.-C.}\ \bibnamefont {Guo}},\ }\bibfield  {title} {\enquote
  {\bibinfo {title} {{Integrated optical circulator by stimulated Brillouin
  scattering induced non-reciprocal phase shift}},}\ }\href {\doibase
  10.1364/OE.23.025118} {\bibfield  {journal} {\bibinfo  {journal} {Opt.
  Express}\ }\textbf {\bibinfo {volume} {23}},\ \bibinfo {pages} {025118}
  (\bibinfo {year} {2015})}\BibitemShut {NoStop}%
\end{thebibliography}%

\end{document}